\documentclass[aps,nofootinbib,endfloatsamsmath,amsmath,amssymb,twocolumn,superscriptaddress]{revtex4-1}

\ifdefined\isdraft
\usepackage[draft]{graphicx}
\else
\usepackage{graphicx}
\fi
\usepackage{dcolumn}
\usepackage{bm}
\usepackage{bm, color}
\usepackage{lineno}
\usepackage{verbatim}

\usepackage[unicode=true,pdfusetitle,
 bookmarks=true,bookmarksnumbered=true,bookmarksopen=true,bookmarksopenlevel=1,
 breaklinks=true,pdfborder={0 0 1},backref=false,colorlinks=true]
 {hyperref}
\hypersetup{
 linkcolor=red, citecolor=black, urlcolor=blue, filecolor=blue, pdfpagelayout=OneColumn, pdfnewwindow=true, pdfstartview=XYZ, plainpages=false}
\usepackage{breakurl}

\newcommand{\bok}{\boldsymbol{k}}
\newcommand{\boR}{\boldsymbol{R}}
\newcommand{\bor}{\boldsymbol{r}}
\newcommand{\bos}{\boldsymbol{s}}
\newcommand{\boc}{\boldsymbol{c}}
\newcommand{\boq}{\boldsymbol{q}}
\newcommand{\boG}{\boldsymbol{G}}
\newcommand{\boA}{\boldsymbol{A}}
\newcommand{\boF}{\boldsymbol{F}}
\newcommand{\bobek}{\boldsymbol{\beta}_{\boldsymbol{k}}}
\newcommand{\mex}{\mathrm{e}}
\newcommand{\blambda}{\overline{\lambda}}

\begin{document}

\title{Symmetry Conserving Maximally Projected Wannier Functions}

\author{K. Koepernik}
\affiliation{Institute for Theoretical Solid State Physics, IFW Dresden, 01069 Dresden, Germany}
\author{O. Janson}
\affiliation{Institute for Theoretical Solid State Physics, IFW Dresden, 01069 Dresden, Germany}
\author{Yan Sun}
\affiliation{Max Planck Institute for Chemical Physics of Solids, 01187 Dresden, Germany}
\author{J. van den Brink}
\affiliation{Institute for Theoretical Solid State Physics, IFW
  Dresden, 01069 Dresden, Germany}
\affiliation{Institute for Theoretical Physics and
  W{\"u}rzburg-Dresden Cluster of Excellence ct.qmat, Technische
  Universit{\"a}t Dresden, 01069 Dresden, Germany}

\date{\today}

\begin{abstract}

To obtain a local description from highly accurate density functional theory codes that are based on modified plane wave bases, a transformation to a local orthonormal Wannier function basis is required.
In order to do so while enforcing the constraints of the space group symmetry the Symmetry Conserving Maximally Projected Wannier Functions (SCMPWF) approach has been implemented in the Full-Potential-Local-Orbital code, FPLO.
SCMPWFs represent the zeroth order approximation to maximally localized Wannier functions, projecting a subset of wave functions onto a set of suitably chosen local trial-functions with subsequent orthonormalization.
The particular nature of the local orbitals in FPLO make them an ideal set of projectors, since they are constructed to be a chemical basis. While in many cases projection onto the FPLO basis orbitals is sufficient, the option is there to choose particular local linear combinations as projectors, in order to treat cases of bond centered Wannier functions. 
This choice turns out to lead to highly localized Wannier functions, which obey the space group symmetry of the crystal by construction. 
Furthermore we discuss the interplay of the Berry connection and position operator and especially its possible approximation, symmetries and the optimal choice of Bloch sum phase gauge in cases where the basis is not explicitly known. 
We also introduce various features which are accessible via the FPLO implementation of SCMPWFs, discuss and compare performance and provide example applications.

\end{abstract}

\maketitle
\tableofcontents

\section{Introduction}
\label{sec:intro}

\subsection{Wannier functions and FPLO}
Natural basis states for the treatment of extended systems are
extended states which carry a pseudo-momentum quantum number. On the
other hand, chemistry is more intuitively described by local basis
states, which resemble atomic orbitals, while extended model
Hamiltonians are often built from (implicit) atomic entities
(tight-binding).  Although several numerical methods exist, which
treat molecules and solids with an atom like basis, many contemporary
highly accurate density functional theory (DFT) codes are based on
modified plane wave bases. To gain back the local description a
transformation to a local orthonormal Wannier function basis is desirable.

The usefulness of Wannier functions has gained new momentum in the
context of topological properties of extended systems, for whose
determination a downfolding of the whole system's Hamiltonian onto a
smaller model is very helpful. While smaller models reduce the often
heavy burden of calculating topological quantities, numerical
accuracy, especially symmetry conservation, becomes an important issue
as well, in particular for higher order transport properties, which
involve taking derivatives (e.g.~Berry curvature dipole) or for
tensorial properties.

Wannier functions (WFs) can be defined in many ways due to the gauge
freedom in choosing a unitary transformation among the wave functions
to be Wannierized. One way of fixing the gauge, now widely used, is
the requirement of maximum localization\cite{Mar97,Sou01,Piz20} (ignoring the
constraints of the space group symmetry). This is an rather intricate
algorithm which becomes even more complicated, if space group symmetry
is added\cite{Sak13,Run14}.

Extending the ideas of Ref.~\onlinecite{Ku02} to retain full symmetry
information we implemented a different scheme into the
Full-Potential-Local-Orbital code, FPLO\cite{Koepernik99}, as outlined
in Ref.~\onlinecite{Eschrig09}. This scheme, which we call Symmetry
Conserving Maximally Projected Wannier Functions (SCMPWF), basically
represents the zeroth order approximation of the maximally localized
approach\cite{Mar97}, in projecting a subset of wave functions onto a set of
suitably chosen local trial-functions with subsequent
orthonormalization.

The particular nature of the local orbitals in FPLO make them an ideal
set of projectors, since they are constructed to be a ``chemical''
basis. In many cases projection onto the FPLO basis orbitals is
sufficient, while the option is there to choose particular local linear
combinations (molecular orbitals [MO]) as projectors, in order to
treat cases of bond centered Wannier functions. It turns out that our
choice leads to highly localized Wannier functions, which obey the
space group symmetry of the crystal by construction. In cases where
the resulting WFs are not well-localized it is due to a bad choice and
can be fixed.

Many similar projective schemes have been proposed before, often
including localization criteria (Refs.~\onlinecite{Spo94,Qia08} and
Ref.~\onlinecite{Mar12} and references therein) as well as including
symmetry considerations\cite{Smi01}.
The main difference to our method is that ours is straightforward and
simple as the approach of Ref.~\onlinecite{Ku02}. On top it conserves
symmetry and is generally applicable to entangled bands with very few
adjustable parameters.

The main challenge of our approach is that the user has to decide
where the WFs shall sit and which local symmetry they shall
have. However, this is often exactly what one wants to do, when
constructing a Wannier model for a given subset of bands.  The strong
connection of the FPLO basis and the resulting WFs and the
comparatively small basis size allows for automatic Wannierization,
which is nothing but constructing the L{\"o}wdin orthogonalized FPLO
basis.  In principle, everything which can be expressed in WFs could
be transformed back into the non-orthogonal FPLO basis, which opens
the possibility to circumvent WFs altogether, although this is less
efficient.

The method described here is used in conjunction with the FPLO density
functional (DFT) code, but it can also be used whenever a Hamiltonian
is given in a local basis (tight-binding [TB] models ), even if
the basis is not explicitly known. This can for instance be helpful
to further reduce the dimensionality of a TB-model.

The structure of a local basis code also requires modifications to the
way the Berry connection/curvature need to be calculated. The
connection of these quantities to the position operator was discussed
in great detail in Ref.~\cite{Blo62} before the recognition of their
fundamental nature for topological properties of mater.  The local
basis formulation of the Berry operators presented here offers
insights into the correct way of neglecting the position operator
matrix elements and into the correct phase choice of Bloch sums,
especially when the Wannier basis is not explicitly known, in particular for tight-binding models.

\subsection{Outline}
\label{sec:outline}

This paper thus serves as a reference to the Wannier module of FPLO and describes the algorithm underlying it as well as the post-processing tools, including semi-infinite slab calculations, Weyl point search and calculation of Z$_2$ topological indices.
In Sec.~\ref{sec:method} we outline in detail the construction of Symmetry Conserving Maximally
Projected Wannier Functions. In
Sec.~\ref{sec:methodtopo} the position operator matrix elements are
discussed. Sec.~\ref{sec:features}
introduces various features which are accessible via the FPLO
implementation of SCMPWF and Sec.~\ref{sec:performance} discusses and
compares performance and gives example applications, in particular on the construction of tight-binding models and band disentangling,  the calculation of the Berry curvature and anomalous Hall conductivity.


\section{Formalism and Method}
\label{sec:method}

\subsection{Wannier Orbitals in FPLO}\label{sec:methodwf}

The FPLO basis consists of local atom like orbitals
$\Phi_{\boR\bos\nu}(\bor)$ localized at site
$\bos$ in unit cell $\boR$, having atom like
quantum numbers (qns) $\nu=nlm\sigma$ without spin orbit coupling and
$\nu=nlj\mu$ in 4-component full-relativistic mode, where $n$ is the
main quantum number, $lm$ are the qns. of real spherical harmonics,
$\sigma$ is the spin index and $lj\mu$ are the qns. of a standard
(complex) spherical spinor (see Sec.~\ref{sec:FPLObasis}).

From these orbitals Bloch sums can be formed
\begin{equation}
\Phi_{\bos\nu}^{\bok}\left(\bor\right)=\frac{1}{\sqrt{N}}\sum_{\boR}\mex^{i\bok\left(\boR+\lambda
  \bos\right)}\Phi_{\boR\bos\nu}\left(\bor\right)\label{eq:blochsum}.
\end{equation}
where $N$ is the number of unit cells in the Born-von-K{\'a}rm{\'a}n
(BvK) torus. The chosen normalization is further discussed in
Sec.~\ref{sec:Blochsums}. The parameter $\lambda$ allows to choose the
phase gauge and is of interest later. The default gauge in FPLO and
hence in constructing the Wannier functions is the \textit{relative}
gauge ($\lambda=1$) in which matrix elements of an operator between
two Bloch sums only depend on the relative vector connecting the two
orbital locations in each term of the sum. In post processing modules
which use the WFs the \textit{periodic} gauge ($\lambda=0$) can be
chosen, in which the Bloch sums are periodic in $\bok$-space.

The FPLO orbitals from different sites are in general non-orthogonal,
which leads to a non-trivial overlap matrix $S^{\bok}$.
This avoids complicated orbital shapes away from the
atomic core region and in fact the SCMPWF transformation is essentially
nothing but the orthogonalization of (a subset of) our local orbitals (LO). 
Introducing the row vector
$\Phi^{\bok}$ of all orbitals and
the matrix of wave function coefficients $C^{\bok}$ the
overlap and Hamiltonian matrices read
\begin{eqnarray}
S^{\bok} & = & \left\langle \Phi^{\bok}\mid\Phi^{\bok}\right\rangle \nonumber\\
H^{\bok} & = & \left\langle
               \Phi^{\bok}\mid\hat{H}\mid\Phi^{\bok}\right\rangle 
\label{eq:hamLOmatrix}
\end{eqnarray}
and the eigenvalue problem becomes
\begin{equation}
  H^{\bok}C^{\bok}=S^{\bok}C^{\bok}\varepsilon^{\bok},\quad C^{\bok+}S^{\bok}C^{\bok}=1
\end{equation}
which gives the row vector of full wave functions
\begin{equation}
  \Psi^{\bok}=\Phi^{\bok}C^{\bok}.
\end{equation}
as a linear combination of the Bloch sums of LOs.
In the following we will often suppress the orbital/WF indices (and
sometimes site indices). The corresponding expressions must then be
understood as sub-blocks of vectors-of-functions or matrices.

In order to obtain Wannier functions one needs to Fourier back transform
the extended wave functions $\Psi^{\bok}$ including
some unitary matrix $\mathcal{U}^{\bok}$ which constitutes a
general gauge choice
\begin{equation}
w_{\boR\boc}=\frac{1}{\sqrt{N}}\sum_{\bok}\mex^{-i\bok(\boR+\lambda\boc)}\left(\Psi^{\bok}\mathcal{U}^{\bok}\right)_{\boc}\label{eq:generalWFbacktransform}
\end{equation}
where the normalization was chosen such that with
$\mathcal{U}_{\boc^{\prime}}^{\bok+}\mathcal{U}_{\boc}^{\bok}=\delta_{\boc^{\prime}\boc}$
also
$\left\langle w_{\boR^{\prime}\boc^{\prime}}\mid
  w_{\boR\boc}\right\rangle
=\delta_{\boR^{\prime}\boR}\delta_{\boc^{\prime}\boc}$
holds.  The phase gauge ($\lambda$-term) is there for convenience and
can be thought of as a part of $\mathcal{U}$.  The difference between
various methods to calculate Wannier functions is the choice of
$\mathcal{U}$. In the following we will describe our choice and argue
why this is a natural and good one.  It is a matter of experience
through many calculations that a certain setup seems to guarantee
localization if done right. Basically, if one knows from band
characters and symmetry conditions which orbitals engender a certain
band complex then the WFs obtained from projection onto these orbitals
will be localized.  In essence, SCMPWF as implemented in FPLO are a
symmetry and chemistry based basis reduction. Note, that the resulting
WFs cannot in general be more localized than the corresponding
projector orbitals, due to orthogonalization tails.

In order to obtain Wannier functions we need to select a physically
motivated subset of wave functions (bands) and construct an unnormalized
Hilbert subspace
\begin{equation}
\mathcal{H}=\sum_{k}\left|\Psi^{\bok}\right\rangle h^{\bok}\left\langle \Psi^{\bok}\right|  \label{eq:hilbertsubspace}
\end{equation}
from them, where $h^{\bok}$ is an energy dependent diagonal matrix
with values between zero and one, which discards the unwanted part of
the band structure. This Hilbert space must contain at least as many
bands as the number of desired Wannier functions. It can contain more
bands, however, of which some get discarded by the projection onto
trial functions by the fact that the overlap of the trial functions with
some wave-functions will be small. 

If properly chosen this Hilbert space is mainly spanned by a well
defined linear combination of a subset of the original LOs. For
generality we consider local linear combinations of LOs, which can be
called molecular orbitals (MOs). Note that orthogonality of the MOs in
this stage is not required.  In detail we construct MOs
\begin{equation}\label{eq:defMO}
\phi_{\boc
  i}=\sum_{\boR\bos}\Phi_{\boR\bos\nu^{\prime}}U_{\boR\bos\nu^{\prime},\boc
  i}  
\end{equation}
sitting at an imagined Wannier center $\boc$, having an index $i$ from
LOs $\Phi_{\boR\bos\nu^{\prime}}$ in the vicinity of $\boc$. In
practice the input $U_{\boR\bos\nu^{\prime},\boc i}$ consists of a few
lines specifying the center $\boc$, the contributing site numbers,
difference vectors from the center to these sites and weights.  in the
simplest case $U_{\boR\bos\nu^{\prime},\boc
  i}=\delta_{\boR+\bos,\boc}\delta_{\nu^{\prime},i}$ for a subset of
orbitals. $U$ can also contain a transformation onto different local
quantization axes in which the LOs are rotated, which
helps to yield WFs adapted to certain local symmetries.

At this step we insist that the set of MOs transforms properly under
the spacegroup.  A space group operation with point group matrix
$\alpha$ and translation $\boldsymbol{\tau}$ is given in Seitz
notation by $g=\left\{\alpha\mid\boldsymbol{\tau}\right\}$. The atom
positions transform as
\begin{equation}
g\left(\boR+\bos\right)=\alpha\left(\boR+\bos\right)+\boldsymbol{\tau}=\alpha\boR+g\bos
\end{equation}
The transformed site can be backfolded onto the original set of sites via
\begin{equation}
  g\bos=\bos_{g}+\boR_{g,s}
\end{equation}
where $\bos_{g}$ is the site in the original set of sites
which $\bos$ gets mapped onto by $g$ and 
$\boR_{g,s}$ is a lattice vector. This backfolding is
needed since Bloch sums of the same site in two different unit cells
can differ in phase and since sites are indices to the Bloch sums.
Hence, consistency requires that we always consider the functions in
the original set of sites. Of course also the Wannier centers
$\boc$ transform this way. Note, that in our formalism the
Wannier centers are not an output of the calculation but an input.
The most general transformation property of local orbitals or MOs is
given by (see Sec.~\ref{sec:spacegroupsym})
\begin{equation}\label{eq:symMO}
  \left\{ \alpha\mid\tau\right\} \phi_{\boc i}=\sum_{i^{\prime}}\phi_{\alpha\boR+\boR_{g,c},\boc_{g}i^{\prime}}D_{\boc,i^{\prime}i}\left(g\right)
\end{equation}
where $D_{\boc,i^{\prime}i}$ is a matrix which mixes the MOs
at center $\boc$ such that MOs at the transformed center $\boc_{g}$ are
obtained. In essence, if a MO at some center is picked, all MOs which
are symmetry related at the same center as well as all equivalent MOs
at symmetry related centers must be included into the set of
projectors.  The sets at related centers can be arbitrarily unitary
mixed.  Knowing the transformation properties of the FPLO orbitals and
the MO matrix $U$ in Eq.~(\ref{eq:defMO}), $D$ in Eq.~(\ref{eq:symMO})
is fully defined.  By constructing the MOs this way we have explicit
information about the symmetry properties of the resulting Wannier
functions.
In most cases the MOs are identical to a subset of our LOs.

Using Eq.~\ref{eq:overlapBlochMO} in Sec.~\ref{sec:Blochsums}, we can
calculate the projection of the Hilbert subspace
Eq.~\ref{eq:hilbertsubspace} onto a MO as
\begin{eqnarray}
\mathcal{H}\left|\phi_{\boldsymbol{R}\boc}\right\rangle 
&=&\sum_{\bok}\left|\Psi^{\bok}\right\rangle h^{\bok}\left\langle
    \Psi^{\bok}\mid\phi_{\boldsymbol{R}\boc}\right\rangle \nonumber\\
&=&\sum_{\bok}\left(\left|\Psi^{\bok}\right\rangle
    h^{\bok}C^{\bok+}S^{\bok}U^{\bok}\frac{1}{\sqrt{N}}\right)_{\boc}\mex^{-i\bok\left(\boR+\lambda\boc\right)}\nonumber\\
&=&\frac{1}{\sqrt{N}}\sum_{\bok}\tilde{w}_{\boc}^{\bok}\mex^{-i\bok\left(\boR+\lambda\boc\right)}  
\label{eq:hilbertprojMO}
\end{eqnarray}
which defines the Bloch sums 
\begin{equation}
\tilde{w}_{\boc{}i}^{\bok}
=\left(\left|\Psi^{\bok}\right\rangle
  h^{\bok}C^{\bok+}S^{\bok}U^{\bok}
\right)_{\boc{}i}\label{eq:rawWF}
\end{equation}
of the raw (not orthonormal) Wannier functions as well as the phase
gauge to be used in the Fourier back transformation into real space.

This simple projection completely defines $\mathcal{U}^{\bok}$
in Eq.~(\ref{eq:generalWFbacktransform}) (besides orthonormality) and is
quite intuitive: define the desired subset of bands by specifying
$h^{\bok}$ and project out the part which has overlap with suitably
selected MOs (trial functions). If these MOs span the most part of
this Hilbert space the result must be the Wannier functions describing
$\mathcal{H}$.  The raw WF contains the weights
$C^{\bok+}S^{\bok}U^{\bok}$ which can be interpreted as the ``square
root'' of the band weights of the MOs in $\Psi^{\bok}$; a form of band
characters (fat bands) of the considered orbitals, only with the phase
information retained. So, we basically sum wave functions according to
MO-weights contained in them.

We apply the L{\"o}wdin procedure to Eq.~(\ref{eq:rawWF})
in order to obtain orthonormal WFs, which are
guaranteed to have the smallest least square deviation from the
raw functions: we divide by the square root of the overlap matrix
of the raw WFs
\begin{equation}
O^{\bok}=\left\langle  \tilde{w}^{\bok}\mid \tilde{w}^{\bok}\right \rangle
= 
U^{\bok+}S^{\bok}C^{\bok}\left(h^{\bok}\right)^{2}C^{\bok+}S^{\bok}U^{\bok}
\label{eq:rawoverlap}
\end{equation}
and obtain orthonormal Bloch sums of maximally projected WFs
\begin{equation}
  w_{\boc i}^{\bok}=\left(\tilde{w}^{\bok}\frac{1}{\sqrt{O^{\bok}}}\right)_{\boc i}.\label{eq:maxprojWFs}
\end{equation}
The inverse square root of the overlap is
calculated using the eigen decomposition
$O^{\bok}Z^{\bok}=Z^{\bok}o^{\bok}$ according to
($O^{\bok})^{-1/2}=Z^{\bok} \frac{1}{\sqrt{o^{\bok}}} Z^{\bok+}$.
Furthermore, since we know the symmetry
transformation $U^{\bok}=U^{\bok\prime}D$ we also know the transformation
property  $O^{\bok}=D^{+}O^{\bok\prime}D$, which gives $Z^{\bok}=D^{+}Z^{\bok\prime}$
and $\left(O^{\bok}\right)^{-\frac{1}{2}}=D^{+}\left(O^{\bok\prime}\right)^{-\frac{1}{2}}D$.
So, the WFs $w_{\boc i}^{\bok}$ transforms as the MOs themselves.
Finally, we Fourier back transform into real space by inserting
Eq.~(\ref{eq:maxprojWFs}) into Eq.~(\ref{eq:hilbertprojMO})
\begin{equation}
w_{\boR\boc i}=\frac{1}{\sqrt{N}}\sum_{\bok}w_{\boc i}^{\bok}\mex^{-i\bok\left(\boR+\lambda\boc\right)}\label{eq:WFinRealSpace}
\end{equation}

In summary our gauge fixing matrix when transforming wave functions
into Wannier functions is obtained by basically picking linear
combinations of wave functions of a Hilbert subspace which have
maximal projection (band weight) for a number of MOs/LOs. The result
then is orthogonalized and Fourier back transformed into real space to
obtain the Wannier functions, which by construction have maximum
resemblance to the MOs. Since the projectors are identical to or are
constructed out of our original LOs, which form the basis of the wave
functions to begin with, the projection result is optimal as compared
to the original idea of Ref.~\cite{Ku02} of projecting onto hand-made
atom like trial functions, which are not a basis of the wave function
formation.

In the extreme case, where we calculate a Wannier basis as large as
the FPLO basis (automatic mode), we can use $\phi=\Phi$ for all orbitals and
$h^{\bok}=1$ (all bands) which gives $U=1$ and with
$C^{\bok}C^{\bok+}=\left(S^{\bok}\right)^{-1}$:
\begin{equation}
  \tilde{w}^{\bok}=\Phi^{\bok}
\end{equation}
which results in the orthonormalized WFs
\begin{equation}
  w^{\bok}=\Phi^{\bok}\frac{1}{\sqrt{\left\langle\Phi^{\bok}\mid\Phi^{\bok}\right\rangle}}  \label{eq:weqphi}
\end{equation}
which is nothing but the L{\"o}wdin orthonormalized FPLO basis. From
this it is also clear why we need to use the same $\lambda$ in the
Fourier back transformation as in the definition of the orbital Bloch
sums Eq.~(\ref{eq:blochsum}). If $\Phi^{\bok}$ were orthonormal the
Fourier back transformation of Eq.~(\ref{eq:blochsum}) would just
yield the (assumed to be orthonormal) local orbitals, i.e.~Wannier
functions.

Eqs.~(\ref{eq:rawWF},\ref{eq:maxprojWFs}) can be written in matrix
form as
\begin{eqnarray}
  w^{\bok}&=&\Psi^{\bok}\mathcal{U}^{\bok}\nonumber\\
&=&\Phi^{\bok}a^{\bok}\label{eq:blochwffromblochLO}
\end{eqnarray}
which emphasizes either the connection to the wave
functions or the basis.
While
\begin{equation}
  \mathcal{U}^{\bok}=h^{\bok}C^{\bok+}S^{\bok}U^{\bok}
\frac{1}{\sqrt{O^{\bok}}}
\end{equation}
 is projective unitary $\mathcal{U}^{\bok+}\mathcal{U}^{\bok}=1$,
\begin{equation}
 a^{\bok}=C^{\bok}\mathcal{U}^{\bok}
\end{equation}
 fulfills $a^{\bok+}S^{\bok}a^{\bok}=1$  since the basis
 has nontrivial overlap.
The Hamiltonian in the Wannier basis then becomes
\begin{eqnarray}
H^{(w)\bok}
&=&\left\langle w^{\bok}\mid\hat{H}\mid w^{\bok}\right\rangle\nonumber\\
&=&\mathcal{U}^{\bok+}\varepsilon^{\bok}\mathcal{U}^{\bok}\nonumber\\
&=&a^{\bok+}H^{\bok}a^{\bok}\label{eq:wanHam}
\end{eqnarray}
with $H^{\bok}$ according to Eq.~(\ref{eq:hamLOmatrix}).
A general operator $\hat{B}$ with matrix elements between LO Bloch
sums $B^{\bok}$ then reads in Wannier basis $B^{(w)\bok}=\left\langle w^{\bok}\mid\hat{B}\mid w^{\bok}\right\rangle =a^{\bok+}B^{\bok}a^{\bok}$.

\subsection{Accuracy and localization}

If the Wannier transformation is well set up the eigenvalues of
Eq.~(\ref{eq:wanHam}) will reproduce the band structure of the chosen
band subspace. In cases of an isolated band complex with a WF basis of the
same dimension the Wannier and FPLO band structure are usually
identical. If band disentangling is needed, the dimension of the
Hilbert subspace is usually larger than the WF basis dimension.
Then $h^{\bok}$ can be chosen such that the agreement of the two
band structures is highly satisfactory for the targeted subset of bands.

For post processing the wannierized operators need to be stored in
their real space representation which is obtained from
\begin{equation}
B_{\boldsymbol{0}\boc^{\prime},\boR\boc}^{(\mathrm{TB})}=\sum_{\bok}f_{\bok}\mex^{-i\bok\left(\boR+\lambda\left(\boc-\boc^{\prime}\right)\right)}
B_{\boc^{\prime},\boc}^{(w)\bok}\label{eq:opWFRealSpace}
\end{equation}
where in the simplest case $f_{\bok}=\frac{1}{N}$.
The superscript stands for tight-binding (TB).
From the tight binding representation the WF Bloch representation is
recovered via
\begin{equation}
  B_{\boc^{\prime}\boc}^{(\mathrm{BL})\bok}=\sum_{\boR}\mex^{i\bok\left(\boR+\lambda\left(\boc-\boc^{\prime}\right)\right)}B_{\boldsymbol{0}\boc^{\prime},\boR\boc}^{(\mathrm{TB})}\label{eq:opWFTB}
\end{equation}

Besides the TB-representation of the operators also the basis
representation of the WFs themselves is useful. Inserting
Eqs.~(\ref{eq:blochsum},\ref{eq:blochwffromblochLO}) into
Eq.~(\ref{eq:WFinRealSpace}) one gets
\begin{equation}
w_{\boR\boc}=\sum_{\boR^{\prime}}\Phi_{\boR^{\prime}\bos}a_{\boR^{\prime}\bos,\boR\boc}\label{eq:WffromLO}
\end{equation}
with the coefficients
\begin{equation}
a_{\boldsymbol{0}\bos,\boR-\boR^{\prime},\boc}=a_{\boR^{\prime}\bos,\boR\boc}=\frac{1}{N}\sum_{\bok}\mex^{-i\bok\left(\boR+\lambda\boc-\boR^{\prime}-\lambda\bos\right)}a_{\bos\boc}^{\bok}  
\end{equation}
These coefficients are a direct measure of localization.

At this step we have two possibilities. Either a real space cutoff
$\rho$ is defined such that all matrix elements from a WF center
$\boc^{\prime}$ to another center at $\boR+\boc$ are discarded if
$\mid\boR+\boc-\boc^{\prime}\mid>\rho$, or one chooses the maximum
possible set of center-pairs $\boc^{\prime},\boR\boc$ which are
consistent with the $\bok$-mesh used in the Fourier back
transformation, which is the mesh used in the self consistent DFT
calculation (new since FPLO version 19.00). A cutoff is useful to
reduce storage space, to speed up the calculation or to aim at a WF
model with minimal number of parameters(not as accurate, of course).

Operators recovered via Eq.~(\ref{eq:opWFTB}) at a post processing
stage are in general different from their exact Wannier transform
$B^{(w)\bok}$ for a $\bok$-point not included in the $\bok$-mesh.  This
is the very essence of Wannier interpolation. Hence, the resulting
band structure after this transformation through the real space
representation will also differ from the band structure obtained
directly from the exact transform Eq.~(\ref{eq:wanHam}).  The size of
this error determines the quality of the Wannier fit.

If the error is unacceptable there are two possible reasons: either
the cutoff removed too much information and needs to be increased or
the maximum possible cutoff is not big enough, which is equivalent to
saying that the underlying $\bok$-mesh is not fine enough. 
In practice a cutoff between 25-40 Bohr radii is sufficient to reduce
this error to a satisfactory degree. An exception is the automatic mode
in which all FPLO orbitals become Wannier functions. Since, the
higher lying states are spanned by polarization orbitals which have a 
larger extent than the valence orbitals the corresponding WFs are also
more extended.

To further elucidate this issue we discuss the Fourier back
transformation in more detail. If symmetry was of no concern the
acceptable $\boR$-mesh in
Eqs.~(\ref{eq:opWFRealSpace},\ref{eq:opWFTB}) is determined entirely
by the $\bok$-mesh\cite{Yat07} . The $\bok$-mesh is defined as a
regular grid defined via $N_i$ subdivisions of the three primitive
reciprocal lattice vectors. The $\boR$-mesh then is the dual
mesh, i.e.~the smallest parallelepiped supercell in real space for
which $\exp \left(i\bok \boR\right)\ne1$. This supercell is the reciprocal cell
of the smallest grid micro cell. This mesh is not optimal, instead one
folds the vectors of this mesh back to form a set of vectors
surrounding the origin and having smallest possible length.
Localization of the Wannier functions ensures that the vectors of
larger length become more and more unimportant. 

If $\boR$-vectors were included which lie outside of the supercell the
Wannier functions will acquire replica features: the contributions of
the local orbitals to a Wannier function as measured by
$a_{\boldsymbol{0}\bos,\boR\boc}$ first decrease exponentially with
increasing distance of the orbital to the Wannier center until
distances are reached which are comparable to multiples of the real
space length corresponding to the inverse of the smallest $\bok$-mesh
distance.  At these points the orbital contributions start increasing
again, i.e.~the Wannier functions are quasi-periodic objects in real
space with periods defined by the inverse of the smallest $\bok$-mesh
distances.

Since we construct symmetry conserving Wannier functions, our
$\boR$-mesh is ideally chosen to reflect the symmetry. This is
achieved by backfolding the supercell vectors into the smallest
possible spherical volume around the origin for each pair of WF
centers $\boc^{\prime},\boR\boc$. Then for all vectors of this set all
vectors additionally obtained by symmetry are added to the
set. Finally, weights are attached to all vectors such that the
weights for symmetry related vectors are identical and that the sum of
the weights of all vectors which are identical by supercell
translations is one. This way we include vectors which violate the
supercell condition, but replica do not occur due to the weighting.
This algorithm determines the maximal number of real space vectors
if no cutoff is used.

In order to reduce computation time the $\bok$-summation is done for
the irreducible part of the mesh, which changes the effective weights
to $f_{\bok}=\frac{m_{\bok}}{N}$ by multiplying with the
multiplicity of the $\bok$-point. This also requires an explicit
symmetrization of the resulting matrix elements
Eq.~(\ref{eq:opWFRealSpace}).  This can easily be done, since the
transformation properties of the WFs are explicitly known.

The symmetry which we consider is the full space group or in spin polarized
full relativistic mode the group formed by operations which do
not invert the magnetic field and by products of time reversal with
the operations which invert the field. In non spin polarized full
relativistic mode time reversal is added as an extra symmetry.

Without spin orbit coupling the FPLO orbitals are real, which means
that WFs which are obtained by projecting onto these orbitals are also
real, if inversion symmetry is present. In relativistic mode the
angular parts of the orbitals are spherical spinors, which are
inherently complex, and so are the Wannier functions.

\subsection{Practical application of the method}

For application of this method the input needs to be discussed. We
outlined above that, unless automatic mode is used, a Hilbert subspace of
the full band structure needs to be chosen first. This is usually the
bands around the Fermi level. Once this decision is made it has to be
determined which orbitals contribute to these bands, which can be
achieved by inspecting the orbital character of the band structure. At
this point one has to make sure that at each point in $\bok$-space the
number of bands highlighted by the orbital character is not smaller
than the dimension of the desired Wannier basis.  If this condition is
not fulfilled is is an indication that other orbitals contribute
essentially to the targeted bands and most importantly it means that
at these $\bok$-points the raw WF overlap Eq.~(\ref{eq:rawoverlap})
will be singular. In cases where bond centered WFs are expected,
simple molecular orbitals can be constructed from the relevant
orbitals, otherwise the relevant orbitals are the projectors themselves.

In full relativistic mode the option exists to project onto spherical
spinors with $lj\mu$ quantum numbers (Sec.~\ref{sec:LOsymemtry},
Eq.~(\ref{eq:fourspinor})) or onto orbitals which are 
transformed into pseudo non-relativistic symmetry with $lm\sigma$ qns
(Sec.~\ref{sec:LOsymemtry}, Eq.~(\ref{eq:pseudoNRELLOtransform})).
The latter are still four spinors but their large components are
mostly resembling real spherical harmonics and they transform as
real spherical harmonics. Additionally, local quantization axes can be
chosen separately for the orbital angular momentum part as well as the
spin part, which facilitates the construction of specialized models
and band structure analysis.

The automatic mode can be used in two ways: either all
basis orbitals are Wannierized or a reduced set is Wannierized, where
all deep lying orbitals (semi-core and deep lying valence orbitals),
which do not contribute to the essential band structure are removed.

The last step is the choice of $h^{\bok}$. We will discuss the case of
entangled bands, since this is the most common case. Isolated band
complexes are a trivial sub case. The targeted group of bands has a
certain energy window $\left[E_{\min},E_{\max}\right]$ in which it is located. When these bands are
entangled with other bands it means that the projector's character is
also appearing to a certain degree in the entangled irrelevant bands
outside of the core energy window while character of the irrelevant
bands flows into the targeted bands inside the core window due to
hybridization. By choosing $h^{\bok}=1$ for the core energy window
one ensures that all targeted bands are fully included in the Hilbert
subspace $\mathcal{H}$. At points where the targeted bands go outside
the core energy window $\mathcal{H}$ will suffer a sudden collapse of
some of its dimensions. Hence, we apply smooth Gaussian tails
\begin{equation}
  h^{\bok}=\exp\left(
-\left(\frac{
\varepsilon^{\bok}-E_{\min/\max}
}{\Delta_{\min/\max}}\right)^2
\right)\label{eq:Gaussiantails}
\end{equation}
at the lower and upper end of the energy window, where
$\varepsilon^{\bok}$ is the band energy of the considered band.  The
width parameter $\Delta{}E$ and the core window boundaries
$E_{\min/\max}$ need to be adjusted to achieve several
objectives. First, the resulting wannierized band structure should not
be pulled to lower or higher energies. Secondly,
Eq.~(\ref{eq:rawoverlap}) must not be singular and lastly the
projector weight which flows into entangled bands outside the core
window must be sufficiently sampled by the Gaussian tails. 

It turns out that a core window which is smaller than the targeted
window supplied with relatively wide tails are the best recipe in most
entangled cases.  This choice also increases the localization of the
resulting WFs. In cases like bcc iron where the $d$-bands are
hybridized with $sp$-bands of several tens of electron volts band
width the upper tail must be large enough (order 10 eV) to capture
some $d$-weight which flows far up into the $sp$-bands. The resulting
WFs are astonishingly good (see Sec.~\ref{sec:banddisentangling}).
The main reason why the tails can be
larger than expected is the fact that we project onto the very
orbitals which are the basis of the band structure and hence get large
weights from the $d$-projectors in the core window while the tails
collect only small parts of $d$-character.
In cases of isolated band complexes the tails can be chosen very small
and the core window set to the energy window of the isolated band complex.

When picking isolated band complexes for Wannierization it must be
avoided that the gaps above and below are topological. Otherwise, the
WFs cannot be localized due to topological obstruction. This is
usually flagged by the fact that it will be hard to find a set of
projectors with basis size equal to the number of targeted bands.  In
such cases molecular orbitals might look like a possible
basis. However, the result would either be singular or not localized
(the other option is breaking of symmetry, which we exclude by
construction).  Essentially, one has to avoid to pick parts of split
elementary band representations according to the concepts of
Topological Quantum Chemistry\cite{Bra17}.

There is a general recipe which circumvents this issue: always pick a
set of projectors which forms a chemical basis, i.e.~for instance a
$4s4p3d$ basis for a transition metal or a $5s5p$/$6s6p$-basis for
many cases with heavy main group elements (sometimes even a simple
$5p$/$6p$-basis might work). In compounds the collection of these
minimal chemical bases for all atoms contributing to the targeted
bands must be chosen as projectors. This way the probability to pick
split elementary band representations  is drastically lowered, since the size of
the projector set also increases the number of the targeted bands and
chemical basis sets are likely to form whole elementary band representations.
This strategy increases the size of the resulting WF model, but for
many applications one actually wants the whole set of valence bands
and the low lying conduction bands, which is actually formed by the
chemical basis.

The algorithm described here can also be applied in the molecule mode
of FPLO 
 (which is a genuine 0d mode without the need of a simulation
box) 
to construct basis-reduced Hamiltonian models.


\section{Topological aspects}
\label{sec:methodtopo}

This section is not essential for the basic understanding of the
general method of calculating SCMPWFs. It discusses the position
operator matrix elements in the framework of local basis methods,
which enter a complete description of the Berry connection and
curvature.

\subsection{Position operator}\label{sec:methodposop}

The Berry connection and its relation to the position operator $\bor$
has been discussed widely, especially in the context of Wannier
functions where it is given as the Fourier back transform of the
$\bor$-matrix elements between a WF in the first cell and all other
WFs. The representation of the position operator in Bloch and Wannier
basis has also has been extensively discussed in
Ref.~\onlinecite{Blo62}, where all expressions for Berry connection
and curvature are provided although without the realization of their
benefit in the study of 
topological aspects of electronic structures\cite{Vanderbilt18}.

Here we will reiterate the representation of the position operator and
introduce a basis invariant formulation which allows for consistent
and arbitrary basis transformations and naturally lends itself to
application in local orbital frameworks. A natural gauge dependent
approximation for the Berry connection will be given, which is
important in TB frameworks, where the basis is only implicitly known.
In the following we will deal with whole connection/curvature matrices
to retain the transformation properties. The final Berry
connection/curvature is a sub-space trace of these matrices.

We start with the obvious observation that the position operator
is not translation invariant, which makes its lattice Fourier
transform $\bor^{\boq\bok}$ non-diagonal in $\bok$. Furthermore, the
limit $\boq\to\bok$ is badly defined (see
Eqs.~(\ref{eq:rqk},\ref{eq:rkk}) in Sec.~\ref{sec:blochtheorem}).

In the space of extended (Bloch) functions let us introduce the
operator identity
\begin{equation}
  \bor =\mex^{i\bok\bor} i\nabla_{\bok} \mex^{-i\bok\bor}
-i\nabla_{\bok}
\end{equation}
which defines the Berry operator (discussed below)
\begin{equation}
 \bobek =\mex^{i\bok\bor} i\nabla_{\bok} \mex^{-i\bok\bor}\label{eq:berryoperator}
\end{equation}
Both equations are differential operators with respect to $\bok$
for all terms which are multiplied from the right.
The action of $\bobek$ and $\nabla_{\bok}$ hence must be understood as
\begin{subequations}
\label{eq:rchainrules}
\begin{eqnarray}
  \bobek\left|\Phi^{\bok}\right\rangle
&=&\left|\bobek\Phi^{\bok}\right\rangle
+\left|\Phi^{\bok}\right\rangle{}i\nabla_{\bok}\label{eq:betachainrule}\\
  i\nabla_{\bok}\left|\Phi^{\bok}\right\rangle
&=&\left|i\nabla_{\bok}\Phi^{\bok}\right\rangle
+\left|\Phi^{\bok}\right\rangle{}i\nabla_{\bok}
\end{eqnarray}
\end{subequations}
(using the chain rule), where
$\left|\bobek\Phi^{\bok}\right\rangle$ is the isolated action of
$\bobek$ on $\Phi^{\bok}$.

Matrix elements of $\bor$ between Bloch functions (i.e.~Bloch sums or
linear combinations thereof) can be written as
\begin{subequations}
  \begin{eqnarray}
\left\langle \Phi^{\boq}\mid\bor\mid\Phi^{\bok}\right\rangle 
&=&\left\langle \Phi^{\boq}\mid\bobek\mid\Phi^{\bok}\right\rangle 
-\left\langle \Phi^{\boq}\mid
    i\nabla_{\bok}\mid\Phi^{\bok}\right\rangle \label{eq:rmatelnotconfined}\\
 \left\langle \Phi^{\boq}\mid\bor\mid\Phi^{\bok}\right\rangle 
 &=&  \left\langle
   \Phi^{\boq}\mid\bobek\Phi^{\bok}\right\rangle
 -\left(i\nabla_{\bok}S_{\Phi}^{\boq\bok}\right)
\label{eq:rasbetagrad}
\end{eqnarray}
\end{subequations}

where in Eq.~(\ref{eq:rmatelnotconfined}) the differentiations are not
confined in both terms, while in Eq.~(\ref{eq:rasbetagrad}) the
differentiation is confined, i.e.~these are normal matrices. The
unusual shape of Eq.~(\ref{eq:rasbetagrad}) is formally correct in
that it transforms properly under basis changes such that with
$\Psi^{\bok}=\Phi^{\bok}C^{\bok}$ one gets
\begin{equation}
\left\langle \Psi^{\boq}\mid\bor\mid\Psi^{\bok}\right\rangle 
=  \left\langle \Psi^{\boq}\mid\bobek\Psi^{\bok}\right\rangle
-\left(i\nabla_{\bok}S_{\Psi}^{\boq\bok}\right)\label{eq:rasbetagradpsi}
\end{equation}
by bracketing Eq.~(\ref{eq:rasbetagrad}) with $C^{\bok}$ from both
sides and applying the chain rule Eq.~(\ref{eq:rchainrules}) to both
terms on the rhs.  At this stage it is vital to always consider the
full non-diagonal form.  Note, that the last term of
Eq.~(\ref{eq:rasbetagrad}) reduces to
$\left(i\nabla_{\bok}\delta_{\boq,\bok}\right)$ for orthonormal
$\Phi^{\bok}$, which is basis independent. (For correct transformation
behavior the overlap matrix has to be kept until the end result,
however.) Hence, {\it the essential part of the position operator is the
Berry connection matrix} discussed in the following.

\subsection{Berry connection}

Using the Berry operator defined above it
can be shown (Sec.~\ref{sec:blochtheorem},
Eq.~(\ref{eq:berryOpPsiBlochTheo})) 
that
$\left| \bobek\Phi^{\bok}\right\rangle$ fulfills the Bloch theorem and
hence $\left\langle \Phi^{\boq}\mid\bobek\Phi^{\bok}\right\rangle$ is
diagonal in $\bok$.  Each Bloch function can be written as
$\Phi^{\bok}=\mex^{i\bok\bor}u_{\Phi}^{\bok}$ which leads to
\begin{equation}
\left\langle \Phi^{\boq}\mid\bobek\Phi^{\bok}\right\rangle =\delta_{\boq\bok}\left\langle u_{\Phi}^{\bok}\mid i\nabla_{\bok}u_{\Phi}^{\bok}\right\rangle   
\end{equation}
which contains the matrix underlying the Berry connection
\begin{equation}
\boA_{\Phi}^{\bok} =\left\langle u_{\Phi}^{\bok}\mid i\nabla_{\bok}u_{\Phi}^{\bok}\right\rangle\label{eq:berryconnection}
\end{equation}

Consequently, the position operator matrix elements in a Bloch basis
consist of a $\bok$-diagonal well defined part, the Berry connection matrix,
and the gradient of the overlap matrix (delta function for orthonormal
bases) (which also contains diagonal elements, implicitly) and this is
true in all bases.  

An illustrative example of the meaning of
Eq.~(\ref{eq:rasbetagradpsi}) is to formally calculate the dipole integral
$\boldsymbol{D}=\int\bor n\left(\bor\right)d^{3}r$ of the density
$n\left(\bor\right)=\sum_{\bok{}n}f_{\bok{}n}\left|\Psi_n^{\bok}\right|^{2}$
which using Eq.~(\ref{eq:rasbetagradpsi}) becomes
\begin{equation}
\boldsymbol{D}=\sum_{\bok{}n}f_{\bok{}n}\left[A_{\Psi,nn}^{\bok}-\lim_{q\to
    k}\left(i\nabla_{\bok}\delta_{\boq\bok}\right)\right]d^{3}r  
\end{equation}
of which the first term, the essential part of $\bor$, is the
polarization $\boldsymbol{P}^{\lambda}$ of Ref.~\onlinecite{Kin93} up to
constant factors.

The Berry connection matrix is Hermitian if the
Bloch basis is orthonormal
\begin{equation}
\boA_{\Phi}^{\bok+}=\boA_{\Phi}^{\bok}-i\nabla_{\bok}S_{\Phi}^{\bok} \label{eq:Akhermitianconj}
\end{equation}
e.g.~for Wannier functions and Hamiltonian
eigen functions, which remedies its asymmetric formulation
(derived via $i\nabla_{\bok}S^{\bok}=i\nabla_{\bok}\left\langle \Phi^{\bok}\mid\Phi^{\bok}\right\rangle =i\nabla_{\bok}\left\langle \mex^{-i\bok\bor}\Phi^{\bok}\mid\mex^{-i\bok\bor}\Phi^{\bok}\right\rangle$).

The basis change $\Psi^{\bok}=\Phi^{\bok}C^{\bok}$ of the $\bok$-diagonal Berry connection matrix in
Eq.~(\ref{eq:rasbetagradpsi}) reads explicitly
\begin{equation}
\boA_{\Psi}^{\bok}=C^{\bok+}\boA_{\Phi}^{\bok}C^{\bok}+C^{\bok+}S_{\Phi}^{\bok}i\nabla_{\bok}C^{\bok}  
\label{eq:APsi}
\end{equation}
If one interprets $\Phi$ as the Bloch sums of local orbitals or
Wannier functions and $\Psi$ as the eigen functions of the Hamiltonian,
the Berry connection consists of a term $C^+\boA_{\Phi}C$, 
where $\boA_{\Phi}$ shall be called {\it basis-connection},
 and the usual gradient-of-coefficients term (with an
overlap matrix for the non-orthonormal case). The latter term is always
available while the former is not known in most situations involving
tight binding models. Below, a proper approximation for the
basis-connection will be introduced.  Practically, Eq.~(\ref{eq:APsi})
is not useful due to the random phase issue in
$\nabla_{\bok} C^{\bok}$. It will be useful in deriving corrections
to the Berry curvature later on.

For a plane wave basis
$\Psi^{\bok}=\sum_{\boG}\mex^{i\left(\bok+\boG\right)\bor}C_{\bok+\boG}$
the basis functions are the exponentials and the basis connection
matrix is zero:
\begin{equation}
\boA_{\boG^{\prime}\boG}^{\bok}
=\left\langle 
\mex^{i\boG^{\prime}\bor} \mid i\nabla_{\bok}\mex^{i\boG\bor}
\right\rangle =0  
\end{equation}

For a local basis we introduce vector valued functions 
\begin{equation}
\left(\bor\Phi\right)_{\boR\bos}=\left(\bor-\boR-\bos\right)\Phi_{s}\left(\bor-\boR-\bos\right)\label{eq:vectorbaluedPhi}
\end{equation}
with Bloch sums
\begin{equation}
\left(\bor\Phi\right)_{\bos}^{\bok}=\frac{1}{\sqrt{N}}\sum_{\boR}\mex^{i\bok\left(\boR+\lambda\bos\right)}\left(\bor\Phi\right)_{\boR\bos}  
\end{equation}
which leads to ($\blambda=1-\lambda$)
\begin{equation}
\left(\bobek\Phi_{\bos}^{\bok}\right)
=\left(\bor\Phi\right)_{\bos}^{\bok}+\blambda\Phi_{\bos}^{\bok}\bos
\end{equation}
and the basis connection
\begin{eqnarray}
\boA_{\Phi,\bos^{\prime}\bos}^{\bok}
&=&\left\langle \Phi_{\bos^{\prime}}^{\bok}\mid\bobek\Phi_{\bos}^{\bok}\right\rangle \nonumber\\
&=&\left\langle
    \Phi_{\bos^{\prime}}^{\bok}\mid\left(\bor\Phi\right)_{\bos}^{\bok}\right\rangle
    +\blambda{}S_{\bos^{\prime}\bos}^{\bok}\bos  
\label{eq:basiconnectionfromrelativer}
\end{eqnarray}
The $\lambda$-term is zero for the relative gauge, but non-zero for
the periodic gauge of the Bloch sums.
The first term on the rhs., which shall be called {\it reduced position operator},
can be straightforwardly evaluated as 
\begin{equation}
\left\langle \Phi_{\bos^{\prime}}^{\bok}\mid\left(\bor\Phi\right)_{\bos}^{\bok}\right\rangle   
=\sum_{\boR}\textnormal{e}^{i\bok\left(\boR+\lambda\left(\bos-\bos^{\prime}\right)\right)}\left\langle \Phi_{\boldsymbol{0}\bos^{\prime}}\mid\left(\bor\Phi\right)_{\boR\bos}\right\rangle \label{eq:reducedrblochsums}
\end{equation}
from local orbital matrix elements, which are translation
invariant:
\begin{equation}
  \left\langle
    \Phi_{\boR^{\prime}\bos^{\prime}}\mid\left(\bor\Phi\right)_{\boR\bos}\right\rangle
  =\left\langle
    \Phi_{\boldsymbol{0}\bos^{\prime}}\mid\left(\bor\Phi\right)_{\boR-\boR^{\prime},\bos}\right\rangle \label{eq:LOrelative_r_matels}
\end{equation}
and Hermitian if $\Phi$ is orthonormal.

Eq.~(\ref{eq:LOrelative_r_matels}) is very useful due to its
invariance, which makes them transferable on a lattice.  It is the
ideal form to store position operator matrix elements (also for the
Wannier functions).

The relation to the position operator follows by expanding the
$\bor-\boR-\bos$ part of Eq.~(\ref{eq:vectorbaluedPhi}) in 
Eq.~(\ref{eq:basiconnectionfromrelativer}), using 
\begin{eqnarray}
i\nabla_{\bok}S_{\bos^{\prime}\bos}^{\bok}
&=&
    -\sum_{\boR}\mex^{i\bok\left(\boR+\lambda\left(\bos-\bos^{\prime}\right)\right)}S_{\boldsymbol{0}\bos^{\prime},\boR\bos}\cdot\nonumber\\
&&\cdot\left(\boR+\lambda\left(\bos-\bos^{\prime}\right)\right)\nonumber
\end{eqnarray}
and reads
\begin{eqnarray}
\boA_{\Phi,\bos^{\prime}\bos}^{\bok}
&=&\frac{1}{N}\sum_{\boR}\textnormal{e}^{i\bok\left(\boR+\lambda\left(\bos-\bos^{\prime}\right)\right)}\left\langle
  \Phi_{\boldsymbol{0}\bos^{\prime}}\mid\bor\mid\Phi_{\boR\bos}\right\rangle
\nonumber\\
&&+\left(i\nabla_{\bok}-\lambda\bos^{\prime}\right)S_{\bos^{\prime}\bos}^{\bok}  
\end{eqnarray}
In the periodic gauge $\lambda=0$ and for WFs
$S_{\bos^{\prime}\bos}^{\bok}=\delta_{\bos^{\prime}\bos}$ this is the
expression usually given. However,
Eq.~(\ref{eq:basiconnectionfromrelativer}) is the simpler and arguably
the more natural choice.

In order to calculate the basis-connection
Eq.~(\ref{eq:basiconnectionfromrelativer}) for the Wannier functions
one needs to express the reduced position operator matrix
Eq.~(\ref{eq:reducedrblochsums}) of the WFs in terms of the basis
orbitals. Note, that in our implementation the relative gauge
$\lambda=1$ is used and hence the last term in
Eq.~(\ref{eq:basiconnectionfromrelativer}) is zero and
basis-connection and reduced position operator matrix are identical.

Our Wannier functions are related to the LOs via
Eq.~(\ref{eq:blochwffromblochLO}). The Berry connection in the Wannier
basis hence is obtained by a basis change Eq.~(\ref{eq:APsi}) when
letting $\Psi\to{}w$ and $C\to{}a$:
\begin{equation}
\boA_{w}^{\bok}=a^{\bok+}\boA_{\Phi}^{\bok}a^{\bok}+a^{\bok+}S_{\Phi}^{\bok}i\nabla_{\bok}a^{\bok}  \label{eq:WFberryconnection}
\end{equation}
The first term on the rhs.~is given by
Eq.~(\ref{eq:basiconnectionfromrelativer}) with $\Phi$ being our LOs,
while the gradient needs special treatment.  It turned out that a
direct calculation via the real-space coefficients
$a_{\boldsymbol{0}\boc^{\prime},\boR\boc}$ is not very
accurate. Instead we use a numerical differentiation technique akin to
the one described in Ref.~\onlinecite{Mar97}. Finally, $\boA_{w}^{\bok}$
is Fourier back transformed which gives
$\left\langle
  w_{\boldsymbol{0}\boc^{\prime}}\mid\left(\bor{}w\right)_{\boR,\boc}\right\rangle$
in real space which is stored together with the other WF information
for post processing. From this the Wannier basis connection
$\boA_{w}^{\bok}$ can be retrieved via
Eq.~(\ref{eq:basiconnectionfromrelativer},\ref{eq:reducedrblochsums})
with $\Phi\to{}w$,
$S_{\bos^{\prime}\bos}\to{}\delta_{\bos^{\prime}\bos}$ in either Bloch
sum gauge.

The reduced position operator matrix elements
Eq.~(\ref{eq:LOrelative_r_matels}), especially for orthonormal bases,
are quite intuitive. The on-site elements are the integral of two
orbitals multiplied with the position vector from the orbital
origin. For two equal orbitals which are eigenfunctions of inversion
this integral is always zero due to parity. For other cases and also
for the off-site case one can argue that the orthonormality of the
orbitals/WFs will somewhat transfer to these matrix element to make
them rather small.

Also the reduced elements are Hermitian (for orthonormal bases) and
transform according to all symmetries
(see.~Sec.~\ref{sec:symmetryBerry}). Hence, they are a reasonable
choice to be considered as the physically important matrix. By
approximating it by zero we get
\begin{equation}
\boA_{w,\bos^{\prime}\bos}^{\bok}
\approx \blambda
\delta_{\bos^{\prime}\bos}
\bos   \label{eq:approxA}
\end{equation}
for orthonormal bases. Note, that this is nonzero in the periodic
gauge, in which the Bloch sums are $\bok$-periodic.

The basis connection can only be zero in one gauge, which forces a
choice which matrix to put to zero.  We argue that the relative gauge,
which is usually not mentioned in the context of Wannier functions is
the one in which the WF basis-connection can be neglected. We will
show numerical evidence for this later.  This might be important in
tight-binding models, where the basis is not explicitly known and
hence the basis connection is missing. Furthermore, approximation
Eq.~(\ref{eq:approxA}) is needed to preserve symmetry properties
(Sec.~\ref{sec:symmetryBerry}), which further indicates the
correctness of this choice.

\subsection{Berry curvature}

In this section we derive the Berry curvature matrix 
\begin{equation}
\boF_{\Psi}^{\bok}=\nabla_{\bok}\times\boA_{\Psi}^{\bok}  
\end{equation}
in eigenstate representation. Starting from the (Wannier) basis
connection
$\boA_{w}^{\bok}=\left\langle w^{\bok}\mid \bobek{}w^{\bok}
\right\rangle$
(usually reassembled from the reduced position operator in WF basis
during post processing) we make a basis change
$\Psi^{\bok}=w^{\bok} C^{\bok}$ using Eq.~(\ref{eq:APsi}) with
$\Phi\to{}w$
\begin{equation}
\boA_{\Psi}^{\bok}=C^{\bok+}\boA_{w}^{\bok}C^{\bok}+C^{\bok+}S_{w}^{\bok}i\nabla_{\bok}C^{\bok}
\label{eq:AwC}
\end{equation}
where the overlap matrix is the unit matrix for orthonormal WFs.
Let's introduce some short hands
\begin{eqnarray}
\left\langle \boA\right\rangle _{C}
&=&C^{\bok+}\boA_{w}^{\bok}C^{\bok}\label{eq:defAC}\\
\left\langle Si\nabla\right\rangle _{C}
&=&C^{\bok+}S_{w}^{\bok}i\nabla_{\bok}C^{\bok}  \label{eq:defSgradC}\\
\left\langle\boldsymbol{f}\right\rangle _{C}
&=& C^{\bok+}\left(\nabla_{\bok}\times\boA_{w}^{\bok}\right)C^{\bok}
\end{eqnarray}
Differentiation of the normalization
$C^{\bok+}S_{w}^{\bok}C^{\bok}=\boldsymbol{1}$ gives
\begin{equation}
\left\langle Si\nabla\right\rangle _{C}^{+}
=\left\langle Si\nabla\right\rangle _{C}
+C^{\bok+}\left(i\nabla_{\bok}S_{w}^{\bok}\right)C^{\bok}
\end{equation}
and  Eq.~(\ref{eq:Akhermitianconj}) gives
\begin{equation}
\left\langle \boA\right\rangle _{C}^{+}
=\left\langle \boA\right\rangle
_{C}-C^{\bok+}\left(i\nabla_{\bok}S_{w}^{\bok}\right)
C^{\bok}
\end{equation}
Then using
$\nabla\times u\boldsymbol{v}w=\left(\nabla
  u\right)\times\boldsymbol{v}w+u\left(\nabla\times\boldsymbol{v}\right)w-u\boldsymbol{v}\times\left(\nabla
  w\right)$,
where $u$, $w$ are matrices and $\boldsymbol{v}$ is a vector of
matrices, and the identity
$C^{\bok}C^{\bok+}S_{w}^{\bok}=\boldsymbol{1}$ and commuting
non-vector matrices through the $\times$-symbol one gets
\begin{subequations}
\label{eq:AandF}
\begin{eqnarray}
  \boA_{\Psi}^{\bok}
  &=& \left\langle \boA\right\rangle _{C} + \left\langle{}
      Si\nabla\right\rangle _{C}\label{eq:Aaux}\\  
  \boF_{\Psi}^{\bok} 
  &=& \left\langle \boldsymbol{f}\right\rangle _{C}-i\left\langle
      \boA\right\rangle _{C}^{+}\times\left\langle \boA\right\rangle
      _{C}
      +i\boA_{\Psi}^{\bok+}\times\boA_{\Psi}^{\bok}\label{eq:BerryCurvature}\\
  & =&  \left\langle \boldsymbol{f}\right\rangle _{C}+i\left\langle
       \boA\right\rangle _{C}^{+}\times\left\langle
       Si\nabla\right\rangle _{C}+i\left\langle Si\nabla\right\rangle
       _{C}^{+}\times\left\langle \boA\right\rangle _{C}\nonumber\\
&&+i\left\langle Si\nabla\right\rangle _{C}^{+}\times\left\langle
   Si\nabla\right\rangle _{C} \label{eq:BerryCurvatureAD}
\end{eqnarray}
\end{subequations}
Note,  that the cross product expressions are matrix products at the
same time
($\left(\boldsymbol{a}\times\boldsymbol{b}\right)_{nm,k}=\sum_{lij}\varepsilon_{ijk}\boldsymbol{a}_{nl,i}\boldsymbol{b}_{lm,j}$).
The second term grouping of $\boF_{\Psi}^{\bok}$ corresponds to the
result of Ref.~\onlinecite{Wan06} while the first emphasizes the symmetry
between the curvature in both bases and is discussed in the appendix
of Ref.~\onlinecite{Wan06}.

Due to Eq.~(\ref{eq:Akhermitianconj}) $\boA_{\Psi}^{\bok}$ is
Hermitian, since $\Psi^{\bok}$ is orthonormal, and the basis curvature
$\boldsymbol{f}^{\bok}=\nabla\times\boA_{w}^{\bok}$ is Hermitian since $S_{w}^{\bok}$ is
smooth and $\nabla\times \nabla S=0$. This makes
$\left\langle\boldsymbol{f}\right\rangle_{C}$ and hence also
$\boF_{\Psi}^{\bok}$ Hermitian.  Eqs.~(\ref{eq:AandF}) are true for
any basis $w^{\bok}$ and hence could also be used directly in the
non-orthogonal FPLO basis.  
(If additionally $\Psi^{\bok}$ were to be non-orthonormal
$\left(S_{\Psi}\right)^{-1}$ needs to be inserted before or after all
$\times$-symbols in Eq.~(\ref{eq:AandF}).)

In practice,
$\left\langle{}Si\nabla\right\rangle _{C}$ is obtained in the parallel
transport gauge\cite{Wan06}, which is not a periodic gauge and hence
makes Eq.~(\ref{eq:Aaux}) useless for topological applications but it
allows to calculate Eq.~(\ref{eq:BerryCurvature}).  In detail
non-degenerate perturbation theory of first order gives
\begin{eqnarray}
\left\langle Si\nabla\right\rangle
_{C,mn}
&=&i\frac{C_{m}^{\bok+}\left(\nabla_{\bok}H_{w}^{\bok}\right)
C_{n}^{\bok}-C_{m}^{\bok+}\left(\nabla_{\bok}S_{w}^{\bok}\right)
C_{n}^{\bok}\varepsilon_{n}^{\bok}}
{\varepsilon_{n}^{\bok}-\varepsilon_{m}^{\bok}}\cdot\nonumber\\
&& \cdot\left(1-\delta_{mn}\right)  \label{eq:DfromPT}
\end{eqnarray}
where the gradient of the Hamiltonian in Wannier basis reads
\begin{eqnarray}
\nabla_{\bok}H_{w}^{\bok}
&=&
    i\sum_{\boR}\mex^{i\bok\left(\boR+\lambda\left(\bos-\bos^{\prime}\right)\right)}
\left(\boR+\lambda\left(\bos-\bos^{\prime}\right)\right)\cdot\nonumber\\
&& \cdot\left\langle w_{\boldsymbol{0}\bos^{\prime}}\mid\hat{H}
\mid w_{\boR\bos}\right\rangle   .
\end{eqnarray}
For an orthonormal basis the gradient of the overlap vanishes.  The
basis curvature $\boldsymbol{f}^{\bok}$ is obtained by directly
taking the curl of the exponential in Eq.~(\ref{eq:reducedrblochsums})
($\Phi\to{}w$) and inserting into
Eq.~(\ref{eq:basiconnectionfromrelativer}). (The curl of the overlap
can be obtained in a similar way for non-orthonormal bases.)

We need to discuss the non Abelian (degenerate) case, in which the  Berry
curvature matrix obtains a covariant correction in each degenerate
subspace\cite{Cul05}. If $P_i$ denotes the projector onto the $i$-th
subspace (degenerate or not)  one needs to write
\begin{equation}
P_{i}\boF_{\Psi nA}^{\bok}P_{i}=P_{i}\left(\boF_{\Psi}^{\bok}-i\boA_{\Psi}^{\bok}P_{i}\times P_{i}\boA_{\Psi}^{\bok}\right)P_{i}\label{eq:nAF}
\end{equation}
for each subspace. For a degenerate subspace
$\boA_{\Psi,ii}^{\bok}=P_{i}\boA_{\Psi}^{\bok}P_{i}$ is a single
vector-valued matrix element and hence
$\boA_{\Psi,ii}^{\bok}\times\boA_{\Psi,ii}^{\bok}=0$ and the
correction is zero. In a degenerate subspace only the average of the
trace is physically meaningful. In the case of an orthonormal basis
this allows to introduce a unitary transformation, which ultimately
leads to
$i \mathrm{Tr}P_{i}\left\langle \boA\right\rangle _{C}P_{i}\times
P_{i}\left\langle \boA\right\rangle _{C}P_{i}=0$
and permits to discard such block diagonal terms (if only degenerate
subspace averages are considered).  Inserting
Eq.~(\ref{eq:BerryCurvature}) into Eq.~(\ref{eq:nAF}) and introducing
$Q_{i}=1-P_{i}$ one gets for orthonormal bases
\begin{eqnarray}
\lefteqn{P_{i}\boF_{\Psi nA}^{\bok}P_{i}}\nonumber\\
&&=P_{i}\left(\left\langle \boldsymbol{f}\right\rangle
   _{C}-i\left\langle \boA\right\rangle _{C}Q_i\times{}Q_i\left\langle
   \boA\right\rangle _{C}+i\boA_{\Psi}^{\bok}Q_{i}\times
   Q_{i}\boA_{\Psi}^{\bok}\right)P_{i}\nonumber\\
\label{eq:nAFdetail}
\end{eqnarray}
This expression contains $\left\langle Si\nabla\right\rangle$
bracketed between $Q_{i}$ and $P_{i}$ and hence only matrix elements
between different subspaces such that the energy denominator in
Eq.~(\ref{eq:DfromPT}) is never zero. This solves the division by zero
issue.  
In essence the non-Abelian Berry curvature is obtained from
Eqs.~(\ref{eq:BerryCurvature},\ref{eq:BerryCurvatureAD}) by removing
the diagonal blocks of all subspaces from
$\left\langle\boA\right\rangle _{C}$ 
and $\left\langle{}S{}i\nabla\right\rangle _{C}$.

Ref.~\onlinecite{Wan06} discusses the terms of $F^{\bok}$ and argues
that the
$\left\langle Si\nabla\right\rangle\times\left\langle
  Si\nabla\right\rangle$-term
of Eqs.~(\ref{eq:BerryCurvatureAD}) is by far the leading term. This
would allow to ignore the basis connection/curvature terms, especially
if they are not obtainable.  However, this is not necessarily
correct. In Sec.~\ref{sec:symmetryBerry} it is shown that the term
$\left\langle Si\nabla\right\rangle\times\left\langle
  Si\nabla\right\rangle$
does not transform properly under symmetry in the periodic gauge.
However, if $\boA_w^{\bok}$ is approximated by Eq.~(\ref{eq:approxA})
in Eqs.~(\ref{eq:defAC},\ref{eq:Aaux}) proper symmetry transformation
is restored. This shows that the term grouping
Eqs.~(\ref{eq:BerryCurvature},\ref{eq:nAFdetail}) is to be preferred
and in fact under the approximation the only term left is
$iP\boA_{\Psi}^{\mathrm{appr}}Q\times{}Q\boA_{\Psi}^{\mathrm{appr}}P$
with approximated $\boA_{\Psi}^{\mathrm{appr}}$, which reduces to the
$\left\langle Si\nabla\right\rangle$-only term in relative gauge.

Another argument for the grouping Eq.~(\ref{eq:BerryCurvature}) is the
fact that both $iP\boA_{\Psi}Q\times{}Q\boA_{\Psi}P$ and
$\left\langle \boldsymbol{f}\right\rangle _{C}-i\left\langle
  \boA\right\rangle _{C}Q_i\times{}Q_i\left\langle \boA\right\rangle
_{C}$
are invariant under symmetry and with respect to the Bloch sum gauge
choice controlled by $\lambda$ (see
Sec.~\ref{sec:blochsumgaugeinvariance}) while the other grouping is
not, especially not the approximation of only taking the
$\left\langle Si\nabla\right\rangle\times\left\langle
  Si\nabla\right\rangle$-term.
Finally, if the $\Psi$-basis is complete $\left\langle \boldsymbol{f}\right\rangle _{C}-i\left\langle
  \boA\right\rangle _{C}Q_i\times{}Q_i\left\langle \boA\right\rangle
_{C}$ will go to zero and $\boF_{\Psi{}nA}^{\bok}=iP\boA_{\Psi}Q\times{}Q\boA_{\Psi}P$
is the complete Berry curvature.

In FPLO before version 19 only the periodic gauge was
implemented; now both gauges as well as the full basis connection are
available.  Furthermore, for historical reasons the FPLO implementation
contains an overall minus sign for the Berry connection/curvature as
if we defined
$\boA^{\bok}=\left\langle u^{\bok}\mid -i\nabla_{\bok}
  u^{\bok}\right\rangle$ instead of Eq.~(\ref{eq:berryconnection}).
Taking this sign into account we are consistent with the data
presented in Ref.~\onlinecite{Wan06}.

On a final note, in cases of TB-models for which no explicit WFs but
the nature/symmetry of these functions are known, instead of
approximation Eq.~(\ref{eq:approxA}) one could derive parametrized
analytic expressions for leading matrix elements in
Eq.~(\ref{eq:LOrelative_r_matels}) based on the nature of the WFs
which inserted into Eq.~(\ref{eq:basiconnectionfromrelativer}) give a
reasonable approximation for the basis connections.


\section{Features of Maximally Projected Wannier Functions}
\label{sec:features}

The examples shown here are not discussed in terms of their DFT setup.
More details can be found in Sec.~\ref{sec:structures}.

\subsection{Work flow}
The WF creation is controlled by the choice of projectors $\phi$, the energy
window, which can be chosen individually for each projector (although
most of the time a global window is used) and the real space
cutoff $\rho$ and a coefficient threshold (to save space), below which 
WF data are discarded. The latter might need to be changed from
its default if the targeted bands are very flat.

The projectors can be single FPLO orbitals, for which only a site
number, an orbital descriptor and optionally local quantization axes
need to be defined.  In full relativistic mode, the projectors can be
in four-spinor quantum numbers $nlj\mu$ or in pseudo-non-relativistic
qns. $nlm\sigma$, in which case separate local spin axes can be
defined.  Also unnormalized linear combinations of LOs (MOs) can be
chosen as projectors, which are defined by picking an imagined
Wannier center, from which difference vectors to the contributing
sites must be defined.  Furthermore, for each contribution a site
number, a weight and local axes must be specified.  The main
constraint is that the set of projectors must be closed under all
symmetry operations.  Band disentangling and the quality of the fit is
controlled by proper choice of projectors and energy window.

Alternatively, there is an automatic mode, which picks either all FPLO
orbitals as projectors or only the subset which does not clearly form
semi-core states. The FPLO basis is relatively small with of 15 to 35
orbitals per atoms for most of the periodic table, which avoids a too
large resulting WF basis. Removal of semi-core states further
decreases the basis size. The advantage of this mode is application in
automated calculations and the fact that no energy window needs to be
defined. The cutoff $\rho$ however might need to be increased
(preferably to it's maximum) due to
the larger extent of higher energy WFs.

\begin{figure}[h]
\includegraphics[width=0.70\columnwidth]{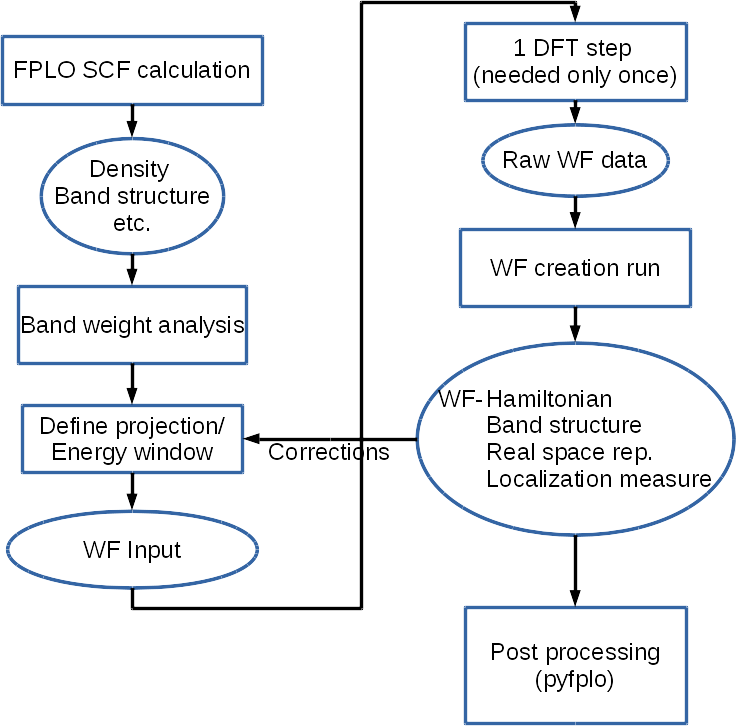}
\caption{
The work flow to construct and use SCMPWFs in the FPLO implementation.
\label{fig:workflow}
}
\end{figure}

To construct WFs with FPLO a WF input file needs to be created, which
can be done by hand or with the help of a python module, which allows
to efficiently create single-orbital projectors for any set of
sites/orbitals as well as molecular orbital projectors.  Starting from
a converged calculation and with the input file present an FPLO-run
will produce a file with all the data needed to construct the Wannier
functions in a second run (see Fig.~\ref{fig:workflow}). At this stage
the real space Wannier representation, the symmetry operations and all
demanded operators (like reduced position operator, full relativistic
spin operators $\beta\vec{\Sigma}$ and the exchange-correlation (xc)
magnetic field) are written to a separate file for post
processing. Additionally, WF output on a real space grid for
visualization purposes can be requested.

\subsection{Model extraction}

\begin{figure}[h]
\includegraphics[width=0.99\columnwidth]{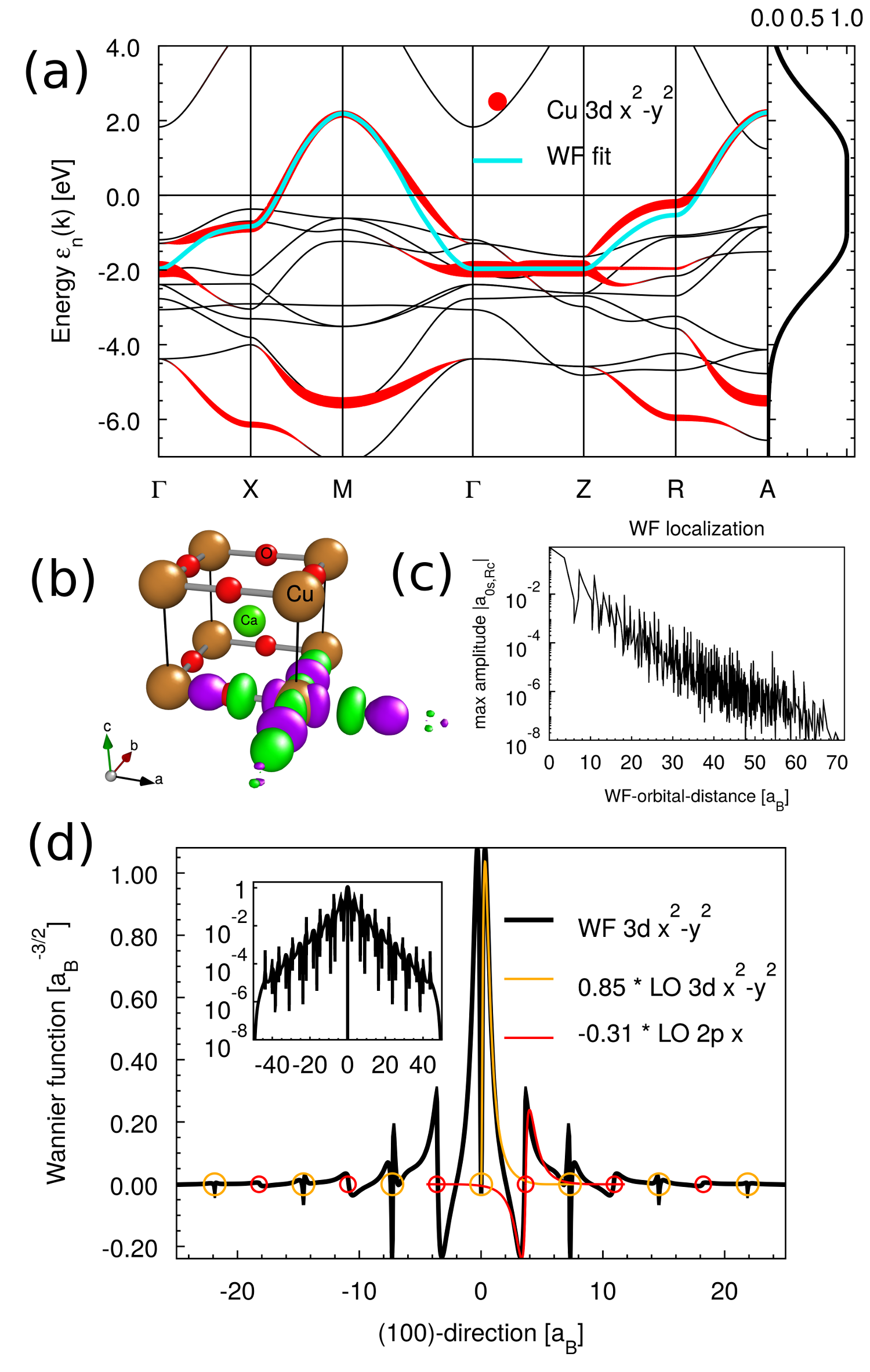}
\caption{(color online) 
(a) the band structure (black) of CaCuO$_2$, the
FPLO orbital $3d_{x^2-y^2}$ band character (red) and the single band
Wannier fit (cyan). (b) the unit cell containing the Wannier
function at isovalue 0.04. (c) the maximum of orbital
contributions to the WF in logarithmic scale as a function of 
WF-orbital distance. (d) the WF printed along the (100)-axis through
the WF center (black), the Cu $3d_{x^2-y^2}$ (orange) and the
O $2p_{x}$ (red) orbitals scaled according to their amplitude.
\label{fig:CaCuO2}}
\end{figure}

The first example demonstrates the 
construction of a single band model of the anti-bonding
(AB) Cu $3d_{x^2-y^2}$ band for the infinite layer cuprate CaCuO$_2$. The
band structure (Fig.~\ref{fig:CaCuO2}) contains the prototypical AB
band in the energy interval $[-2,2]$~eV around the Fermi level
highlighted by the FPLO-orbital band character.  The bonding part
below -3~eV, which is more diffuse, is of course also present. 

The input (essentially a site number and an orbital name) for the
Wannierization contains a single projector consisting of the
$3d_{x^2-y^2}$ orbital sitting at the Cu site. 
The $\bok$-mesh has $12^3$ points.
The core energy window
is chosen as $[-1,1]$ eV with symmetrical Gaussian tails,
Eq.~(\ref{eq:Gaussiantails}), of width $\Delta=2$~eV as indicated on
the right side of the band structure. This window is considerably
overlapping with the bonding part, yet the resulting WF (cyan) follows the
anti-bonding part, where $h^{\bok}$ is dominant. The fit is not
perfect, owed to the fact that there are lots of hybridizations
interrupting the unbroken flow of the band character. It however
samples the essence of the targeted band. The WF fit was calculated
from WF data with a real space cutoff $\rho=25 a_{\mathrm{B}}$. To
illustrate the exponential localization of the WF the maximum of the
amplitudes $\max\left(|a_{\boR\bos,\boldsymbol{0}\boc}|\right)$ of
contributing orbitals around the WF, Eq.~(\ref{eq:WffromLO}), as a
function of WF-orbital-distance are shown in Fig.~\ref{fig:CaCuO2}c for
a cutoff $\rho=70 a_{\mathrm{B}}$. Note, the logarithmic scale.  

Fig.~\ref{fig:CaCuO2}b shows the crystal structure and the WF, which
clearly has $x^2-y^2$ symmetry and sizable oxygen hybridization tails
as to be expected for this case. The WF along the (100)-direction
through the WF center is plotted in
Fig.~\ref{fig:CaCuO2}d. Additionally, the Cu $3d_{x^2-y^2}$- and O
$2p_{x}$-orbitals scaled according to their amplitude
$a_{\boR\bos,\boldsymbol{0}\boc}$ is shown. The circles show the
position of the atoms in the same color as the orbitals.  This WF is
rather extended since it needs to describe a Cu-O hybridized band via
a single WF. It is however exponentially localized as indicated by the
inset which shows $\left|w\right|$ in logarithmic scale.  Note, that the
projector consists only of the Cu $3d_{x^2-y^2}$-orbital (orange), all
other contributions to the WF are pulled in due to the projection process.

\subsection{Band disentangling}\label{sec:banddisentangling}

A more complex example which demonstrates band disentangling is the
$3d$-only WFs for spin polarized fully relativistic bcc iron (see
Sec.~\ref{sec:bccFe}). Spin is no longer a good qn. although spin
orbit coupling and hence spin-mixing is quite small.  The complication
is the hybridization with the $4sp$-bands which have a band width of
several tens of electron volts, which means that bands which start as
$d$-bands at some $\bok$-point can mutate into free electron bands
away from the point. Hence, the predominantly $3d$-character Hilbert
space is not restricted to a finite energy window and the
corresponding WFs cannot follow the $d$-band structure at all points.

\begin{figure}[h]
\includegraphics[width=0.99\columnwidth]{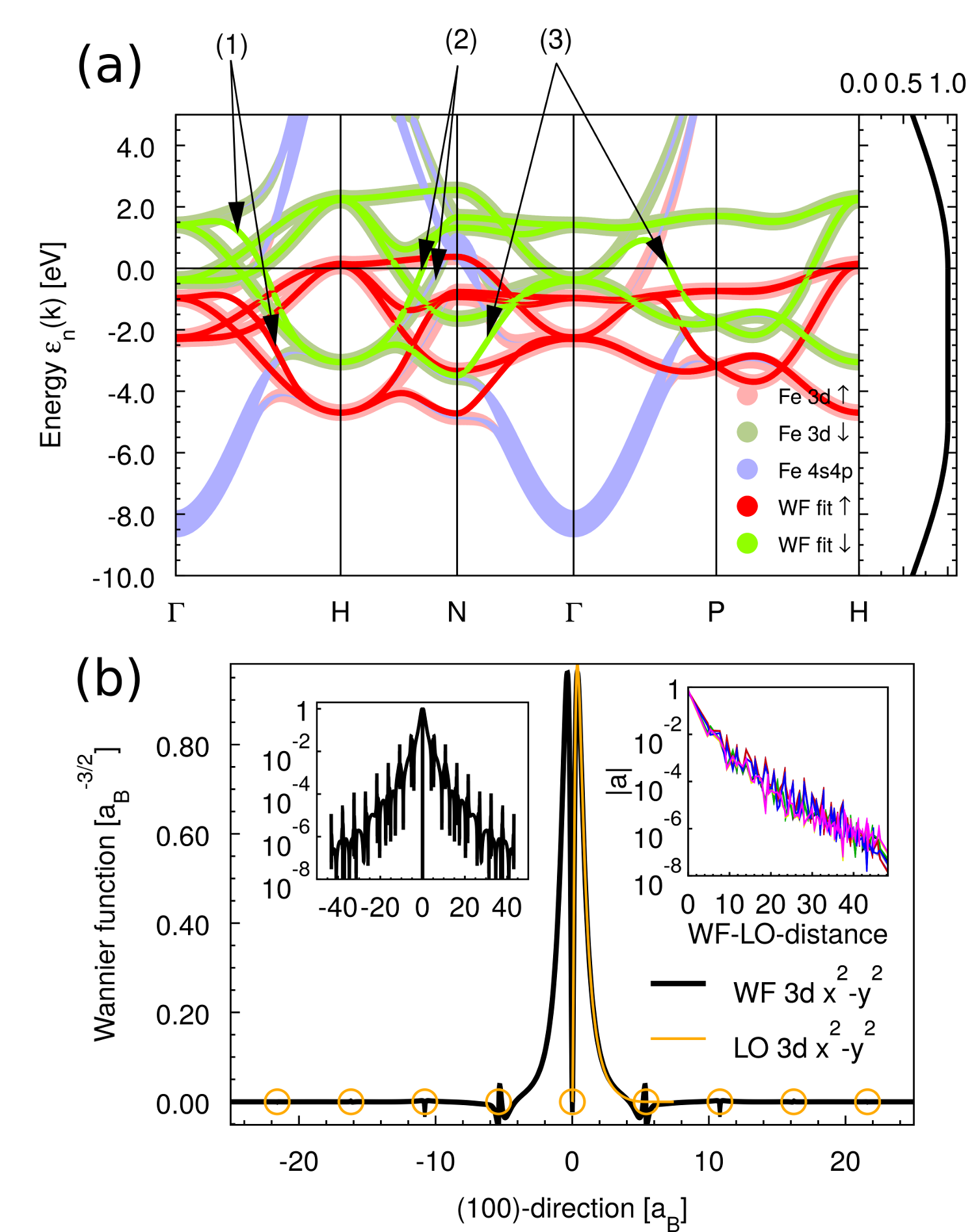}
\caption{(color online) 
(a) band weights of the FPLO calculation in pastel colors and Fe $3d$ 
WF fit in bright colors (see text). (b) the $3d_{x^2-y^2}$ WF printed along the (100)-axis
through the WF center (black), the Fe $3d_{x^2-y^2}$ local orbital (orange).
Left inset shows $\left|w\right|$ along the same
  path in logarithmic scale.
Right inset shows the maximal orbital contributions to the WFs in 
logarithmic scale as a function of WF-LO distance.
\label{fig:Fe3d}}
\end{figure}

We performed a Wannier fit using $3d$ orbitals with non-relativistic
symmetry as projectors (see Sec.~\ref{sec:LOsymemtry},
Eq.~(\ref{eq:pseudoNRELLOtransform})). These are obtained from a
linear combination of the four spinors Eq.~\ref{eq:fourspinor}, such
that the resulting orbital transforms as a real Harmonic (which is an
inbuilt option besides projection onto the relativistic four spinors
themselves). The energy window as indicated at the right side of
Fig.~\ref{fig:Fe3d}a has a core region $[-5,0]$~eV and Gaussian tails
of width $\Delta=7$~eV.  We used a $\bok$-mesh with $16^3$ points in
the primitive reciprocal cell and a cutoff $\rho=25 a_{\mathrm{B}}$
for the real space representation of the WFs. Although, the energy
window is quite large, the resulting Wannier fit (bright green and
red) follows the bands with $3d$-character (milky green and red) very
closely at most parts of the Brillouin zone (BZ). The WFs also resolve
the original spin character.

Fig.~\ref{fig:Fe3d}a shows three regions marked by numbers. In region
1 the highest minority $d$-band at $\Gamma$ becomes a free electron
band (milky blue) as it progresses towards the H-point. At the same
time an $sp$-band enters the $d$-region from below. So, the WFs
must reconnect the manifold at $\Gamma$ to the manifold at H by
interpolating through some dispersion not existing in the DFT band
structure (the same happens for the majority spin bands).

One of the lowest (degenerate) minority bands (-3~eV) at H bends
upwards and turns into an $sp$-band at N, as marked by the right arrow
in region 2, while the missing $d$-character flows in from above 2~eV.
At the N-point all $d$-weight sits in the two bands at 1-1.5~eV.
Consequently, the WF (left arrow) must deviate from the DFT band to
smoothly interpolate the pure $d$-bands between H and N.

A similar situation as in region 1 occurs in region 3 between the H-
and $\Gamma$-point and the $\Gamma$- and P-point,
respectively. $d$-weight flows to lower energies from N to $\Gamma$
and to higher energies between $\Gamma$ and P, which forces the WFs to
deviate from the original band structure.  Between, P and H the second
and forth lowest Wannier band deviate from the DFT bands by $180$~meV
(not visible in the fat-band plot).  Besides these regions the WF fit
follows the original bands rather closely.

Fig.~\ref{fig:Fe3d}b plots the WF along the $(100)$-direction through
the Wannier center (black) together with the $3d_{x^2-y^2}$ orbital
(orange). The circles denote the atom positions. There are some
notable orthogonalization tails at first and second neighbor, but
otherwise the WF and the local orbital are nearly identical.  The
right inset shows $\max\left(|a_{\boR\bos,\boldsymbol{0}\boc}|\right)$
(Eq.~(\ref{eq:WffromLO})) as a function of WF-LO distance and the left
inset shows the WF along the same direction as the main panel with
logarithmic $y$-axis. Clearly, the expansion coefficients as well as
the WF itself are exponentially localized and the localization is
similar to that of the $d$-orbital.

\begin{figure}[h]
\includegraphics[width=0.99\columnwidth]{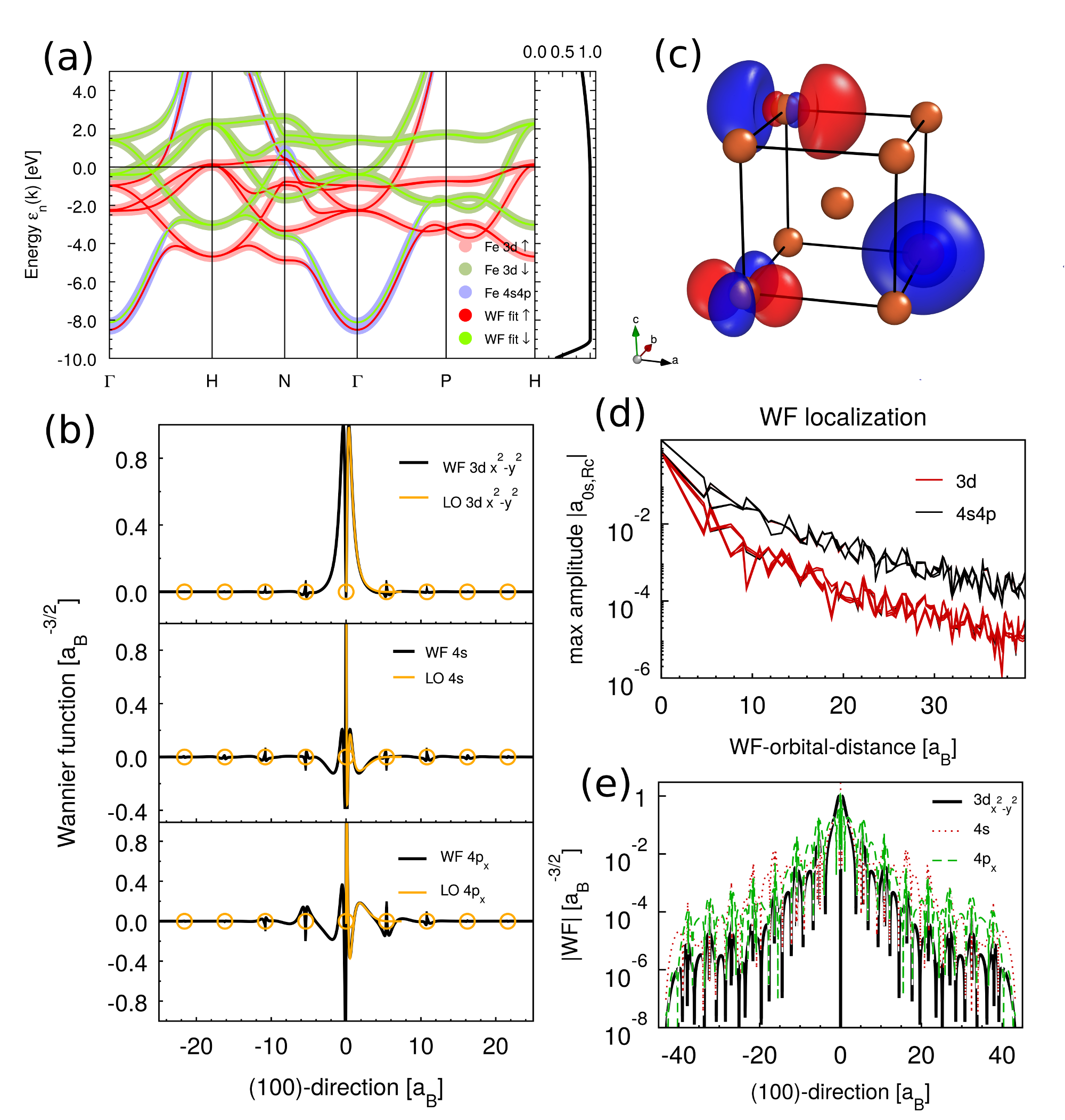}
\caption{(color online) 
(a) band weights of the FPLO calculation in pastel colors and the
WF fit in bright colors. (b) top-down: the $3d_{x^2-y^2}$, $4s$ and
$4p$ WFs printed along the (100)-axis
through the WF center (black), the FPLO local orbitals (orange).
(c) the corresponding WF isosurfaces for isovalue 0.09.
(d) shows the maximal orbital contributions to the WFs in 
logarithmic scale as a function of WF-LO distance.
(e) shows $\left|w\right|$ along the same  path
 in logarithmic scale.
\label{fig:Fe3d4s4p}}
\end{figure}

To get a useful $3d$ WF-fit, the $4s4p$ orbitals have to be included,
which represents the minimum chemical basis for bcc Fe. The
corresponding projectors are all the $4s4p3d$-orbitals sitting at
their respective site, i.e.~$(000)$. The energy window needs to have a
rather large upper Gaussian tail to sample all the needed
representations of which especially the ``free-electron''-states lie
at higher energies. The core window is $[-9,-1]$ and the upper tail
has $\Delta=15$~eV (the lower does not
matter). A real space cutoff $\rho=40 a_{\mathrm{B}}$ was used.
Fig.~\ref{fig:Fe3d4s4p}a shows the resulting WF-fit. The DFT
results are shown in pastel colors with larger width while on top the
WF-fit is plotted with a smaller width. The two sets of bands are on
top of each other to within $20$~meV error. Fig.~\ref{fig:Fe3d4s4p}b
shows the WFs along the $(100)$-direction through the origin for the
$3d_{x^2-y^2}$, $4s$ and $4p_{x}$-orbitals together with the
corresponding FPLO basis functions (orange and only plotted towards
the positive axis). The WFs follow the orbitals very closely within
the first neighbor region, while the ``free-electron''-WFs have
sizable hybridization tails which is related to their large band
width.  Fig.~\ref{fig:Fe3d4s4p}c shows the (suitably translated)
isosurfaces of these three functions for isovalue $0.09$, while (d)
shows the maximum orbital contributions
$\max\left(|a_{\boR\bos,\boldsymbol{0}\boc}|\right)$ to all WFs in
logarithmic scale, which shows exponential localization, albeit less
localized then the pure Fe-$3d$-fit discussed above. Note, that the
$3d$ orbitals are more localized than the rest as is to be expected.
Finally Fig.~\ref{fig:Fe3d4s4p}e shows the absolute value of the WFs
from (b) in logarithmic scale along the same path, which also proves
localization.

\subsection{Berry curvature, Anomalous Hall Conductivity}
\label{sec:AHC}

\begin{figure}[h]
\includegraphics[width=0.99\columnwidth]{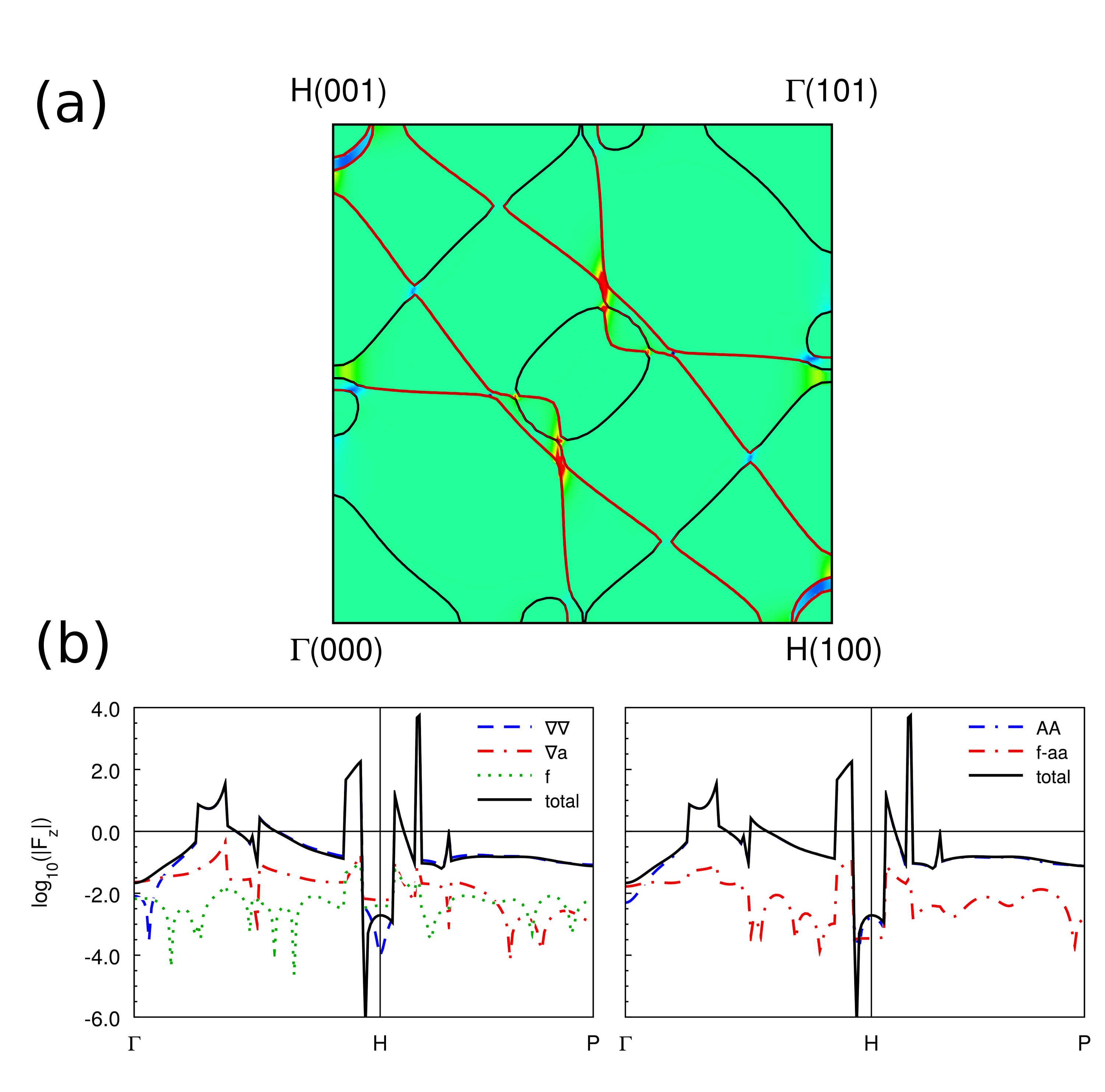}
\caption{(color online) 
Berry curvature of spin polarized full relativistic bcc Fe.
(a) Thin lines: Fermi surface contour in the $k_{y}=0$ plane.
Color plot: $-\boF_z$.
(b) The Berry curvature components along the path $\Gamma$HP
(see text). The left panel shows term-grouping of
Ref.~\onlinecite{Wan06}: $\nabla\nabla=i\left\langle
  Si\nabla\right\rangle _{C}\times\left\langle Si\nabla\right\rangle
_{C}$, $\nabla{}a=i\left\langle Si\nabla\right\rangle
_{C}\times\left\langle \boA\right\rangle _{C}+\text{h.c.}$ and $f=\left\langle \boldsymbol{f}\right\rangle _{C}$
The right panel shows our preferred grouping
$AA=i\boA_{\Psi}\times\boA_{\Psi}$ and $f-aa=\left\langle \boldsymbol{f}\right\rangle _{C}-i\left\langle \boA\right\rangle _{C}\times\left\langle \boA\right\rangle _{C}$.
\label{fig:BerrycurvFe}}
\end{figure}

To compare to Ref.~\onlinecite{Wan06} (Wang06) and to illustrate the
Berry curvature terms and their grouping we show the results for the
Berry curvature of full-relativistic spin-polarized bcc Fe in
Fig.~\ref{fig:BerrycurvFe} (for the $3d4s4p$-Wannier model of
Sec.~\ref{sec:banddisentangling}). By comparing the Fermi surface
contour to Ref.~\onlinecite{Wan06} it becomes clear that the two band
structure codes do not give the exact same result, which is not
surprising given that the cited results use a pseudo-potential code
and treats the relativistic effects differently from our full
4-component treatment. It is likely that our results compare better
for slightly different lattice constants. Nevertheless, the Berry
curvature plot and the curvature plotted along the Brillouin zone path
$\Gamma(000)$-H$(010)$-P$\left(\frac{1}{2}\frac{1}{2}\frac{1}{2}\right)$
compare well. Fig.~\ref{fig:BerrycurvFe}b shows the two different ways
of grouping the terms. The left panel uses the grouping of Wang06,
which corresponds to Eq.~(\ref{eq:BerryCurvatureAD}) while the right
panel shows the grouping Eq.~(\ref{eq:BerryCurvature}) (also discussed
in the appendix of Wang06), preferred by us for symmetry, gauge
invariance and approximation reasons.  In bcc Fe there is only one
site at $(000)$ and hence the gauge dependent corrections are all zero
and periodic and relative gauge identical.  The term-grouping
according to the Wang06 shows that the
$\left\langle{}Si\nabla\right\rangle\times\left\langle{}Si\nabla\right\rangle$
is indeed the dominant term while
$\left\langle{}Si\nabla\right\rangle\times\left\langle{}\boA\right\rangle_{C}$
and $\left\langle{}f\right\rangle_{C}$ are quite small. The
approximation Eq.~(\ref{eq:approxA}) is zero and hence the
approximated curvature is the
$\left\langle{}Si\nabla\right\rangle\times\left\langle{}Si\nabla\right\rangle$
term itself.

Our preferred term grouping in the right panel shows that
$\boA_{\Psi}\times\boA_{\Psi}$ is even closer to the total result
while
$\left\langle{}f\right\rangle_{C}-\left\langle{}A\right\rangle_{C}\times\left\langle{}A\right\rangle_{C}$
is rather small.

We also integrated the Berry curvature over the irreducible part of
the BZ (with subsequent symmetrization) to obtain anomalous Hall
conductivity (AHC). To speed up this quite time consuming calculation
we reduced the real space cutoff to $\rho=25 a_{\mathrm{B}}$ which
leads to small band energy errors of maximal $200$~meV around the
N-point, but yields a smaller Wannier model.  For a mesh with
$300^{3}$ k-points in the total BZ we obtain
$715.4\Omega\text{cm}^{-1}$ and $720.5\Omega\text{cm}^{-1}$ for the
$\nabla\times\nabla$-only and the full Berry curvature, respectively.
For $600^{3}$ k-points this becomes $717.9\Omega\text{cm}^{-1}$ and
$722.90\Omega\text{cm}^{-1}$.  This is a bit smaller than the 
$756.8\Omega\text{cm}^{-1}$ reported by Ref.~\onlinecite{Wan06}, which 
used a finer adaptive integration. However, their smallest mesh which
compares to our largest also deviates by about the same amount, which is most
likely due to the method/band structure differences discusses above.

\begin{figure}[h]
\includegraphics[width=0.99\columnwidth]{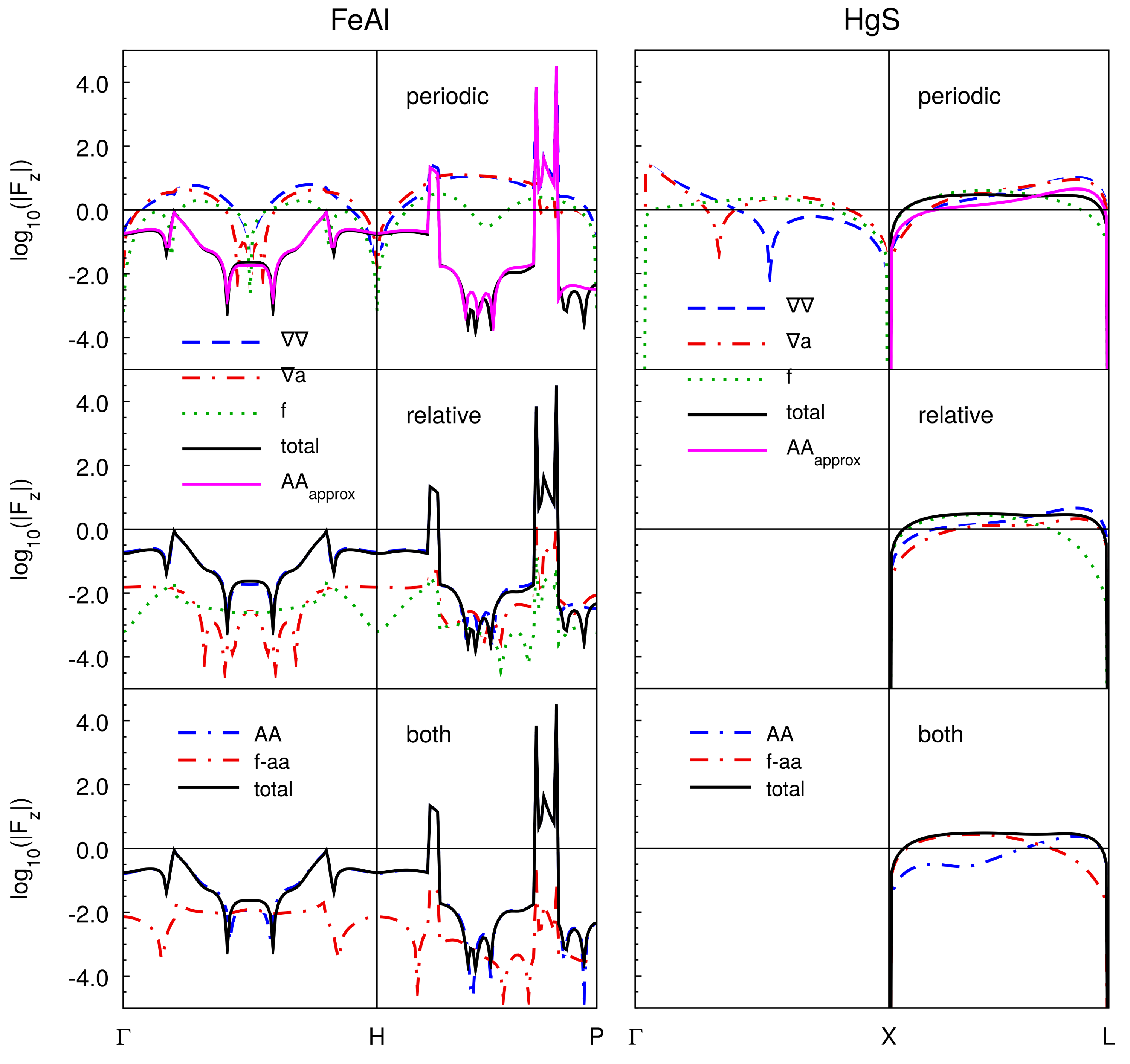}
\caption{(color online) 
  Left: Berry curvature of spin polarized full relativistic B2 FeAl.
  Right: Berry curvature of non spin polarized full relativistic HgS.
  Top: term-grouping of Ref.~\onlinecite{Wan06} for periodic gauge.
  Middle: term-grouping of Ref.~\onlinecite{Wan06} for relative gauge.
  Bottom: preferred term-grouping.
  The terms are explained in Fig.~\ref{fig:BerrycurvFe} except for
$AA_{\text{approx}}$ which is the $AA$-term with approximated basis connection.
\label{fig:Berrycurv}}
\end{figure}

To further assess the validity of the general arguments of
Sec.~\ref{sec:methodposop} in favor of our preferred term grouping we
show the Berry curvature results for two more cases. The left panel of
Fig.~\ref{fig:Berrycurv} shows the Berry curvature along a BZ-path for
spin-polarized full-relativistic B2 FeAl, which is obtained from bcc
Fe by replacing the iron at the unit cell center by Al (see
Sec.~\ref{sec:B2FeAl}), while the right panel shows $\boF$ for
non-spin-polarized full-relativistic HgS along the same path
(different high symmetry point names).  The top and middle rows show
the results for the term-grouping of Wang06 in the periodic and
relative gauge, respectively, while the bottom row shows the preferred
term-grouping which is independent of the Bloch sum
gauge. Additionally, in the top row the
$\boA_{\Psi}^{\mathrm{appr}}\times{}\boA_{\Psi}^{\mathrm{appr}}$-term
is shown within the approximation Eq.~(\ref{eq:approxA}), which is
distinct in the periodic gauge and equivalent to the
$\left\langle Si\nabla\right\rangle _{C}\times\left\langle
  Si\nabla\right\rangle _{C}$-term in relative gauge.

Clearly, for FeAl in the periodic gauge the individual terms in the
Wang06 grouping are larger than the total which demonstrates the
necessity of including the basis connection terms.  Interestingly, 
the approximated $\boA_{\Psi}^{\mathrm{appr}}\times{}\boA_{\Psi}^{\mathrm{appr}}$-term in periodic gauge
(which equals the $\nabla\times\nabla$-term in relative gauge) is
quite close to the total. The preferred grouping shows as for bcc Fe
that the $\boA_{\Psi}\times{}\boA_{\Psi}$-term is largely dominant.

For HgS the situation is less satisfactory. We first note that all
individual terms in the Wang06 grouping in periodic gauge (top right
panel) are nonzero along $\Gamma$-X, although the total as well as the
approximated $\boA_{\Psi}^{\mathrm{appr}}\times{}\boA_{\Psi}^{\mathrm{appr}}$-term are zero due to symmetry
(non-spin polarized).  In the relative gauge (middle right panel) as
well as in the preferred grouping (bottom right panel) the symmetry is
preserved, which confirms the general arguments of
Sec.~\ref{sec:methodposop}. On the line X-L neither term grouping has
a dominant term.

\subsection{Molecular orbital projectors}

\begin{figure}[h]
\includegraphics[width=0.99\columnwidth]{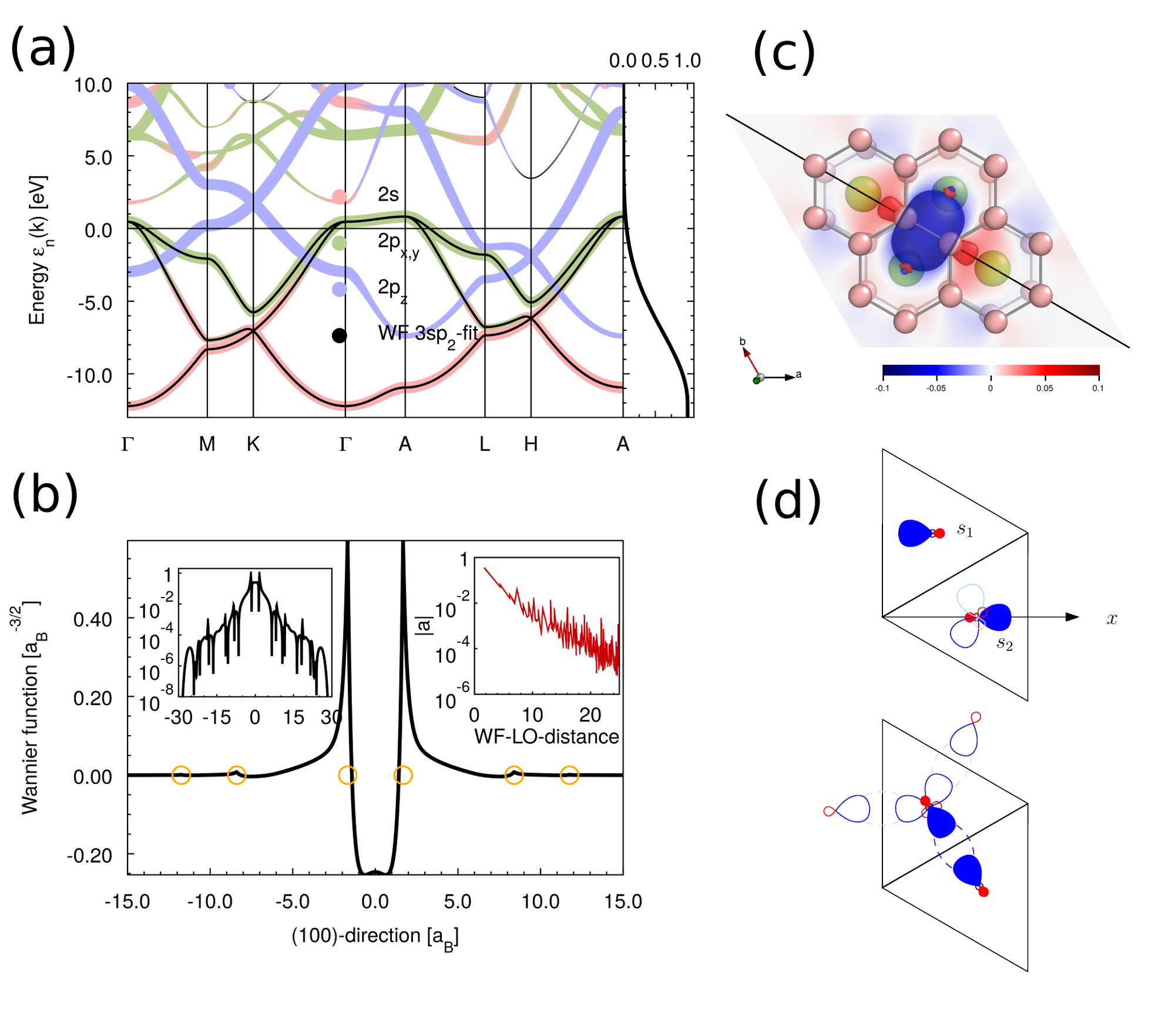}
\caption{(color online) 
(a) the FPLO band weights in pastel colors compared to the 
$sp^2$-hybrid Wannier fit in black and the energy window plotted at
the right side. (b) the WF profile along a path through the Wannier
center as indicated by the thin line in c. The left inset shows $|w|$
along the same path in a logarithmic plot, while the right inset shows 
$\max\left(|a_{\boR\bos,\boldsymbol{0}\boc}|\right)$, i.e.~the orbital
contributions to the WF, also in a logarithmic plot. (c) shows the 
isosurface of the WF for isovalue $0.4$. (d) a schematic of the
MO-projectors used to obtain the WFs.
\label{fig:MgB2}
}
\end{figure}

Up to now we always projected onto single orbitals, which creates WFs
which look like these orbitals in the Fe-case or which acquire sizable
hybridization tails for CaCuO$_2$. Especially in cases of isolated
band complexes (and if these complexes are Wannier representable) bond
centered WFs might be needed. To illustrate this we perform a Wannier
fit for MgB$_2$, which has alternating triangular Mg- and hexagonal
boron-layers. The occupied band structure consists of 3
boron-$2s2p_{x,y}$ bands, which can be described by $sp^2$-hybrids and
one mostly decoupled boron-$2p_z$ band. We will construct a Wannier
model of the occupied $sp^2$-hybrids. 

Fig.~\ref{fig:MgB2}d shows the two boron sites in the boron plane of
the unit cell viewed down the $c$-axis. On each site we define a
projector containing a $2s$- and a $2p_{x}$-orbital with the local
quantization axis chosen such that the linear combinations as depicted
in the upper panel of Fig.~\ref{fig:MgB2}d (in solid color) are
obtained. The $2s$- and $2p$-orbitals get a weight of $1$ and
$\sqrt{2}$, respectively, which corresponds to the relative weights in
an $sp^{2}$-hybrid. Also depicted are the other two hybrids at site
$s_2$, which are obtained by a $C_3$-rotation. Together with the
rotated hybrids at site $s_1$ these form six MOs, of which the bonding
combinations, shown in the lower panel, form the 3 occupied
$sp^2$-bands. These unnormalized bonding $sp^2$-MOs consist each of
four LOs. A simple python script was used to setup these molecular
orbital projectors, which essentially uses the $C_3$ rotation to
determine the local axes of all contributing $p$-orbitals.

The energy window, shown at the right side of Fig.~\ref{fig:MgB2}a was
chosen such that the most part of the target bands where sampled by a
larger upper Gaussian tail with $\Delta_E=7$~eV. Due to the weak
hybridization with the $2p_{z}$-band away from the high symmetry lines,
the WF band energies at the band bottom along $\Gamma$-A would be
pulled to higher energies, compared to the FPLO bands, if a simple
rectangular window encompassing the target bands would be chosen.

The 3 bands of the WF-fit in Fig.~\ref{fig:MgB2}a accurately follow
the FPLO band structure. Fig.~\ref{fig:MgB2}c shows an isosurface of
one of the bonding $sp^2$ WFs together with a semi-transparent density
plot of the WF in the plane through it's center.  Along a path,
indicated by the thin line through the Wannier center, the WF has a
profile as shown in Fig.~\ref{fig:MgB2}b, where the circles mark the
positions of the boron atoms. The WF is clearly localized as shown by
the logarithmic plot of $|w|$ along the path in the left inset and the
logarithmic plot of the maximal orbital contributions
(Eq.~(\ref{eq:WffromLO})) to the WF in the right inset.

\subsection{Molecule Wannier functions}\label{sec:molWF}

\begin{figure}[h]
\includegraphics[width=0.70\columnwidth]{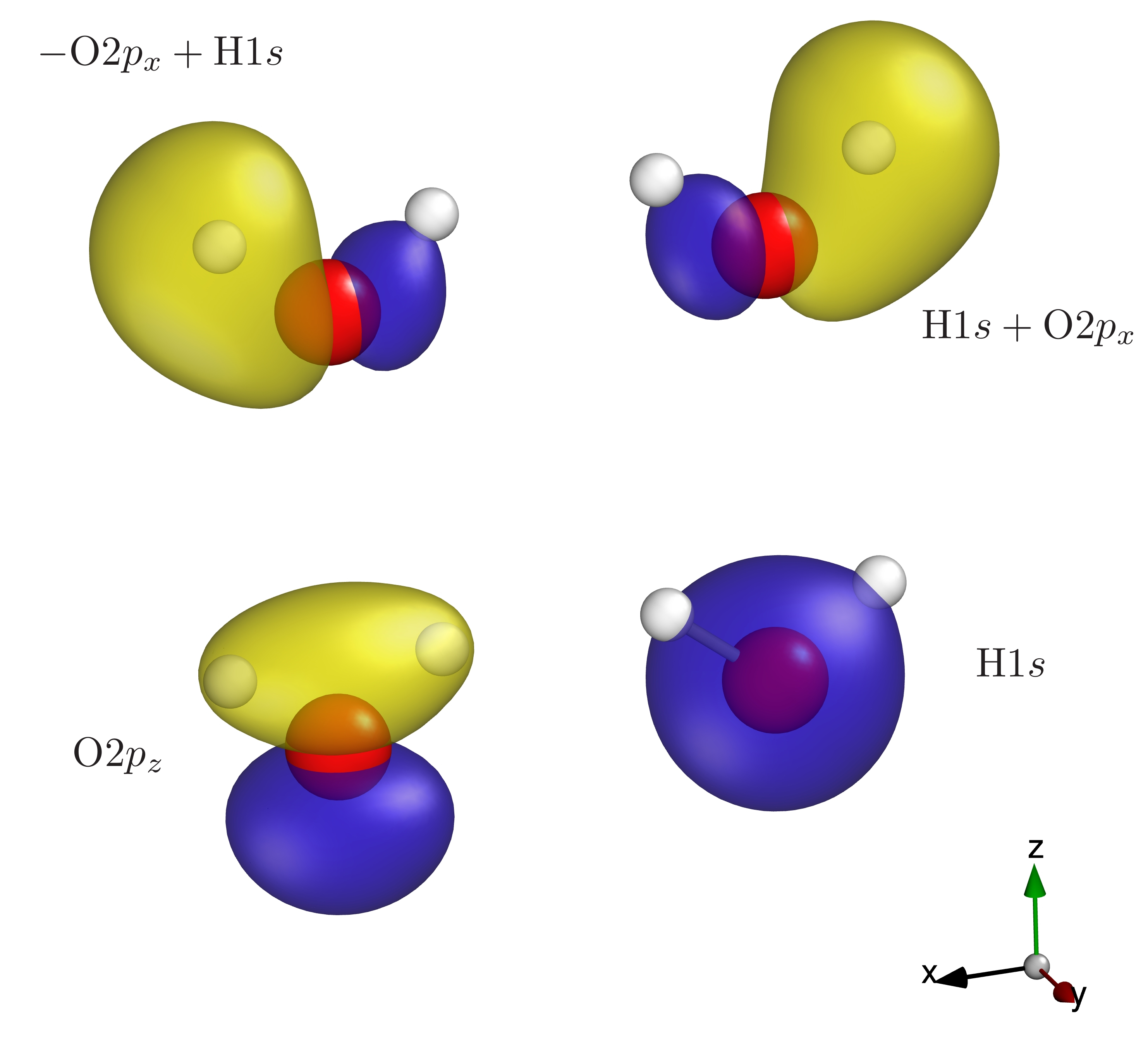}
\caption{
(color online)
Isosurfaces of the four WFs of the minimum model for H$_2$O 
for an isovalue of $0.09$. The character is denoted at the side.
The atoms are shown as white (H) and red (O) spheres.
\label{fig:H2O}
}
\end{figure}

In FPLO molecules are treated without a simulation box, since all
orbitals have a finite compact support.  The algorithm described in
Sec.~\ref{sec:methodwf} is not restricted to bulk materials. The only
places where the extended nature of a sytem comes into play are
$\bok$-sums and by using the $\Gamma$-point only (effectively no sum)
one can also create molecule WFs. To illustrate this we will determine
the smallest model Hamiltonian for an H$_2$O molecule (see
Sec.~\ref{sec:H2O}). There are 4 occupied valence MOs formed by
O-$2s/2p$ and H-$1s$ orbitals of which one is formed to nearly $100\%$
by the O-$2p_{\text{y}}$ orbital, whose lobes point out of the H-O-H
plane and which hence is decoupled by symmetry from the other orbitals
in lower orders, so that this orbital can be neglected in a model. To
obtain WFs we chose the O-$2s$, O-$2p_{z}$ and the two H-$1s$ orbitals
as projectors. The O-$2p_{x}$ orbital will be pulled into the Wannier
basis via hybridization tails at the other WFs by choosing an energy
window $[-24,0.3]$~eV with an upper $\Delta=1$~eV, which encompasses
the four lowest valence MOs and some unoccupied states. This basis will
fit the three mixed occupied MOs and one unoccupied MO. The
corresponding Wannier model Hamiltonian Eq.~\ref{eq:H2OHam} will have
the same eigen energies as the corresponding MOs in the DFT
calculation Eq.~\ref{eq:H2OErg} except for the omitted orbital.  The
Hamiltonian parameters are $h_{2s}=-18.9547$, $h_{2s,2p_{z}}=1.6832$,
$h_{2p_{z}}=-8.3644$, $h_{2s,1s}=6.5549$, $h_{2p_{z},1s}=-2.9232$,
$h_{1s}=-8.9814$ and $h_{1s,1s}=3.5277$.
\begin{equation}
  \label{eq:H2OHam}
H=\left(\begin{array}{cccc}
h_{2s} & h_{2s,2p_{z}} & h_{2s,1s} & h_{2s,1s}\\
h_{2s,2p_{z}} & h_{2p_{z}} & h_{2p_{z},1s} & h_{2p_{z},1s}\\
h_{2s,1s} & h_{2p_{z},1s} & h_{1s} & h_{1s,1s}\\
h_{2s,1s} & h_{2p_{z},1s} & h_{1s,1s} & h_{1s}
\end{array}\right)
\end{equation}
\begin{equation}
  \label{eq:H2OErg}
E=\left(\begin{array}{rl}
-24.398&\\
-12.509&\\
-8.6791&\\
-6.6177&\quad\text{O$2p_{y}$ omitted}\\
 0.3046
\end{array}\right)  
\end{equation}
Isosurfaces of the resulting WFs in real space are shown in Fig.~\ref{fig:H2O}.

\subsection{Post processing}

For post processing of the Wannier function data a python package is
provided.  It reads the Wannier data and maps them onto a chosen
structure. This can be the original unit cell forming the original
lattice, any supercell and corresponding lattice, a semi infinite slab
or a finite slab based on a chosen supercell. If the original lattice
is 2d the resulting slabs are 1-dimensional. The slabs are idealized
in the sense that no surface relaxation effects can be considered; it
is a straightforward mapping of the original hoppings/data onto the
sites forming the slabs.  For the semi-infinite slab the termination
can be chosen and for the finite slab both terminations can be chosen
and an arbitrary number of atoms can be removed to get a rough estimate
of such effects on the surface states.

The python package offers an interface to obtain the Hamiltonian and 
all requested operators for a user specified $\bok$-point, which
provides maximum flexibility.
It also offers ready made procedures to calculate several properties
depending on the chosen structure mapping.  For all but the
semi-infinite mapping the band structure, including band weights, and
Fermi surfaces and Fermi surface cuts can be calculated.  For the
original and super cell mappings a bulk-projected band structure can
be obtained which integrates the band structure along a chosen
$\bok$-direction. This offers the spectral densities with surface
effects excluded.

\begin{figure}[h]
\includegraphics[width=0.90\columnwidth]{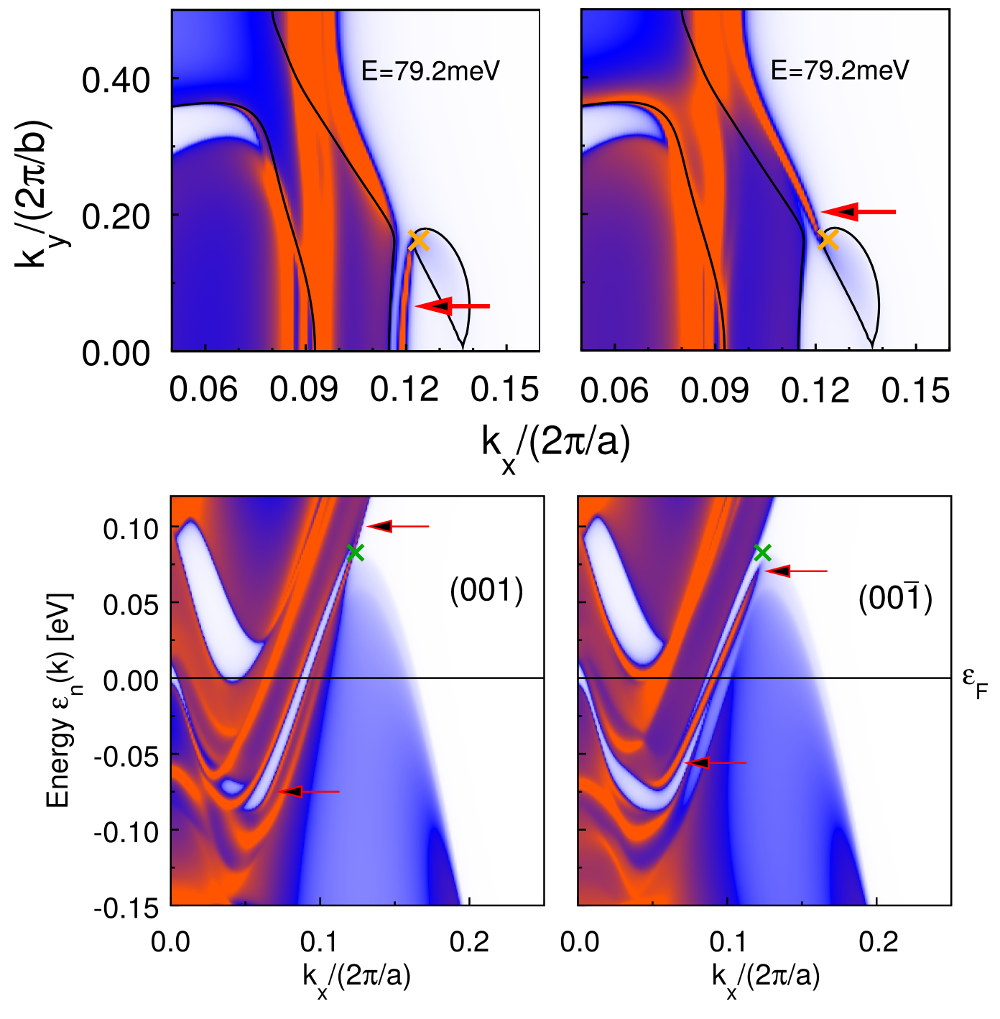}
\caption{(color online) 
Surface spectral densities (SD) for TaIrTe$_4$
of a semi infinite slab with $(001)$- (left column) and $(00\bar{1})$-
(right column) termination. The crosses mark the position of the Weyl point.
Upper row: Fermi surface SD, the arrows mark the Fermi arcs.
Lower row: the corresponding energy resolve SD along a line
$k_{y}=0.125$. The arrows mark the surface bands forming the arcs.
\label{fig:tairteEDC}
}
\end{figure}

For the semi-infinite slabs a Green's function
method\cite{san84,san85} is used to calculate spectral densities as a
function of energy and momentum (band structure plot equivalent) or in
a $\bok$-plane (Fermi surface cut
equivalent). Fig.~\ref{fig:tairteEDC}
shows an example of this for the ternary Type-II Weyl semi-metal
TaIrTe$_4$\cite{Koe16}.

For the mapping onto the original lattice the symmetry operations and
their eigenvalues can be analyzed. Furthermore, an adaptive algorithm
for the search for Weyl points (WP) is available.  We use a 3d version
of the methods of Ref.~\onlinecite{fuk05} to determine the chirality of
each micro-cell of a regular subdivision of the primitive unit
cell. Then in refinement steps for each cell with a non-trivial
chirality the cell is subdivided into 8 sub-cells for which the
chirality again is determined. The subdivision stops if the sub-cell size
falls bellow a specified threshold.  For each found Weyl point it's
veracity can be be checked by calculating an integral of the Berry
curvature over a small spherical shell around the WP, as well as
creating a plot of the Berry curvature field for visual confirmation.
The resulting Fermi arks can be analyzed by calculating the spectral
density for a mapping onto a semi-infinite slab or by calculating the
band structure and Fermi surface cuts for a finite slab.

Also for the original mapping the Z$_2$ topological indices can be
determined for non-centrosymmetric lattices using a Wannier-center
algorithm\cite{Yu11} with automatic determination of the indices
\cite{Sol11}. This algorithm is also directly available from the
FPLO code itself, by internally L{\"o}wdin orthogonalizing the whole
FPLO basis (which results in full-basis Wannier functions), but it is
faster to use a Wannier model due to the resulting basis reduction.
For centrosymmetric lattices FPLO itself calculates the Z$_2$
indices.

The Berry curvature can be calculated band wise for any non-slab
mapping either with approximated reduced position operator as
discussed in Sec.~\ref{sec:methodposop} or in full form, which also
gives access to the anomalous Hall conductivity (Sec.~\ref{sec:AHC}).
By using the symmetry information mirror Chern numbers\cite{Leg14,Fac19} for
topological crystalline insulators can be obtained as well.

Besides Hamiltonian and reduced position operator in
full-relativistic mode the spin operators
and the exchange correlation magnetic field can be extracted. 
This is achieved by starting from the xc-term of the Hamiltonian
\begin{equation}
H_{B}=\left\langle
  \Phi^{\bok}\mid\beta\boldsymbol{\Sigma}\boldsymbol{B}\mid\Phi^{\bok}\right\rangle
\end{equation}
where the $4\times4$-matrices $\beta$ and $\boldsymbol{\Sigma}$ are defined as 
\begin{equation}
  \beta=\left(\begin{array}{cc}
\boldsymbol{1}_{2\times2}\\
 & -\boldsymbol{1}_{2\times2}
\end{array}\right)
,\quad\boldsymbol{\Sigma}=\left(\begin{array}{cc}
\boldsymbol{\sigma}\\
 & \boldsymbol{\sigma}
\end{array}\right),
\end{equation}
$\boldsymbol{B}\left(\bor\right)$ is the xc-field and $\Phi^{\bok}$ is the
column-vector of the FPLO basis orbitals.
Assuming basis completeness the identity 
\begin{equation}
1=\bigl|\Phi^{\bok}\bigr\rangle\frac{1}{S^{\bok}}\bigl\langle\Phi^{\bok}\bigr|
\end{equation}
can be inserted to separate the field followed by the Wannier transformation
Eq.~(\ref{eq:blochwffromblochLO}) using the identity
$1=S_{w}^{\bok}=a^{\bok+}S_{\Phi}^{\bok}a^{\bok}$ which leads to
Wannier representation
\begin{eqnarray}
H_{B}^{w}&=&\left\langle \beta\boldsymbol{\Sigma}\right\rangle _{w}^{\bok}\left\langle \boldsymbol{B}\right\rangle _{w}^{\bok}\\\left\langle \beta\boldsymbol{\Sigma}\right\rangle _{w}^{\bok}&=&\left\langle w^{\bok}\mid\beta\boldsymbol{\Sigma}\mid w^{\bok}\right\rangle \\\left\langle \boldsymbol{B}\right\rangle _{w}^{\bok}&=&\left\langle w^{\bok}\mid\boldsymbol{B}\mid w^{\bok}\right\rangle   
\end{eqnarray}
with separate matrix representation for the vector of spin operators
and of the xc-field.  

The spin operator matrix can be used to add
model magnetic fields as was for instance done in
Ref.~\onlinecite{Bor19} to simulate a canted magnetic field. The python
interface provides model fields which are constant on user defined
subsets of WFs with the definition
\begin{equation}
H_{B}^{w}=\sum_{i}P_{i}\left\langle \beta\boldsymbol{\Sigma}\right\rangle _{w}\boldsymbol{B}_{i}P_{i}  
\end{equation}
where the diagonal of matrix $P_i$ is one for the targeted WFs and zero
otherwise and $\boldsymbol{B}_i$ is a constant vector.

The Wannier functions can also be used in the dHvA-package of FPLO
instead of the full FPLO data\cite{Khim16}. Although the dHvA-package
already uses an adaptive algorithm to sample the Fermi surface the use
of a Wannier model can considerably speed of the process.



\section{Performance and Comparison}
\label{sec:performance}
\subsection{Numerical performance}\label{sec:numperform}
For orientation Table~\ref{tab:timing} shows the timing of various stages of the
calculations of Sec.~\ref{sec:features}. We recorded the times for the
complete self consistent calculation, for the single loop, which dumps
the raw data needed for the WFs, for the actual Wannier function creation
run, for the calculation of the WFs on a real space grid and for post
processing, where it was used. All calculations are done on a single core.

\begin{table}[h]
\begin{tabular}{|c|c|c|c|c|}
\hline 
 & SCF & WF & WF on grid & post processing\tabularnewline
\hline 
\hline 
CaCuO$_{2}$ & 53 & 8+2 & 9 & \tabularnewline
\hline 
Fe ($3d4s4p$) & 39 & 11+10 & 76 & 198\tabularnewline
\hline 
FeAl & 74 & 23+77 & 248 & 32\tabularnewline
\hline 
HgS & 46 & 18+25 & 142 & 14\tabularnewline
\hline 
MgB$_2$ & 13 & 3+2 & 9 & \tabularnewline
\hline 
H$_2$O & 2.9 & 0.2+0.2 & 1.3 & \tabularnewline
\hline 
\end{tabular}

\caption{CPU time in seconds for the cases discussed in Sec.~\ref{sec:features} on
a single core of an Intel Xeon CPU E5-1650 @ 3.5 GHz. SCF: full self consistent
calculation from scratch. WF: calculation of Wannier functions consisting
of dumping the relevant data from a single SCF cycle (first number)
and the actual WF creation run (second number). WF on grid: the WF
creation run can optionally include the calculation of the WF on a
real space grid for visualization, which is not needed in most applications.
Post processing: for Fe the largest time is used for the calculation
of the Berry curvature of Fig.~\ref{fig:BerrycurvFe}a.}
\label{tab:timing}
\end{table}

\subsection{Comparison with Wannier90}\label{sec:compwan90}

At present, the most popular code in the community is Wannier90~\cite{Mos08,
Mos14, Piz20}. 
It supports interfaces with widely used band-structure codes has
a growing number of postprocessing options~\cite{Piz20}. It is therefore
reasonable to use Wannier90 for benchmarking our FPLO results. To this end, we
use band-structure tools from the Quantum ESPRESSO package version
6.7~\cite{Gia20} and the wannierization software provided in Wannier90 version
3.1.0. We use serial (single-core) executables compiled using the Intel's Math
Kernel Library (MKL) version 2020.4.304. In contrast to the full-potential
code FPLO, Quantum ESPRESSO does not calculate the potential of core electrons
explicitly, it is adopted from external pseudopotential files. Here, we used
ultrasoft LDA (Perdew-Wang 91~\cite{PerdewWang92})
Rappe-Rabe-Kaxiras-Joannopoulos (RRKJ) pseudopotenitals~\cite{Rap90} from
pslibrary version 1.0.0~\cite{DCo14}, except for  bcc-Fe calculations for
which projector-augmented-wave pseudopotenitals~\cite{Blo94} were employed.

\begin{figure}[tb]
  \includegraphics[width=8.6cm]{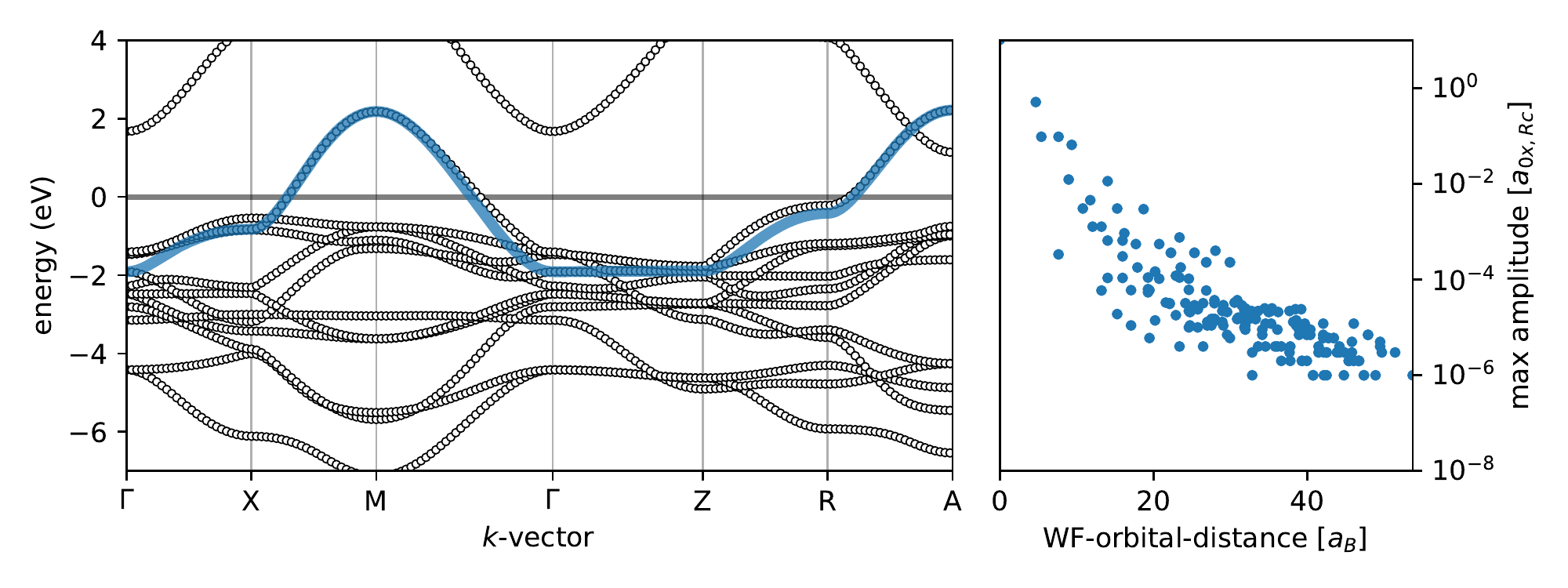}
  \caption{\label{fig:CaCuO2_w90} Left: LDA band structure of CaCuO$_2$
computed with Quantum ESPRESSO and Wannier fit for a one-orbital model
performed using symmetry-adapted scheme implemented in Wannier90. Right:
maximal contributions to the WF in logarithmic scale as a function of
WF-orbital distance.}
\end{figure}

We start with the one-orbital model for CaCuO$_2$. As these calculations do not
include the spin-orbit coupling, we can use symmetry-adapted Wannier
functions~\cite{Sak13} implemented in Wannier90~\cite{Piz20}. The agreement
with the GGA band and the localization of the resulting Wannier functions
(Fig.~\ref{fig:CaCuO2_w90}) are comparable to that in FPLO
(Fig.~\ref{fig:CaCuO2}).

\begin{figure}[tb]
  \includegraphics[width=8.6cm]{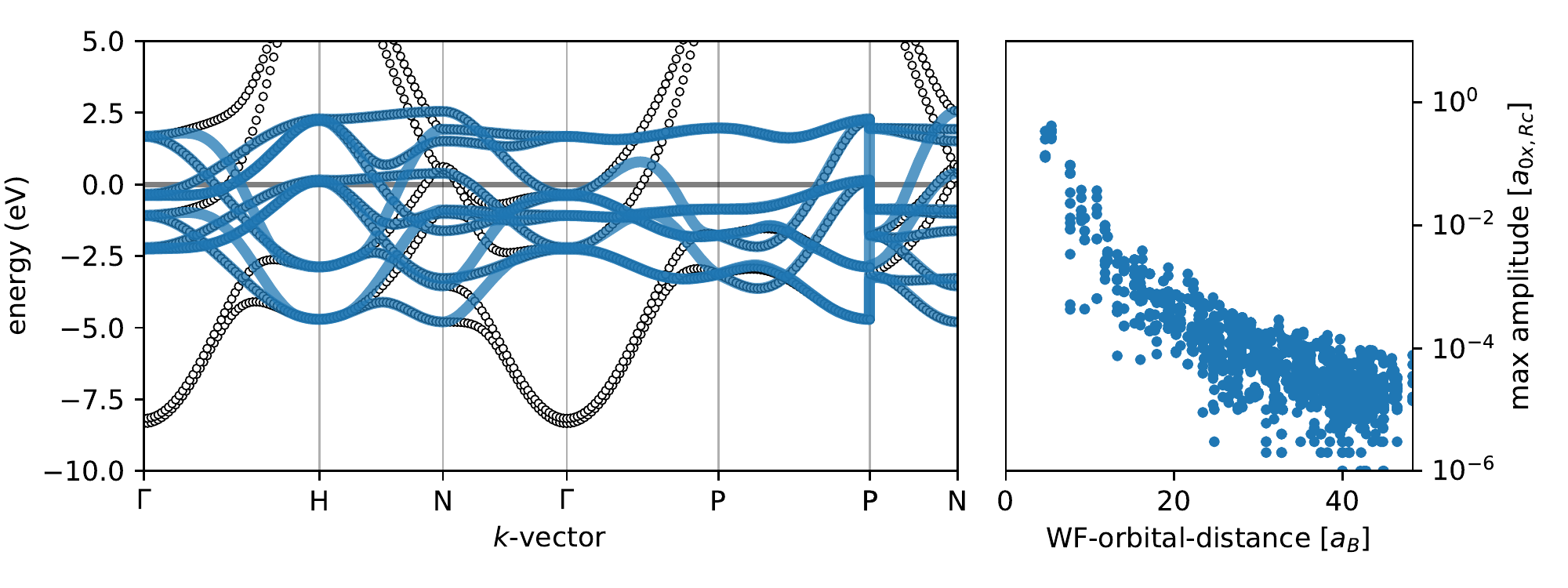}
  \includegraphics[width=8.6cm]{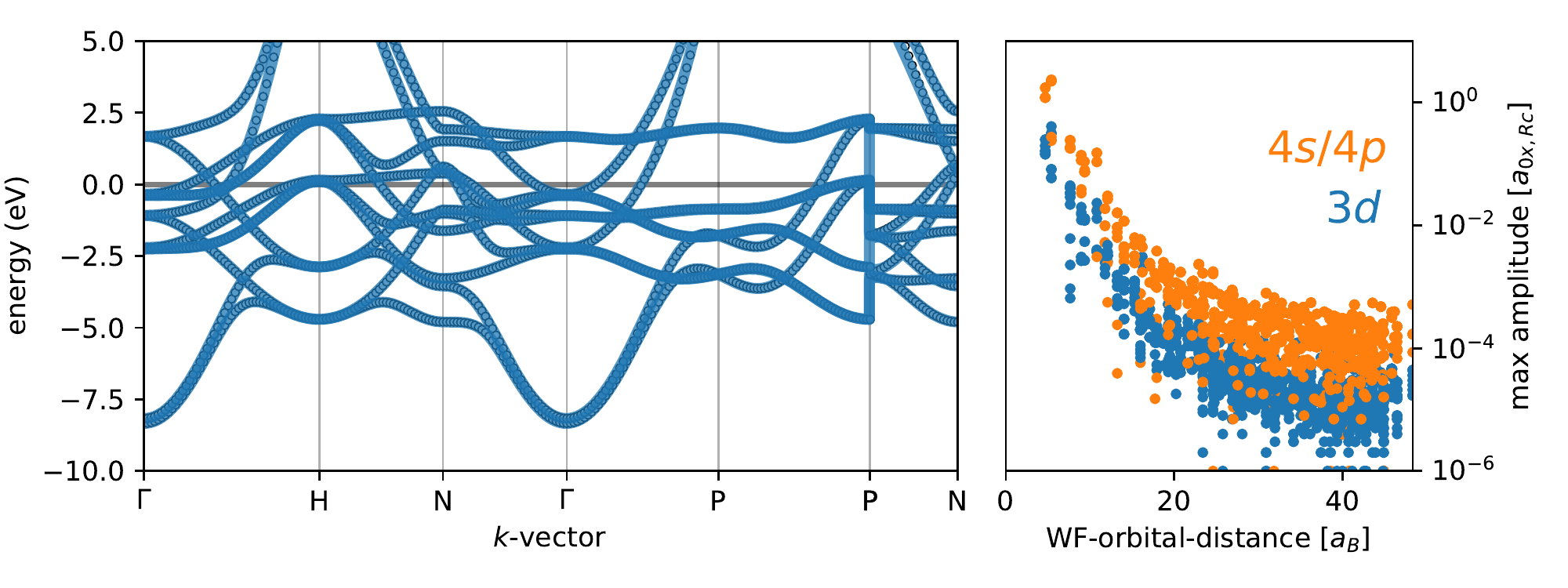}
  \caption{\label{fig:Fe_w90} Top: Wannier fit of the LDA band structure of
bcc Fe computed with Quantum ESPRESSO. The Wannier model is constructed by
projections onto $3d$ states. Right panel shows maximal contributions to the WF
in logarithmic scale as a function of WF-orbital distance. Only the
orbital-diagonal contributions are shown. Bottom: same for the model comprising
$3d$, $4s$, and $4p$ states.}
\end{figure}

\begin{figure}[tb]
  \includegraphics[width=4.5cm]{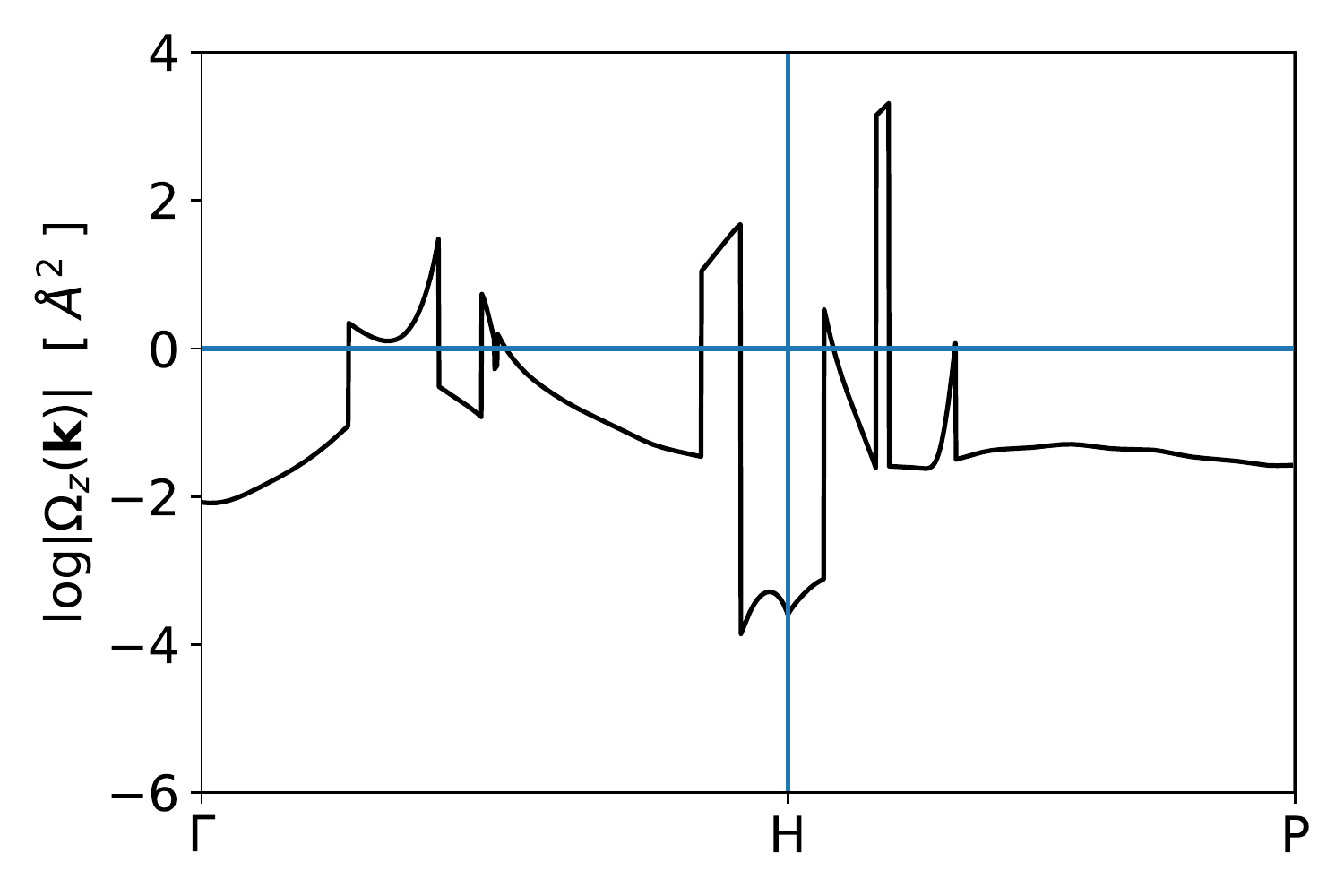}
  \includegraphics[width=3.9cm]{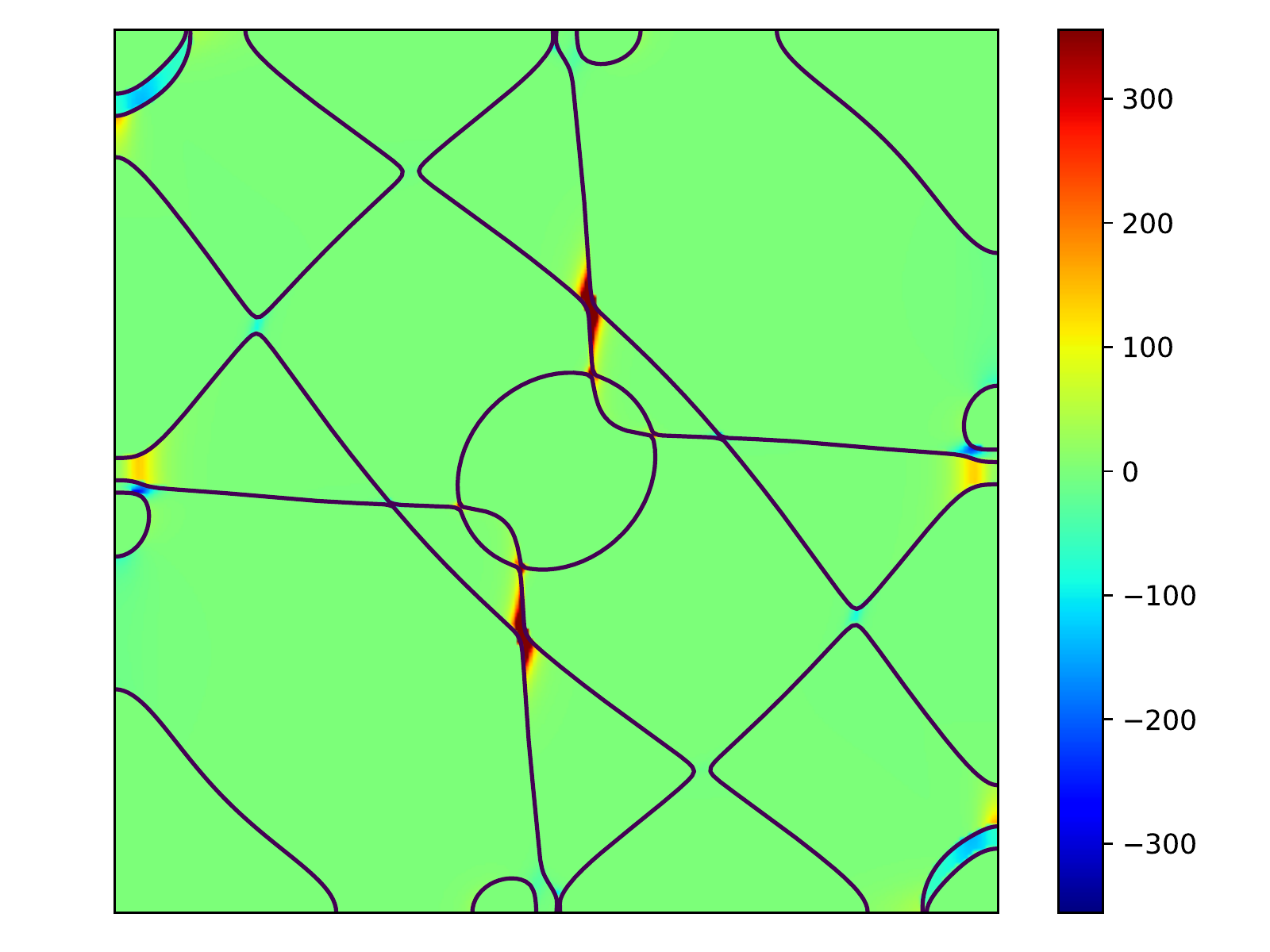}
  \caption{\label{fig:Fe_w90_bc} Berry curvature of spin polarized full
relativistic bcc Fe computed using Wannier90.}
\end{figure}

Next, we consider full-relativistic calculations of bcc-Fe. Again, we
distinguish two models: a ten-bands model which includes only the $3d$ states,
and a 18-bands model, which additionally accounts for $4s$ and $4p$
contributions. The former provides a generally good description of the LDA band
structure, but deviations can be seen with a naked eye (Fig.~\ref{fig:Fe_w90},
top left).  Inclusion of the $4s$ and $4p$ states readily yields an excellent
description of the entire valence band and the low-energy region of the
polarization band (Fig.~\ref{fig:Fe_w90}, bottom left). Also here the degree of
localization is comparable with respective models calculated using FPLO. 
Berry curvature calculations (Fig.~\ref{fig:Fe_w90_bc}) are in excellent
quantitative agreement with the FPLO results (Fig.~\ref{fig:BerrycurvFe}).

\begin{figure}[tb]
  \includegraphics[width=4.25cm]{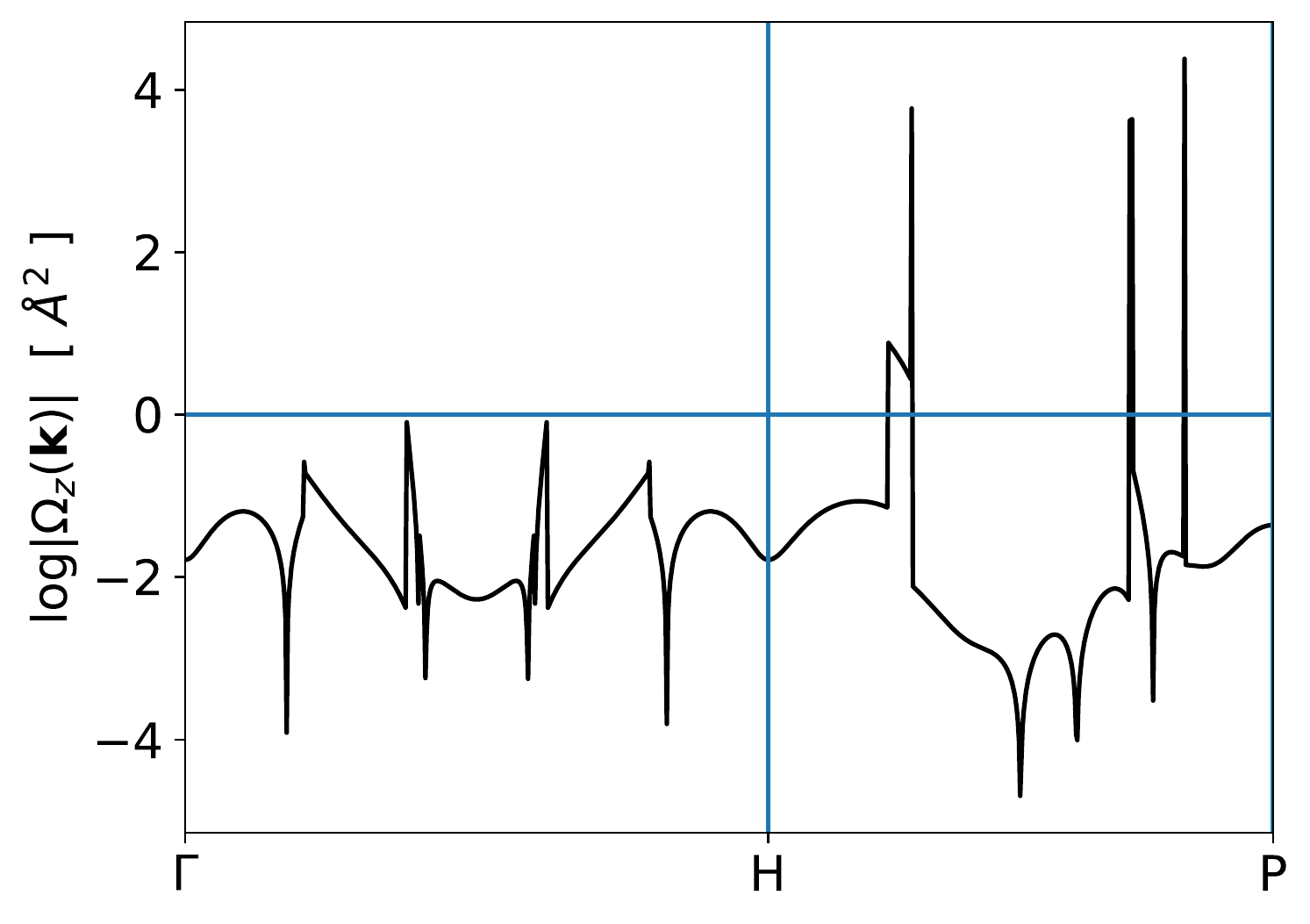}
  \includegraphics[width=4.25cm]{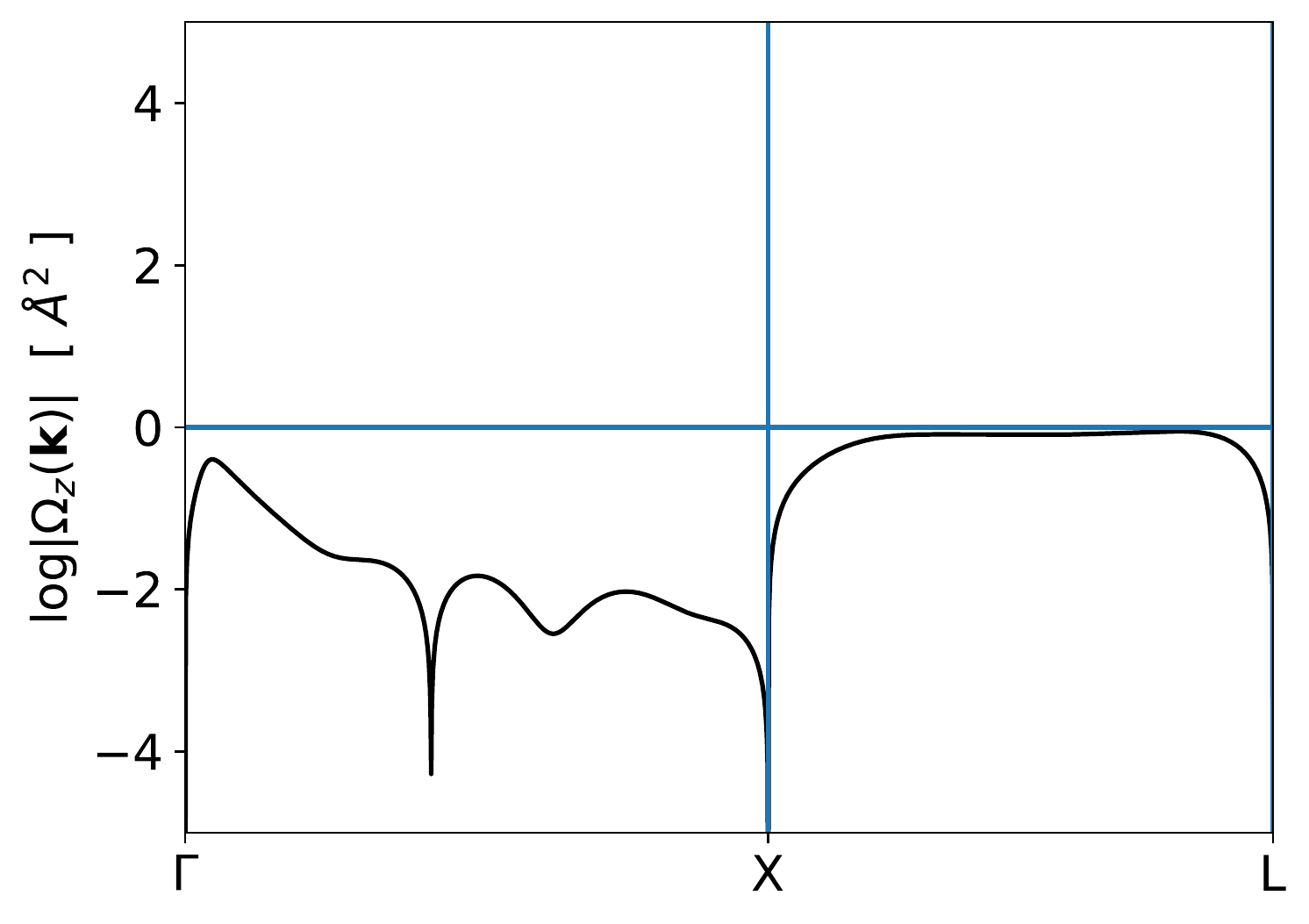}
\caption{\label{fig:w90_bc} Berry curvature of spin polarized full relativistic
FeAl (left) and nonmagentic HgS (right) computed using Wannier90.}
\end{figure}

Berry curvature of FeAl (Fig.~\ref{fig:w90_bc}, left) generally agrees
with the FPLO result (Fig.~\ref{fig:Berrycurv}, left), the
discrepancies likely stem from the RRKJ pseudopotential for Fe. For
HgS, the calculated Berry curvature does not vanish along the
$\Gamma$-X line (Fig.~\ref{fig:w90_bc}, right), in contrast with the
FPLO result (Fig.~\ref{fig:Berrycurv}, right). This discrepancy is
  however quite small and must be due to either symmetry violation
  (due to full relativistic mode) and
  or numerical artefacts, since we argued in Sec.~\ref{sec:AHC} that
  the total contribution must vanish along this line.

\subsection{Comparison with previous reports}\label{sec:prevreports}
We performed calculations of the anomalous Hall effect, the anomalous
Nernst effect, and the spin Hall effect based on the symmetry
conserving Wannier functions implemented in FPLO. Our results for the
anomalous Hall conductivity in Heusler compounds Co$_2$Fe$X$ ($X$=Al,
Ga, Si, and Ge) are around 60, 210, 260, and 64 $S/cm$, respectively,
by choosing the magnetic moment along the $z$ direction. We obtain good
agreement with the previous $ab-initio$ report~\cite{huang2015}. In
addition to calculations, different groups also performed experimental
Hall measurement for Co$_2$FeAl, Co$_2$FeGa, and Co$_2$FeSi. The
calculated results fit well with the experimental measurements in
Co$_2$FeGa and Co$_2$FeGe~\cite{hamaya2012estimation,
  bombor2013half}. Though our results and previous $ab-initio$
reports give similar anomalous Hall conductivity for Co$_2$FeAl, it is
much smaller than the value from Hall measurements~\cite{huang2015},
which might be due to a large extrinsic scattering contribution in the
experimental setup.

For the anomalous Hall effect in typical transition elements such as
Fe, Co, and Ni, our calculations show good agreement with both
experimental measurements and previous $ab-initio$
calculations~\cite{dheer1967galvanomagnetic,
  wang2007fermi,roman2009orientation, wang2007fermi}.  We also made
comparisons with some typical magnetic topological materials from
recent reports, such as Fe$_3$Sn$_2$ which features a massive Dirac cone
~\cite{wang2016anomalous,ye2018massive}, the first experimentally
verified magnetic Weyl semimetal
Co$_3$Sn$_2$S$_2$~\cite{liu2018giant}, and the magnetic nodal line
semimetals Rh$_2$MnGa, Rh$_2$MnAl, and
Fe$_3$GeTe$_2$~\cite{noky2020giant,kim2018large}.  We find good
agreements for all of these magnetic topological systems, see
Tables~\ref{tab:transportAHC}.

\begin{table*}[t]
\caption{
Anomalous Hall Effect.
}

\begin{tabular}{c|c|c|c}
\hline
\hline
Compound & Theory (S/cm)         & experiment (S/cm)                    &  this work (S/cm) \\
\hline
Co$_2$FeAl  & 39~\cite{huang2015}   & 320-360 ~\cite{imort2012anomalous}  & 60  \\
\hline
Co$_2$FeGa  & 181~\cite{huang2015}  & 50-350  ~\cite{hamaya2012estimation} & 210   \\
\hline
Co$_2$FeSi  & 189 ~\cite{huang2015} & 163, 400-600 ~\cite{hamaya2012estimation,bombor2013half} & 260 \\
\hline
Co$_2$FeGe  & 119 ~\cite{huang2015} &               & 64 \\
\hline
Fe$_3$Sn$_2$   &                       & 200~400~\cite{wang2016anomalous,ye2018massive} & 590 \\
\hline
Co$_3$Sn$_2$S$_2$  &                      & 1100 ~\cite{liu2018giant}  & 1130 \\
\hline
Co$_2$MnAl   & 1631~\cite{noky2020giant} & & 1520   \\
\hline
Rh$_2$MnGa   & 1860~\cite{noky2020giant} & & 1760    \\
\hline
Rh$_2$MnAl   & 1720 ~\cite{noky2020giant} & & 1640  \\
\hline
Fe        &                            & 1000 ~\cite{dheer1967galvanomagnetic} & 840\\
\hline
Fe$_3$GeTe$_2$  &                                & 500 ~\cite{kim2018large} & 590 \\
\hline
Ni        & 2200~\cite{wang2007fermi}                &                & 2300    \\
\hline
Co(hcp)        & 470 ~\cite{wang2007fermi}     &     & 480                     \\
\hline
Co (fcc)       & 250 ~\cite{roman2009orientation}     &     & 270               \\
\hline
\hline
\end{tabular}\label{tab:transportAHC}
\end{table*}

Recently, starting with high throughput calculations, strong anomalous Nernst effects were
observed in iron-based ferromagnets Fe$_3$$X$ ($X$=Pt, Ga, Al, and Si) and
Fe$_4$N~\cite{sakai2020iron}. From Table~\ref{tab:transportANE}, one can see
the good agreement between our calculations and recent reports.
We also find good agreements for the anomalous Nernst conductivity
in typical magnetic topological materials Co$_2$MnGa~\cite{guin2019anomalous},
Co$_3$Sn$_2$S$_2$~\cite{guin2019zero}, and Fe$_3$GeTe$_2$~\cite{xu2019large}.
All of them show the reliability of the Wannier functions derived from
FPLO.

  \begin{table*}[t]
--\caption{
  Anomalous Nernst Effect. maximum value with $T<500K$
  for Fe$_3$Pt, Fe$_3$Ga, Fe$_3$Al, Fe$_3$Si and Fe$_4$N.
  }

  \begin{tabular}{c|c|c|c}
  \hline
  \hline
  Compound & Theory (A/Km)  & experiment (A/Km) & this work       \\
  \hline
  Fe$_3$Pt  & 6.2 ~\cite{sakai2020iron} & & 4.0  \\
  \hline
  Fe$_3$Ga   &  3.0 ~\cite{sakai2020iron} &  & 3.0  \\
  \hline
  Fe$_3$Al   & 2.7 ~\cite{sakai2020iron}  & & 3.0  \\
  \hline
  Fe$_3$Si   & 2.5  ~\cite{sakai2020iron} & & 2.7 \\
  \hline
  Fe$_4$N     & 2.4  ~\cite{sakai2020iron} & & 1.1 \\
  \hline
  Co$_2$MnGa & 4 ($E_{F}=E_{0}+0.08$ eV)  & 6.2 (300K) ~\cite{guin2019anomalous} & 4.4 ($E_{F}=E_{0}+0.06$ eV)) \\
  \hline
  Co$_3$Sn$_2$S$_2$ &   & 3 (80K) ~\cite{guin2019zero}  & 2.6 \\
  \hline
  Fe$_3$GeTe$_2$ &   & 0.3 (150 K) ~\cite{xu2019large} & 0.38 \\
  \hline
  \hline
  \end{tabular}\label{tab:transportANE}
  \end{table*}

Another widely studied transport property in the linear response regime is the spin
Hall effect (SHC). We made a systematic comparison between reported results
and that obtained from FPLO. We find that the results for cubic
(bcc and fcc) transition metals from our calculations are in good agreement
with previous $ab-initio$ calculations, see
Tables~\ref{tab:tablePt},\ref{tab:tablefccAu},\ref{tab:tablefccPd},\ref{tab:tablebccTa},
\ref{tab:tablebccNb},\ref{tab:tablebccW},\ref{tab:tablebccMo}.
We also tried to compare our theoretical results to the
experimental reports. However, the experimental values
from different reports differ a lot, due to details of
experimental setup and extrinsic contributions.

\begin{table}[h]
  \centering
  \caption{Pt, SHC in the unit of $(\hbar/e)(S/cm)$
}
    \label{tab:tablePt}
	\begin{tabular} {l|c}
  \hline
  \hline
	  \multicolumn{2}{c}{Pt} \\
  \hline
          \multicolumn{2}{c}{$\sigma_{xy}^{z}$}    \\
  \hline
	   This work         &  2260 \\
  \hline  
	   Ref.($ab-initio$) & 2200~\cite{Guo2008}  \\
  \hline
	   Ref.($ab-initio$) & 2281~\cite{qiao2018}  \\
  \hline 
	   Ref.($ab-initio$) & $\sim$2200~\cite{ryoo2019}  \\
	   \hline
	   Ref.($ab-initio$) & $\sim$2500~\cite{gradhand2011}  \\
	   \hline
	   Ref.(Exp., T=10K) & 1700$\pm$400~\cite{morota2011}  \\
	   \hline
	   Ref.(Exp., room temperature) & $\sim$5100~\cite{ando2008}  \\
	   \hline
	   Ref.(Exp., room temperature) & 310$\pm$50~\cite{mosendz2010} \\
	   \hline
	   Ref.(Exp., room temperature) & 870$\pm$120~\cite{azevedo2011} \\
	   \hline
	   Ref.(Exp., room temperature) & 1750~\cite{ganguly2014} \\
	   \hline
	   Ref.(Exp., room temperature) & 1900~\cite{liu2011} \\
	   
\hline
\hline
  \end{tabular}
\end{table}

\begin{table}[h]
  \centering
  \caption{fcc Au, SHC is in the unit of $(\hbar/e)(S/cm)$
}
    \label{tab:tablefccAu}
        \begin{tabular} {l|c}
  \hline
  \hline
  \multicolumn{2}{c}{Au} \\
  \hline
  \multicolumn{2}{c}{$\sigma_{xy}^{z}$}    \\
  \hline
  This work         &  377 \\
  \hline
  Ref.($ab-initio$) & 400~\cite{guo2009ab}  \\
  \hline
	  Ref.($ab-initio$) & 470~\cite{gradhand2011}  \\
	  \hline
	  Ref.($ab-initio$) & 350~\cite{yao2005}  \\
	  \hline
	  Ref.(Exp., T=4K)  & 11100~\cite{mihajlovic2009}  \\
	  \hline
	  Ref.(Exp., room temperature)  & 880$\pm$80~\cite{mosendz2010}  \\
	  \hline
	  Ref.(Exp., room temperature)  & $\sim$234~\cite{hung2013}  \\
          \hline
          Ref.(Exp., room temperature)  & 500$\pm$100~\cite{vlaminck2013} \\
	  \hline
	  Ref.(Exp., room temperature)  & 42000~\cite{seki2008} \\

\hline
\hline
  \end{tabular}
\end{table}

\begin{table}[h]
  \centering
  \caption{fcc Pd, SHC in the unit of $(\hbar/e)(S/cm)$
}
    \label{tab:tablefccPd}
        \begin{tabular} {l|c}
  \hline
  \hline
  \multicolumn{2}{c}{Pd} \\
  \hline
  \multicolumn{2}{c}{$\sigma_{xy}^{z}$}    \\
  \hline
  This work         &  1180 \\
  \hline
  Ref.($ab-initio$) & 1200~\cite{guo2009ab}  \\
  \hline
  Ref.(Exp., room temperature) & $\sim$350~\cite{watt2020}  \\
  \hline
	  Ref.(Exp., T=10K) & 270$\pm$90~\cite{morota2011}  \\
	  \hline
	  Ref.(Exp., room temperature) & 260$\pm$40~\cite{mosendz2010}  \\
          \hline
	  Ref.(Exp., room temperature) & $\sim$200~\cite{ando2010}  \\
          \hline
	  Ref.(Exp., room temperature) & 300$\pm$70~\cite{kondou2012}  \\
	  \hline
	  Ref.(Exp., room temperature) & 290$\pm$50~\cite{vlaminck2013}  \\

\hline
\hline
  \end{tabular}
\end{table}

\begin{table}[h]
  \centering
  \caption{bcc Ta, SHC in the unit of $(\hbar/e)(S/cm)$
}
    \label{tab:tablebccTa}
        \begin{tabular} {l|c}
  \hline
  \hline
  \multicolumn{2}{c}{Ta} \\
  \hline
  \multicolumn{2}{c}{$\sigma_{xy}^{z}$}    \\
  \hline
  This work         &  -133 \\
  \hline
  Ref.($ab-initio$) & -142~\cite{qiao2018}  \\
  \hline
  Ref.(Exp., T=10K) & 11$\pm$3~\cite{morota2011}  \\
  \hline
          Ref.(Exp., room temperature) & -630~\cite{liu2012}  \\
	  \hline
	  Ref.(Exp., room temperature) & -160$\pm$120~\cite{hahn2013}  \\
\hline
\hline
  \end{tabular}
\end{table}

\begin{table}[h]
  \centering
  \caption{bcc Nb, SHC in the unit of $(\hbar/e)(S/cm)$
}
    \label{tab:tablebccNb}
        \begin{tabular} {l|c}
  \hline
  \hline
  \multicolumn{2}{c}{Nb} \\
  \hline
  \multicolumn{2}{c}{$\sigma_{xy}^{z}$}    \\
  \hline
  This work         &  -94 \\
  \hline
  Ref.(Exp., T=10K) & 100$\pm$20~\cite{morota2011}  \\
\hline
\hline
  \end{tabular}
\end{table}

\begin{table}[h]
  \centering
  \caption{bcc W, SHC in the unit of $(\hbar/e)(S/cm)$
}
    \label{tab:tablebccW}
        \begin{tabular} {l|c}
  \hline
  \hline
  \multicolumn{2}{c}{W} \\
  \hline
  \multicolumn{2}{c}{$\sigma_{xy}^{z}$}    \\
  \hline
  This work         &  -819 \\
  \hline
  Ref.(ab-initio) & -785~\cite{sui2017giant}  \\
	  \hline
	  Ref.(Exp., room temperature) & -1270$\pm$230~\cite{pai2012spin}  \\
\hline
\hline
  \end{tabular}
\end{table}

\begin{table}[h]
  \centering
  \caption{bcc Mo, SHC in the unit of $(\hbar/e)(S/cm)$
}
    \label{tab:tablebccMo}
        \begin{tabular} {l|c}
  \hline
  \hline
  \multicolumn{2}{c}{Mo} \\
  \hline
  \multicolumn{2}{c}{$\sigma_{xy}^{z}$}    \\
  \hline
  This work         &  -276 \\
  \hline
  Ref.(Exp., K=10T) & -230$\pm$50~\cite{morota2011}  \\
          \hline
          Ref.(Exp., room temperature) & -23$\pm$5~\cite{mosendz2010}  \\
\hline
\hline
  \end{tabular}
\end{table}

As for the hexagonal transition metals,
there are some differences between our calculations and that
from previous reports~\cite{freimuth2010} (Table~\ref{tab:tableHex},\ref{tab:tableHex2}),
but overall, the two calculations are in good agreement. The
differences might be due to the
different sets of coordinates for the hexagonal lattice
vectors. 
So far, most of the experiments focused on 
cubic transition metals and some compounds,
whereas, to the best of our knowledge,
there are almost no experimental reports for hexagonal
transition metals. 
In addition to the transition metals,
we also made a comparison for other compounds 
with available experimental reports and $ab-initio$ calculations,
see Table~\ref{tab:tableOthers}, and here good agreements can also be
found.

\begin{table}[h]
  \centering
  \caption{Hexagonal transition	metals from ~\cite{freimuth2010}.
	  The SHC is in the unit of $(\hbar/e)(S/cm)$
}
    \label{tab:tableHex}
        \begin{tabular} {l|c|c|c}
  \hline
  \hline
  \multicolumn{4}{c}{Sc} \\
  \hline
   &$\sigma_{xy}^{z}$ &$\sigma_{yz}^{x}$ &$\sigma_{zx}^{y}$ \\
  \hline
  This work         &-40 &7 &6 \\
  \hline
	  Ref.(ab-initio)   &$\sim$50 &$\sim$-10 &--  \\

  \hline
  \hline
  \multicolumn{4}{c}{Ti} \\
  \hline
   &$\sigma_{xy}^{z}$ &$\sigma_{yz}^{x}$ &$\sigma_{zx}^{y}$ \\
  \hline
  This work         &-2 &-23 &-19 \\
  \hline
  Ref.(ab-initio)   &$\sim$5 &$\sim$-15 &--  \\

  \hline
  \hline
  \multicolumn{4}{c}{Zn} \\
  \hline
   &$\sigma_{xy}^{z}$ &$\sigma_{yz}^{x}$ &$\sigma_{zx}^{y}$ \\
  \hline
  This work         &-70 &-6 &-6 \\
  \hline
  Ref.(ab-initio)   &$\sim$-130 &$\sim$-20 &--  \\

  \hline
  \hline
  \multicolumn{4}{c}{Y} \\
  \hline
   &$\sigma_{xy}^{z}$ &$\sigma_{yz}^{x}$ &$\sigma_{zx}^{y}$ \\
  \hline
  This work         & 109 & 77 &68 \\
  \hline
  Ref.(ab-initio)   & $\sim$140 & $\sim$40 &--  \\

  \hline
  \hline
  \multicolumn{4}{c}{Zr} \\
  \hline
   &$\sigma_{xy}^{z}$ &$\sigma_{yz}^{x}$ &$\sigma_{zx}^{y}$ \\
  \hline
  This work         & -242 & -48 & -19 \\
  \hline
  Ref.(ab-initio)   & $\sim$-300 & $\sim$-60 &--  \\

  \hline
  \hline
  \multicolumn{4}{c}{Tc} \\
  \hline
   &$\sigma_{xy}^{z}$ &$\sigma_{yz}^{x}$ &$\sigma_{zx}^{y}$ \\
  \hline
  This work         & -160 & -18 & -97 \\
  \hline
  Ref.(ab-initio)   & $\sim$-200 & $\sim$-140 &--  \\

  \hline
  \hline
  \multicolumn{4}{c}{Ru} \\
  \hline
   &$\sigma_{xy}^{z}$ &$\sigma_{yz}^{x}$ &$\sigma_{zx}^{y}$ \\
  \hline
  This work         & 90 & 190 & 163 \\
  \hline
  Ref.(ab-initio)   & $\sim$-10 & $\sim$90 &--  \\

  \hline
  \hline
    \end{tabular}
  \end{table}

  \begin{table}[h]
    \centering
  \caption{Continuation of Tab.~\ref{tab:tableHex}.
        The SHC is in the unit of $(\hbar/e)(S/cm)$
  }  
      \label{tab:tableHex2}
          \begin{tabular} {l|c|c|c}

  \hline
  \hline
  \multicolumn{4}{c}{La} \\
  \hline
   &$\sigma_{xy}^{z}$ &$\sigma_{yz}^{x}$ &$\sigma_{zx}^{y}$ \\
  \hline
  This work         & 290 & 316 & 308 \\
  \hline
  Ref.(ab-initio)   & $\sim$150 & $\sim$300 &--  \\

  \hline
  \hline
  \multicolumn{4}{c}{Hf} \\
  \hline
   &$\sigma_{xy}^{z}$ &$\sigma_{yz}^{x}$ &$\sigma_{zx}^{y}$ \\
  \hline
  This work         & -375 & -53 & 112 \\
  \hline
  Ref.(ab-initio)   & $\sim$-800 & $\sim$-600 &--  \\

  \hline
  \hline
  \multicolumn{4}{c}{Re} \\
  \hline
   &$\sigma_{xy}^{z}$ &$\sigma_{yz}^{x}$ &$\sigma_{zx}^{y}$ \\
  \hline
  This work         & -325 & -441 & -519 \\
  \hline
  Ref.(ab-initio)   & $\sim$-500 & $\sim$-700 &--  \\

  \hline
  \hline
  \multicolumn{4}{c}{Cd} \\
  \hline
   &$\sigma_{xy}^{z}$ &$\sigma_{yz}^{x}$ &$\sigma_{zx}^{y}$ \\
  \hline
  This work         & -93 & 2 & -5 \\
  \hline
  Ref.(ab-initio)   & $\sim$-30 & $\sim$-20 &--  \\

  \hline
  \hline
  \multicolumn{4}{c}{Os} \\
  \hline
   &$\sigma_{xy}^{z}$ &$\sigma_{yz}^{x}$ &$\sigma_{zx}^{y}$ \\
  \hline
  This work         & -68 & -162 & -72 \\
  \hline
  Ref.(ab-initio)   & $\sim$-260 & $\sim$-300 &--  \\

\hline
\hline
  \end{tabular}
\end{table}

\begin{table}[h]
  \centering
  \caption{Other compounds, SHC in the unit of $(\hbar/e)(S/cm)$
}
    \label{tab:tableOthers}
        \begin{tabular} {l|c|c|c}
  \hline
  \hline
	        \multicolumn{4}{c}{Pt$_3$O$_4$} \\
  \hline
		&bcc & \multicolumn{2}{c}{Cubic} \\
  \hline
		&$\sigma_{xy}^{z}$ & \multicolumn{2}{c}{$\sigma_{xy}^{z}$}   \\
  \hline
	  This work       &1420     & \multicolumn{2}{c}{215}   \\
 \hline
	  Ref.(ab-initio)~\cite{jadaun2020} &1838     &\multicolumn{2}{c}{244}  \\

  \hline
  \hline
  \multicolumn{4}{c}{PtTe$_2$} \\
  \hline
This work(largest component $\sigma_{yz}^{x}$)       &\multicolumn{2}{c}{122}  \\
 \hline
Ref.(Exp. room temperature)~\cite{xu2020high} & \multicolumn{2}{c}{100-800}  \\

  \hline
  \hline
  \multicolumn{4}{c}{NbSe$_2$} \\
  \hline
This work(largest component $\sigma_{yx}^{z}$)       &\multicolumn{2}{c}{188}  \\
 \hline
Ref.(Exp. room temperature)~\cite{guimaraes2018spin} & \multicolumn{2}{c}{$\sim$150}  \\

  \hline
  \hline
  \multicolumn{4}{c}{WTe$_2$} \\
  \hline
	  &$\sigma_{zy}^{x}$ &\multicolumn{2}{c}{$\sigma_{zx}^{y}$}   \\

	  This work       &34 &\multicolumn{2}{c}{85}   \\
 \hline
	  Ref.(ab-initio~\cite{zhao2020observation} &$\sim$20 &\multicolumn{2}{c}{$\sim$80}  \\
\hline
	  Ref.(Exp., room temperature~\cite{zhao2020observation} &\multicolumn{3}{c}{14-96}   \\

  \hline
  \hline
  \multicolumn{4}{c}{Bi$_2$Se$_3$} \\
  \hline
	  &$\sigma_{xy}^{z}$ &$\sigma_{zx}^{y}$ &$\sigma_{yz}^{x}$  \\

	  This work       & 105 &104 & 96  \\
 \hline
	  Ref.(Exp.,room temperature~\cite{han2017} &\multicolumn{3}{c}{$\sim$80}  \\
\hline
          Ref.(Exp.,room temperature~\cite{mellnik2014} &\multicolumn{3}{c}{500-1000}  \\
\hline
\hline

  \end{tabular}
\end{table}

\subsection{Berry curvature dipole}\label{sec:berrydipole}
 
\begin{figure}[tb]
  \includegraphics[width=0.95\columnwidth]{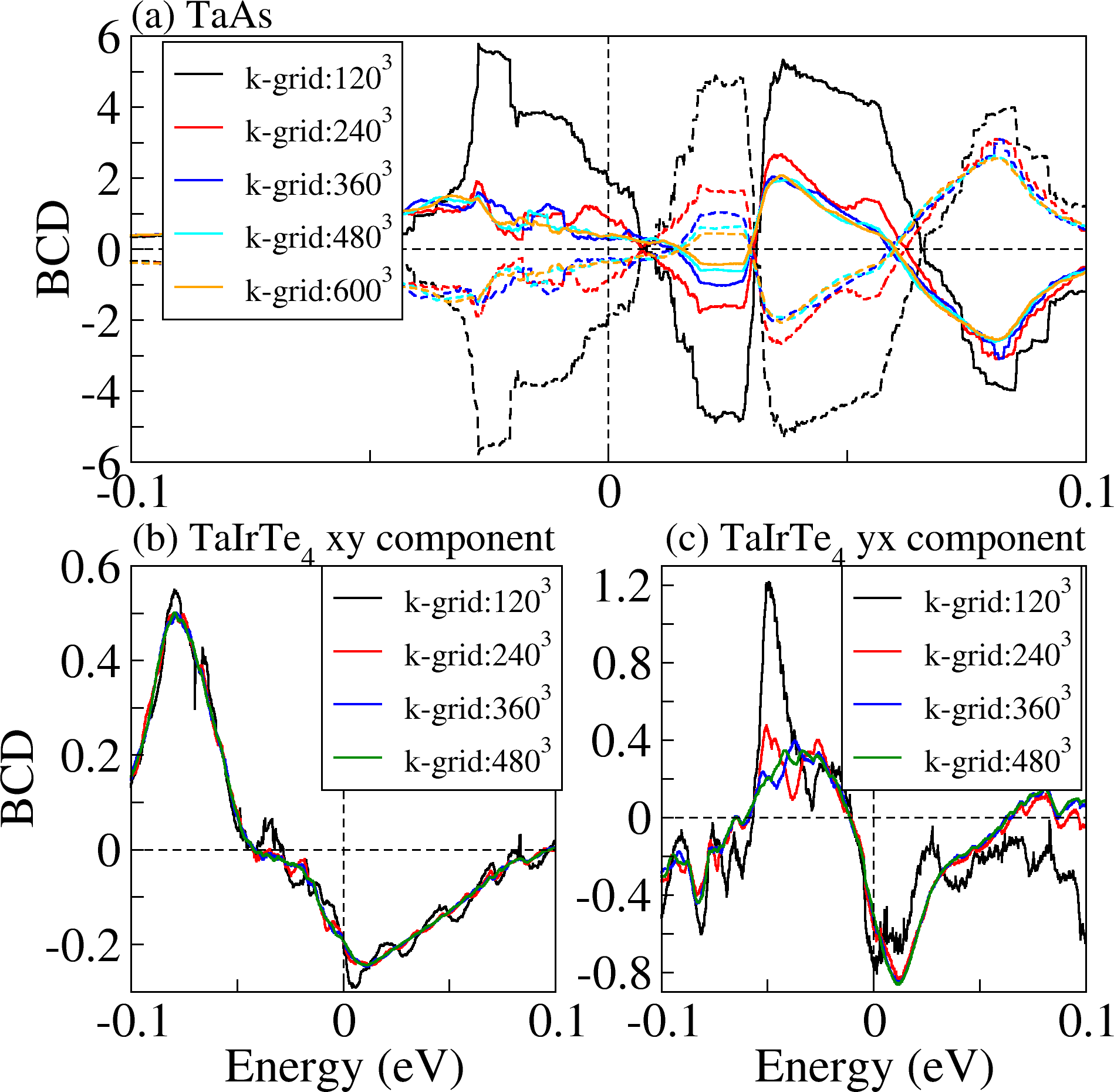}
  \caption{ (a) Berry curvature dipole of TaAs. The solid and dashed
    lines represent the xy- and yx-components, respectively.  The Berry
    curvature dipole of (b) the xy-component and (c) the yx-component of
    TaIrTe$_4$.}
  \label{pic:BCD}
\end{figure}

We also calculated the Berry curvature dipole for the type-I Weyl
semimetal TaAs and the type-II Weyl semimetal TaIrTe$_4$. The crystal
structure of TaAs and TaIrTe$_4$ belong to point group $4mm$ and
$mm2$, respectively. According to the symmetry analysis for these two
point groups, TaAs has only one independent component with
$D_{xy}$=$-D_{yx}$, and TaIrTe$_4$ has two independent components of
$D_{xy}$ and $D_{yx}$. Our calculations fulfill these symmetry
restrictions. The results reach convergence at a dense k-grid of
$720^3$ and $480^3$ for TaAs and TaIrTe$_4$, respectively, see
Fig.~\ref{pic:BCD}. The Berry curvature dipole has a value of about 0.37
for the xy component of TaAs, in good agreement with
Ref.~\onlinecite{zhang2018berry}.  The Berry curvature dipole in
TaIrTe$_4$ can reach up to around -0.19 and -0.50 for the xy and yx
components, respectively.

\cleardoublepage


\section{Summary and Conclusions}

We thoroughly documented the underlying methods of the FPLO package
for the construction of symmetry conserving maximally projected
Wannier functions. 

The method does not enforce maximum localization but rather results
in Wannier functions with (near) maximum projection onto FPLO orbitals
or linear combinations thereof, yet yields a very high degree of
localization due to the nature of the orbitals
while retaining the symmetry properties of the chosen projectors.
The simplicity of this algorithm leads to numerical efficiency and 
a high level of accuracy, which is needed in the context of 
topological properties, where symmetry conservation is of the essence,
especially when taking derivatives (e.g.~for the Berry curvature dipole).

For a complete treatment of topological properties the full Berry
connection/curvature is desirable. To this end we discussed the
position operator matrix elements as they have to be treated in a
local orbital context as well as their symmetry properties in the
context of possible approximations of this operator.  The latter gives
valuable insights for the application of such approximations in tight
binding models with implicit (unknown) Wannier basis.

The method has been applied to a set of compounds to illustrate
various applicational aspects as well as to demonstrate the general
quality of the resulting Wannier models and  the available post
processing tools.

We gave an idea of the efficiency of our
method and repeated some of the calculations using the Wannier90 package
to convince the reader that our method indeed yields comparable results.  We
gave an extensive list of results for the anomalous Hall effect, the
anomalous Nernst effect, and the spin Hall effect compared to results
of other publications.

As a technical reference an extended appendix documents the FPLO basis
orbitals and their transformation properties from which the symmetry
properties of the Wannier functions result.

In summary, we believe to have demonstrated that maximally projected
symmetry conserving Wannier functions based on optimized local orbital
methods have similar localization and flexibility as maximally
localized Wannier functions, however with minimal computational cost
and complexity and explicit symmetry properties.


\section{Acknowledgments}

K. Koepernik acknowledges inspirations by and discussions with Wei Ku,
Helge Rosner, Alexander Tsirlin and Manuel Richter. Ulrike Nitzsche is
acknowledged for the maintenance of the ITF computational cluster and
technical support. We acknowledge financial support by the
Deutsche Forschungsgemeinschaft (DFG, German Research Foundation),
through SFB 1143 project A5 and the W{\"u}rzburg-Dresden Cluster of
Excellence on Complexity and Topology in Quantum Matter- ct.qmat (EXC
2147, Project Id No. 390858490). O.J.~was supported by the
Leibniz Association through the Leibniz Competition.


\appendix
\section{Local basis states}

\subsection{FPLO basis}\label{sec:FPLObasis}

An FPLO basis orbital $\Phi_{\boR\bos\nu}\left(\bor\right)$ is a
solution of a Schr{\"o}dinger/Dirac equation in an atom centered
suitably modified potential with compact support. The boundary
conditions are chosen such that there is no leakage of the orbital
wave function at the compact support radius (this is not possible
exactly in the Dirac case but practically the leakage can be made tiny
and irrelevant). The orbitals sit in the unit cell at lattice vector
$\boR$ at site $\bos$ and have quantum numbers $\nu$ . To be precise
$\Phi_{\boR\bos\nu}\left(\bor\right)=\Phi_{s\nu}\left(\bor-\boR-\bos\right)$
where $\Phi_{s\nu}\left(\bor\right)$ is a function of atom site index $s$
and qns.~$\nu$.

In non- and scalar-relativistic mode the orbitals have the shape
$\Phi_{s nlm\sigma}=\varphi_{s
  nl}\left(r\right)Y_{lm}\left(\hat{\bor}\right)\chi_{\sigma}$ where
$n$ is the main quantum number, $Y_{lm}$ is a real spherical harmonic
with angular momentum qns.~$l$ and $m$ and $\chi_{\sigma}$ is a spin
$\frac{1}{2}$ basis spinor defined via
$\sigma_{z}\chi_{\sigma}=\chi_{\sigma}\sigma$ ($\sigma=\pm1$). The
radial basis functions are not spin dependent which is compensated by
the variational freedom of the chosen basis set which contains
polarization orbitals additionally to the chemical valence (and semi-core)
orbitals.

In full relativistic mode the orbitals are solutions to a 4-component
Dirac equation and are four spinors
\begin{eqnarray}
  \Phi_{s nlj\mu}=\left(\begin{array}{c}
    g_{s nlj}\left(r\right)\chi_{\kappa\mu}\\
    if_{s nlj}\left(r\right)\chi_{-\kappa\mu}
  \end{array}\right)\label{eq:fourspinor}
\end{eqnarray}
with one large component ($g_{s nlj}$) and one small component
($f_{s nlj}$) radial function per $lj$-shell and spherical spinors
$\chi_{\kappa\mu}$, where $\kappa=\left(2j+1\right)\left(l-j\right)$
are the eigenvalues of the spin orbit operator
$\hat{\kappa}=1+\hat{\boldsymbol{\sigma}}\hat{\boldsymbol{L}}$ with
$\hat{\kappa}\chi_{\kappa\mu}=-\chi_{\kappa\mu}\kappa$ (in detail
$\kappa_{j=l-\frac{1}{2}}=l$ and $\kappa_{j=l+\frac{1}{2}}=-l-1$ ) and
$\mu=-j,-j+1,\ldots,j$ is the $\hat{J}_{z}$ eigenvalue.

The orbitals can be made orthonormal at each site. However, off-site
orthonormality is hard to achieve, since this would mean to explicitly
construct Wannier functions from the start. With a proper choice of
orbitals a well conditioned overlap matrix
\begin{equation}
  S_{\boR^{\prime}\bos^{\prime}\nu^{\prime},\boR\bos\nu}=\left\langle
  \Phi_{\boR^{\prime}\bos^{\prime}\nu^{\prime}}\mid\Phi_{\boR\bos\nu}\right\rangle
\nonumber
\end{equation}
can be achieved. Lattice translations
\begin{eqnarray}
  \hat{T}_{\boR^{\prime}}\Phi_{\boR\bos\nu}\left(\bor\right) 
&=& \Phi_{\boR\bos\nu}\left(\bor-\boR^{\prime}\right)\nonumber\\
&=& \Phi_{s\nu}\left(\bor-\boR-\bos-\boR^{\prime}\right)\nonumber\\
&=&\Phi_{\boR+\boR^{\prime},\bos\nu}\left(\bor\right)\label{eq:transLO}
\end{eqnarray}
yield translation invariance according to
\begin{eqnarray}
  S_{\boR^{\prime}\bos^{\prime}\nu^{\prime},\boR\bos\nu} &=& \left\langle
  \hat{T}_{-\boR^{\prime}}\Phi_{\boR^{\prime}\bos^{\prime}\nu^{\prime}}\mid\hat{T}_{-\boR^{\prime}}\Phi_{\boR\bos\nu}\right\rangle
  \nonumber\\
  &=& \left\langle
  \Phi_{\boldsymbol{0}\bos^{\prime}\nu^{\prime}}\mid\Phi_{\boR-\boR^{\prime},\bos\nu}\right\rangle
  \nonumber\\
  &=&
  S_{\boldsymbol{0}^{\prime}\bos^{\prime}\nu^{\prime},\boR-\boR^{\prime},\bos\nu}
\label{eq:transS}
\end{eqnarray}

\subsection{Bloch sums}\label{sec:Blochsums}

In order to form extended basis states Bloch sums of local orbitals
are defined as
\begin{equation}
  \Phi_{\bos\nu}^{\bok}\left(\bor\right)=\frac{1}{\sqrt{N}}\sum_{\boR}\mex^{i\bok\left(\boR+\lambda\bos\right)}\Phi_{\boR\bos\nu}\left(\bor\right)
  \label{eq:LOblochsum}
\end{equation}
where $\lambda=0,1$ ($\blambda=1-\lambda$) picks a particular phase
gauge as discussed in the main text. The normalization contains the
number of unit cells $N$ in the Born-von-K{\'a}rm{\'a}n (BvK) torus,
which is there for formal correctness but never actually appears in
any coded formulas except for the normalization of the Fourier back
transform.  The Fourier transformed overlap matrix is defined as
\begin{eqnarray}
  S_{\bos^{\prime}\nu^{\prime},\bos\nu}^{\bok}&=&\frac{1}{N}\sum_{\boR\boR^{\prime}}\mex^{i\bok\left(\boR+\lambda\bos-\boR^{\prime}-\lambda\bos^{\prime}\right)}S_{\boR^{\prime}\bos^{\prime}\nu^{\prime},\boR\bos\nu}\nonumber\\
  &=&\frac{1}{N}\sum_{\boR\boR^{\prime}}\mex^{i\bok\left(\boR+\lambda\bos-\boR^{\prime}-\lambda\bos^{\prime}\right)}S_{\boldsymbol{0}\bos^{\prime}\nu^{\prime},\boR-\boR^{\prime}\bos\nu}\nonumber\\
  &=&\frac{1}{N}\sum_{\boR^{\prime}}\sum_{\boR}\mex^{i\bok\left(\boR+\lambda\left(\bos-\bos^{\prime}\right)\right)}S_{\boldsymbol{0}\bos^{\prime}\nu^{\prime},\boR\bos\nu}\nonumber\\
  &=&\sum_{\boR}\mex^{i\bok\left(\boR+\lambda\left(\bos-\bos^{\prime}\right)\right)}S_{\boldsymbol{0}\bos^{\prime}\nu^{\prime},\boR\bos\nu}  \label{eq:Sk}
\end{eqnarray}
which motivates the normalization choice in
Eq.~(\ref{eq:LOblochsum}). 
The exponential contains the difference vector between the two orbital
positions if $\lambda=1$, which is why we call it \textit{relative} gauge.

The overlap of a Bloch sum with a single
basis orbital $\Phi_{\boR\bos\nu}$ reads
\begin{eqnarray}
  \left\langle
  \Phi_{\bos^{\prime}\nu^{\prime}}^{\bok}\mid\Phi_{\boR\bos\nu}\right\rangle 
  &=&\frac{1}{\sqrt{N}}\sum_{\boR^{\prime}}\mex^{-i\bok\left(\boR^{\prime}+\lambda\bos^{\prime}\right)}\left\langle
      \Phi_{\boR^{\prime}\bos^{\prime}\nu^{\prime}}\mid\Phi_{\boR\bos\nu}\right\rangle
      \nonumber\\
  &=&\frac{1}{\sqrt{N}}\sum_{\boR^{\prime}}\mex^{-i\bok\left(\boR^{\prime}+\lambda\bos^{\prime}\right)}S_{\boR^{\prime}\bos^{\prime}\nu^{\prime},\boR\bos\nu}\nonumber\\
  &=&\frac{1}{\sqrt{N}}\sum_{\boR^{\prime}}\mex^{-i\bok\left(\boR-\boR^{\prime}+\lambda\bos^{\prime}\right)}S_{\boldsymbol{0}\bos^{\prime}\nu^{\prime},\boR^{\prime}\bos\nu}\nonumber\\
  &=&\frac{1}{\sqrt{N}}\sum_{\boR^{\prime}}\mex^{i\bok\left(\boR^{\prime}+\lambda\left(\bos-\bos^{\prime}\right)\right)}S_{\boldsymbol{0}\bos^{\prime}\nu^{\prime},\boR^{\prime}\bos\nu}\mex^{-i\bok\left(\boR+\lambda\bos\right)}\nonumber\\
  &=&\frac{1}{\sqrt{N}}S_{\bos^{\prime}\nu^{\prime},\bos\nu}^{\bok}\mex^{-i\bok\left(\boR+\lambda\bos\right)}\label{eq:overlapBlochLO}
\end{eqnarray}
using $\boR^{\prime}\to-\boR^{\prime}+\boR$ and
Eqs.~(\ref{eq:transS},\ref{eq:Sk}). Using 
Eq.~(\ref{eq:LOblochsum}) and Eq.~(\ref{eq:overlapBlochLO}) the
overlap of two Bloch functions yields the overlap matrix
\begin{eqnarray*}
  \left\langle
  \Phi_{\bos^{\prime}\nu^{\prime}}^{\bok}\mid\Phi_{\bos\nu}^{\bok}\right\rangle
  & = &
        \frac{1}{N}\sum_{\boR}S_{\bos^{\prime}\nu^{\prime},\bos\nu}^{\bok}\\
  & = & S_{\bos^{\prime}\nu^{\prime},\bos\nu}^{\bok}
\end{eqnarray*}
We introduce local linear combinations of orbitals (MOs) for projection 
purposes
\begin{equation}
  \phi_{\boR\boc i}=\sum_{\boR^{\prime}\bos}\Phi_{\boR^{\prime}\bos\nu^{\prime}}U_{\boR^{\prime}\bos\nu^{\prime},\boR\boc i}  \label{eq:MOfromLO}
\end{equation}
which shall be translation invariant in the same way as the
orbitals Eq.~(\ref{eq:transLO}), .i.e.~$U_{\boR^{\prime}\bos\nu^{\prime},\boR\boc
  i}=U_{\boldsymbol{0}\bos\nu^{\prime},\boR-\boR^{\prime}\boc i}$
and have Bloch sums
\begin{eqnarray}
  \phi_{\boc i}^{\bok} 
  & = & \frac{1}{\sqrt{N}}\sum_{\boR}
        \mex^{i\bok\left(\boR+\lambda\boc\right)}
        \phi_{\boR\boc i}
        \nonumber\\
  & = & \Phi_{\boc i}^{\bok} U_{\bos\nu^{\prime},\boc i}^{\bok}\nonumber\\
  U_{\bos\nu^{\prime},\boc i}^{\bok}
  & = &\sum_{\boR}U_{0\bos\nu^{\prime},\boR\boc i}\mex^{i\bok\left(\boR+\lambda\boc-\lambda\bos\right)}\nonumber
\end{eqnarray}

To calculate the overlap between a MO and an orbital Bloch sum
Eq.~(\ref{eq:overlapBlochLO}) gets then modified according to
\begin{equation}
\left\langle \Phi_{\bos\nu}^{\bok}\mid\phi_{\boR\boc i}\right\rangle 
=\frac{1}{\sqrt{N}}\left(S^{\bok}U^{\bok}\right)_{\bos\nu,\boc i}\mex^{-i\bok\left(\boR+\lambda\boc\right)}
  \label{eq:overlapBlochMO}
\end{equation}

The Fourier back transformation which yields real space objects is
practically obtained by a $\bok$-sum whose underlying mesh implicitly
defines the BvK-torus and $N$
\begin{equation}
  S_{\boldsymbol{0}\bos^{\prime},\boR\bos}=\sum_{\bok}f_{\bok}\mex^{-i\bok\left(\boR+\lambda\left(\bos-\bos^{\prime}\right)\right)}S_{\bos^{\prime}\bos}^{\bok}
  \label{eq:Fourierback}
\end{equation}
with
\begin{equation}
  \sum_{\bok}f_{\bok}=1\nonumber
\end{equation}
If the full mesh is used $f_{\bok}=N^{-1}$, but if only irreducible
$\bok$-points are used $f_{\bok}$ is corrected for the multiplicity of
the irreducible point.
Combining Eqs.~(\ref{eq:Sk},\ref{eq:Fourierback}) one gets
\begin{equation}
  \sum_{\bok}f_{\bok}\mex^{i\bok\left(\boR-\boR^{\prime}\right)}=\delta_{\boR,\boR^{\prime}}
\nonumber
\end{equation}
which is only true modulo a lattice vector which describes the BvK
periodicity.
The back transformation of a Bloch sum itself with correct
normalization reads
\begin{equation}
\Phi_{\boR\bos}=\frac{1}{\sqrt{N}}\sum_{\bok}\mex^{-i\bok\left(\boR+\bos\right)}\Phi_{\bos}^{\bok} \nonumber
\end{equation}
A translation invariant operator in Bloch sum basis (as the overlap)
contains two Bloch sums and is diagonal in $\bok$-space. A full
Fourier back transformation of both sides of the matrix then reduces
to Eq.~(\ref{eq:Fourierback}) with $\frac{1}{N}$ normalization instead
of $\frac{1}{\sqrt{N}}$.

\section{Symmetry}\label{sec:symemtry}

\subsection{Local basis symmetry}\label{sec:LOsymemtry}
The complex spherical harmonics are defined as 
\begin{eqnarray}
\lefteqn{\mathcal{Y}_{lm}\left(\varphi,\theta\right) =}\nonumber\\
&&\left(-1\right)^{\frac{m+\left|m\right|}{2}}\sqrt{\frac{2l+1}{4\pi}\frac{\left(l-\left|m\right|\right)!}{\left(l+\left|m\right|\right)!}}P_{l}^{\left|m\right|}\left(\cos\theta\right)\mex^{im\varphi}  
  \label{eq:CYlm}
\end{eqnarray}
with the associated Legendre functions for positive $m$ as a function
of $z=\cos(\theta)$
\begin{eqnarray}
  P_{l}^{m}\left(z\right) 
&=&
    \left(1-z^{2}\right)^{\frac{m}{2}}\frac{\mathrm{d}^{m}}{\mathrm{d}z^{m}}P_{l}\left(z\right)
  \nonumber\\
P_{l}\left(z\right)&\equiv&   P_{l}^{0}\left(z\right)
= \frac{1}{2^{l}l!}\frac{\mathrm{d}^{l}}{\mathrm{d}z^{l}}\left(z^{2}-1\right)^{l}\nonumber
\end{eqnarray}
From this the real harmonics are obtained  via
\begin{eqnarray}
\lefteqn{Y_{lm}\left(\theta,\varphi\right)=\sum_{m^{\prime}}\mathcal{Y}_{lm^{\prime}}\left(\theta,\varphi\right)U_{mm^{\prime}}^{*}}\label{eq:YeqYU}
\\
&=&\left\{ \begin{array}{lll}
\frac{1}{\sqrt{2}}\left(\left(-1\right)^{m}\mathcal{Y}_{l\left|m\right|}+\mathcal{Y}_{l-\left|m\right|}\right) & \qquad & m>0\\
\mathcal{Y}_{l0} & \qquad & m=0\\
\frac{1}{i\sqrt{2}}\left(\left(-1\right)^{\left|m\right|}\mathcal{Y}_{l\left|m\right|}-\mathcal{Y}_{l-\left|m\right|}\right) & \qquad & m<0
\end{array}\right.\nonumber
\end{eqnarray}

The spherical spinors as introduced after Eq.~(\ref{eq:fourspinor})
are defined as
\begin{equation}
  \chi_{\kappa\mu}=\sum_{s=\pm1}\chi_{s}\mathcal{Y}_{l\mu-\frac{s}{2}}c_{\kappa\mu}^{s}
\label{eq:defchikapmu}
\end{equation}
with basis spinors $\chi_s$, complex Harmonics Eq.~(\ref{eq:CYlm}) and
Clebsch-Gordon coefficients
\begin{equation}
c_{\kappa\mu}^{s}=-\left(\frac{s\left(\kappa+\left|\kappa\right|\right)+\left(\kappa-\left|\kappa\right|\right)}{2\left|\kappa\right|}\right)\sqrt{\frac{1}{2}-s\frac{\mu}{\left(2\kappa+1\right)}}  \nonumber
\end{equation}
which can be written as
\begin{eqnarray}
\chi_{\kappa\mu}
&=&\sum_{s=\pm1,m=-l_{\kappa}}^{l_{\kappa}}\chi_{s}\mathcal{Y}_{l_{\kappa}m}T_{ms,\kappa\mu}^{l_{\kappa}}\label{eq:chieqchiT}\\
\chi_{\kappa\mu}
&=&\sum_{s=\pm1,m=-l_{\kappa}}^{l_{\kappa}}\chi_{s}Y_{l_{\kappa}m}\left(UT\right)_{ms,\kappa\mu}^{l_{\kappa}}\nonumber\\
T_{ms,\kappa\mu}^{l}
&=&\delta_{m,\mu-\frac{s}{2}}c_{\kappa\mu}^{s}\nonumber
\end{eqnarray}
with $l_{\kappa}=\frac{\left|2\kappa+1\right|-1}{2}$ where $T^l$ can be
arranged in a unitary matrix for each $l$-shell.

Inversion $\hat{I}\bor=-\bor$ at the origin of the harmonics with $\hat{I}f\left(\bor\right)=f\left(-\bor\right)$  acts like 
\begin{eqnarray}
\hat{I}\mathcal{Y}_{lm}
&=&\mathcal{Y}_{lm}\left(-1\right)^{l}
                           \nonumber\\
\hat{I}Y_{lm}
&=&Y_{lm}\left(-1\right)^{l}\nonumber\\
\hat{I}\chi_{\kappa\mu}
&=&\chi_{\kappa\mu}\left(-1\right)^{l_{\kappa}}\nonumber\\
\hat{I}\chi_{s}&=&\chi_{s}\nonumber
\end{eqnarray}

A proper rotation $\hat{\alpha}$ with
$\hat{\alpha}f\left(\bor\right)=f\left(\hat{\alpha}^{-1}\bor\right)$
parametrized by the Euler angles is represented by the Wigner-$D$
functions in the $l$ or $lj$ basis which without further details
reads
\begin{subequations}
\label{eq:rotAngular}
\begin{eqnarray}
  \hat{\alpha}\mathcal{Y}_{lm}
  &=&\sum_{m^{\prime}=-l}^{l}\mathcal{Y}_{lm^{\prime}}D_{m^{\prime}m}^l\left(\alpha\right)
\label{eq:rotcalY}\\
\hat{\alpha}Y_{lm}
&=&\sum_{m^{\prime}=-l}^{l}Y_{lm^{\prime}}\tilde{D}_{m^{\prime}m}^{l}\left(\alpha\right)\nonumber\\
  \hat{\alpha}\chi_{\kappa\mu}
&=&\sum_{\mu^{\prime}=-j_{\kappa}}^{j_{\kappa}}\chi_{\kappa\mu^{\prime}}D_{\mu^{\prime}\mu}^{j_{\kappa}}\left(\alpha\right)\label{eq:rotcalYj}\\
\hat{\alpha}\chi_{\sigma}
&=&\sum_{\sigma^{\prime}=\pm1}\chi_{\sigma^{\prime}}D_{\sigma^{\prime}\sigma}^{\frac{1}{2}}\left(\alpha\right)\label{eq:rotspinor}
\end{eqnarray}
\end{subequations}
where $j_\kappa=\frac{2\left|\kappa\right|-1}{2}$ and
$\tilde{D}_{m^{\prime}m}\left(\alpha\right)=\left(UD\left(\alpha\right)U^{+}\right)_{m^{\prime}m}$,
with $U$ from Eq.~(\ref{eq:YeqYU}).

From
Eqs.~(\ref{eq:defchikapmu},\ref{eq:chieqchiT},\ref{eq:rotAngular})
\begin{equation}
  D^{\frac{1}{2}}D^{l}T^{l}=T^{l}D^{j}\label{eq:DDTeqTD}
\end{equation}
where all matrices must be appropriately formed for the complete
$l$-shell (which includes the two $j=l\pm\frac{1}{2}$ subshells
or the single $j=\frac{1}{2}$-shell for $l=0$).
Now, since $j_{-\kappa}=j_{\kappa}$ both spherical spinors in
Eq.~(\ref{eq:fourspinor}) transform with the same $D^{j_\kappa}$ and
Eq.~(\ref{eq:DDTeqTD}) yields
\begin{equation}
\hat{\alpha}\left(\Phi^{l}T^{l+}\right)=\left(\Phi^{l}T^{l+}\right)D^{\frac{1}{2}}D^{l}
 \label{eq:pseudoNRELLOtransform}
\end{equation}
which shows that the totality of all four spinors of an $l$-shell
$\Phi^{l}=\left(\Phi_{j=l-\frac{1}{2}}^{l}\Phi_{j=l+\frac{1}{2}}^{l}\right)$
multiplied by $T^{l+}$ transforms as non-relativistic complex
harmonics $\mathcal{Y}$
(or as real harmonics $Y$ if $\left(UT^l\right)^{+}$ is used).

Finally, the time reversal operator is defined as
$\theta=-i\sigma_yK_0$, where $\sigma_y$ is the $y$-component of the
Pauli matrix vector and $K_0$ is complex conjugation, which gives the
general action
$\theta f_{i}=\sum_{i^{\prime}}f_{i^{\prime}}D_{i^{\prime}i}^{\theta}K_0$
where $D^{\theta}$ can be read off the following relations

\begin{subequations}
\label{eq:TRangular}
\begin{eqnarray}
\theta\mathcal{Y}_{lm}&=&\mathcal{Y}_{l-m}\left(-1\right)^{m}\theta
\label{eq:TRcalY}\\
\theta{}Y_{lm}&=&Y_{lm}\theta\\
\theta\chi_{\kappa\mu}&=&\chi_{\kappa-\mu}\left(-1\right)^{l+j-\mu}K_{0}\\
\theta\chi_{\sigma}&=&\chi_{-\sigma}\sigma{}K_{0}
\label{eq:TRchi}
\end{eqnarray}
\end{subequations}

\subsection{Space group symmetry/Time reversal}
\label{sec:spacegroupsym}

The space group contains the set of operations 
\begin{equation}
  T_{\boR}g=T_{\boR}\left\{ \alpha\mid\boldsymbol{\tau}_{\alpha}\right\} =\left\{ \alpha\mid\boldsymbol{\tau}_{\alpha}+\boR\right\} \nonumber
\end{equation}
where $T_{\boR}$ is a lattice translation and $g=\left\{
  \alpha\mid\boldsymbol{\tau}_{\alpha}\right\}$ is the Seitz symbol containing a
proper/improper rotation $\alpha$ and a non-lattice translation
$\boldsymbol{\tau}_{\alpha}$. The latter depends on the origin choice and a  particular
pick among all the vectors $\boldsymbol{\tau}_{\alpha}+\boR$. 
A real space vector transforms according to
\begin{equation}
T_Rg\bor=\left\{ \alpha\mid\boldsymbol{\tau}+\boR\right\} \bor=g\bor+\boR  =\alpha\bor+\boldsymbol{\tau}+\boR  \nonumber
\end{equation}
This is especially true for a lattice site $\bos+\boR^{\prime}$ for
which one gets
\begin{equation}
  T_{R}g\left(\boR^{\prime}+\bos\right)=\alpha\boR^{\prime}+g\bos+\boR=\alpha\boR^{\prime}+\alpha\bos+\boldsymbol{\tau}+\boR\nonumber
\end{equation}
The inverse is given by 
\begin{equation}
\left\{ \alpha\mid\boldsymbol{\tau}+\boR\right\} ^{-1}=\left\{ \alpha^{-1}\mid-\alpha^{-1}\left(\boldsymbol{\tau}+\boR\right)\right\}\label{eq:ginv}
\end{equation}

When setting up a structure a certain set of sites will be
generated. They do not need to lie in the first unit cell. The
set of transformed sites $g\bos$ will in general not be identical to
this original set. A simple lattice translation
\begin{equation}
g \bos=  \bos_g+R_{g,s}\nonumber
\end{equation}
will map the transformed set back onto the original one. If this is
done consistently, the site numbers $s$, $s_{g}$ can be used as indices to orbitals
and Bloch sums.

The vector of all orbitals (and of Wannier functions) at a site
transforms according to
\begin{subequations}
\label{eq:transPhi}
\begin{eqnarray}
g\Phi_{\boR\bos}\left(\bor\right)
&=&g\Phi_{s}\left(\bor-\boR-\bos\right)\nonumber\\
&=&\Phi_{s}\left(g^{-1}\bor-\boR-\bos\right)\nonumber\\
&=&\Phi_{s}\left(\alpha^{-1}\left(\bor-\boldsymbol{\tau}\right)-\boR-\bos\right)\nonumber\\
&=&\Phi_{s}\left(\alpha^{-1}\left(\bor-\alpha\boR-\alpha\bos-\boldsymbol{\tau}\right)\right)\nonumber\\
&=&\Phi_{s}\left(\bor-\alpha\boR-\bos_{g}-\boR_{g,s}\right)D_s^{\Phi}\left(\alpha\right)\nonumber\\
&=&\Phi_{\alpha\boR+\boR_{g,s},\bos_{g}}\left(\bor\right)D_s^{\Phi}\left(\alpha\right)  
\label{eq:gtransLO}\\
&=&\Phi_{g\left(\boR\bos\right)}\left(\bor\right)D_s^{\Phi}\left(\alpha\right)  
\label{eq:gtransLOshort}
\end{eqnarray}
\end{subequations}
where $D_s^{\Phi}\left(\alpha\right)$ is the appropriate
transformation matrix, which transforms orbitals at site indexed by
$s$ into orbitals at site indexed by $s_{g}$. For the FPLO basis
these matrices are the same at each site and are obtained according to
Sec.~\ref{sec:LOsymemtry}.  The last row,
Eq.~(\ref{eq:gtransLOshort}), is a useful short hand for
Eq.~(\ref{eq:gtransLO}) in deriving symmetry properties since
$g\left(\boR+\bos\right)$ is a complete set of lattice sites which can
serve as summation variable.

In case of molecular orbitals as WF projectors the symmetry properties
of the WFs is an implicit input in our method an they are required to
transform akin to Eq.~(\ref{eq:gtransLO}). The input matrix $U$ in
Eq.~(\ref{eq:MOfromLO}) together with the transformation properties of
the orbitals $D_s^{\Phi}$ determine the transformation properties of
$\phi_{\boc}$ completely. In practice one chooses $U$ in
Eq.~(\ref{eq:MOfromLO}). The code looks if application of symmetry
replicates the set of Wannier functions and if the resulting matrices
$D_c^{\phi}$ are unitary. If not the input is invalid.
This requires the following relations
\begin{eqnarray}
g\phi_{\boc}
&=&\sum_{\boR\bos}g\Phi_{\boR\bos}U_{\boR\bos,\boldsymbol{0}\boc}\nonumber\\
&=&\sum_{\boR\bos}\Phi_{g\left(\boR\bos\right)}D_s^{\Phi}\left(\alpha\right)U_{\boR\bos,\boldsymbol{0}\boc}\nonumber\\
&=&\phi_{g\boc}D_c^{\phi}\left(\alpha\right)\nonumber\\
&=&\sum_{\boR\bos}\Phi_{\boR\bos}U_{\boR\bos,g\boc}D_c^{\phi}\left(\alpha\right)\nonumber\\
&=&\sum_{\boR\bos}\Phi_{g\left(\boR\bos\right)}U_{g\left(\boR\bos\right),g\boc}D_c^{\phi}\left(\alpha\right)  \nonumber
\end{eqnarray}
or
\begin{equation}
  U_{\boR\bos,\boldsymbol{0}\boc}=\left(D_s^{\Phi}\left(\alpha\right)\right)^{+}U_{g\left(\boR\bos\right),g\boc}D_c^{\phi}\left(\alpha\right)  \nonumber
\end{equation}
where again we need to replace $g\left(\boR\bos\right)$ by
$\alpha\boR+\boR_{g,\bos},\bos_g$ and $g\boc$ by
$\boR_{g,\boc},\boc_g$.

Additionally, useful operators need to be lattice periodic
\begin{equation}
T_{\boR}\hat{B}\left(\bor\right)T_{\boR}^{-1}=\hat{B}\left(\bor-\boR\right)=\hat{B}\left(\bor\right)  \label{eq:transB}
\end{equation}
and invariant under the space group
\begin{equation}
g\hat{B}\left(\bor\right)g^{-1}=\hat{B}\left(g^{-1}\bor\right)=\hat{B}\left(\bor\right)D^{B}  \left(\alpha\right)\label{eq:gOp}
\end{equation}
where $D^{B}\left(\alpha\right)$ is the representation matrix of the operator. E.g.~$D^{B}\left(\hat{I}\alpha\right)=\pm\alpha$ if $\hat{B}$ is a polar
 ($+\alpha$) or axial ($-\alpha$) vector and $\alpha$ a proper
 rotation.
This does however not apply to the position operator $\hat{\bor}$,
since it is not translational invariant. For all point group
operations which leave the origin invariant
$\hat{\alpha}\hat{\bor}\hat{\alpha}^{-1}=\alpha^{-1}\bor=\bor\alpha$ holds, but for a
general space group operation one gets $g\hat{\bor} g^{-1}=\alpha^{-1}\left(\bor-\boldsymbol{\tau}\right)=\left(\bor-\boldsymbol{\tau}\right)\alpha$.

If time reversal applies it yields
\begin{equation}
\theta\hat{B}\left(\bor\right)\theta^{-1}=\hat{B}\left(\bor\right)
D^{B}\left(\theta\right)\label{eq:TRMatEl}
\end{equation}
where $D^B\left(\theta\right)$ is e.g.~$+1$ for the Hamiltonian and
$-1$ for the spin and magnetic field. In full relativistic spin
polarized mode products of time reversal with operations, which flip
the magnetic field, are group elements. We apply the latter only to the
subset of the actual space group, which preserves the magnetization
axis, i.e.~we do not implement full magnetic groups.

Using the shorthand Eq.~(\ref{eq:gtransLOshort}) with its meaning
Eq.~(\ref{eq:gtransLO}) matrix elements transform as
\begin{eqnarray}
\lefteqn{B_{\boR^{\prime}\bos^{\prime},\boR\bos}
=\left\langle \Phi_{\boR^{\prime}\bos^{\prime}}\mid\hat{B}\mid\Phi_{\boR\bos}\right\rangle
}\nonumber\\
&&=\left\langle g\Phi_{\boR^{\prime}\bos^{\prime}}\mid g\hat{B}g^{-1}\mid g\Phi_{\boR\bos}\right\rangle \nonumber\\
&&=D_{s^{\prime}}^{\Phi+}\left\langle \Phi_{g\left(\boR^{\prime}\bos^{\prime}\right)}\mid\hat{B}\mid\Phi_{g\left(\boR\bos\right)}\right\rangle D_s^{\Phi}D^{B}\nonumber\\
&&=D_{s^{\prime}}^{\Phi+}B_{g\left(\boR^{\prime}\bos^{\prime}\right),g\left(\boR\bos\right)}D_s^{\Phi}D^{B}  
\label{eq:gMatElRealSpace}
\end{eqnarray}
where $D^B$ applies to the degrees of freedom  of the internal tensor
structure of $B$.

With the general behavior (details in
Eq.~(\ref{eq:TRangular}))
\begin{subequations}
\begin{eqnarray}
\theta \Phi &=& \Phi D^{\Phi\theta}K_0\label{eq:TRLO}\\
\Phi^{+}\theta^{-1}&=&K_{0}D^{\Phi\theta+}\Phi^{+}\label{eq:TRLOplus}
\end{eqnarray}
\end{subequations}
one gets the general behavior of matrix elements under time reversal
\begin{eqnarray}
B_{\boR^{\prime}\bos^{\prime},\boR\bos}
&=&\left\langle \Phi_{\boR^{\prime}\bos^{\prime}}\mid\theta^{-1}\theta\hat{B}\theta^{-1}\theta\mid\Phi_{\boR\bos}\right\rangle \nonumber\\
&=&K_{0}D_{s^{\prime}}^{\Phi\theta+}\left\langle \Phi_{\boR^{\prime}\bos^{\prime}}\mid\hat{B}D^{B\theta}\mid\Phi_{\boR\bos}\right\rangle D_s^{\Phi\theta}K_{0}\nonumber\\
&=&\left(D_{s^{\prime}}^{\Phi\theta+}\hat{B}_{\boR^{\prime}\bos^{\prime},\boR\bos}D_s^{\Phi\theta}D^{B\theta}\right)^{*}.\label{eq:TRMatElRealSpace}
\end{eqnarray}

Now, the transformation of Bloch sums of LOs/WFs is obtained from
Eqs.~(\ref{eq:LOblochsum},\ref{eq:gtransLO}) where the use of
backfolded sites $\bos_g$ is of the essence
\begin{eqnarray}
g\Phi_{\bos}^{\bok}
&=&\frac{1}{\sqrt{N}}\sum_{\boR}\mex^{i\bok\left(\boR+\lambda\bos\right)}g\Phi_{\boR\bos}\nonumber\\
&=&\frac{1}{\sqrt{N}}\sum_{\boR}\mex^{i\bok\left(\boR+\lambda\bos\right)}\Phi_{\alpha\boR+\boR_{g,s},\bos_{g}}D_s^{\Phi}\nonumber\\
&=&\frac{1}{\sqrt{N}}\sum_{\boR}\mex^{i\alpha\bok\left(\alpha\boR+\lambda\alpha\bos\right)}\Phi_{\alpha\boR+\boR_{g,s},\bos_{g}}D_s^{\Phi}\nonumber\\
&=&\frac{1}{\sqrt{N}}\sum_{\boR}\mex^{i\alpha\bok\left(\boR+\lambda\bos_{g}+\lambda\alpha\bos-\boR_{g,s}-\lambda\bos_{g}\right)}\nonumber\\
&&\cdot\Phi_{\boR\bos_{g}}D_s^{\Phi}\nonumber\\
&=&\Phi_{\bos_{g}}^{\alpha\bok}D_s^{\Phi}\mex^{i\alpha\bok\left(\lambda\alpha\bos-\boR_{g,s}-\lambda\bos_{g}\right)}\nonumber\\
&=&\left(\Phi^{\alpha\bok}D^{\Phi\bok}\right)_{\bos}  \label{eq:gBloch}
\end{eqnarray}
with 
\begin{equation}
D_{s^{\prime}s}^{\Phi\bok}=\delta_{s^{\prime}s_g}D_{s}^{\Phi}\mex^{-i\alpha\bok\left(\blambda\left(g\bos-\bos_{g}\right)+\lambda\boldsymbol{\tau}\right)}  \label{eq:DBloch}
\end{equation}

Forming matrix elements of $\hat{B}$ with Bloch sums, inserting pairs
of $gg^{-1}$ and using Eqs.~(\ref{eq:gOp},\ref{eq:gBloch}) yields
\begin{eqnarray}
\lefteqn{\left\langle \Phi_{\bos^{\prime}}^{\bok}\mid\hat{B}\mid\Phi_{\bos}^{\bok}\right\rangle }\nonumber\\
&=&\left\langle g\Phi_{\bos^{\prime}}^{\bok}\mid g\hat{B}g^{-1}\mid g\Phi_{\bos}^{\bok}\right\rangle \\
&=&\mex^{i\alpha\bok\blambda\left(g\bos^{\prime}-\bos_{g}^{\prime}\right)}\cdot\nonumber\\
&&\cdot D_{s^{\prime}}^{\Phi+}\left\langle \Phi_{\bos_{g}^{\prime}}^{\alpha\bok}\mid\hat{B}\mid\Phi_{\bos_{g}}^{\alpha\bok}\right\rangle D_s^{\Phi}D^{B}\cdot\nonumber\\
&&\cdot\mex^{-i\alpha\bok\blambda\left(g\bos-\bos_{g}\right)}
\label{eq:gMatElBloch}
\end{eqnarray}
from this one sees that the relative gauge $\blambda=0$ leads to
simpler expressions, since all the phase factors in
Eq.~(\ref{eq:gMatElBloch})
vanish.

Eqs.~(\ref{eq:LOblochsum},\ref{eq:TRLO}) gives the time reversal of 
the orbital Bloch sums
\begin{eqnarray}
\theta\Phi_{\bos}^{\bok}
&=&\theta\frac{1}{\sqrt{N}}\sum_{\boR}\mex^{i\bok\left(\boR+\lambda\bos\right)}\Phi_{\boR\bos}\nonumber\\
&=&\frac{1}{\sqrt{N}}\sum_{\boR}\mex^{-i\bok\left(\boR+\lambda\bos\right)}\theta\Phi_{\boR\bos}\nonumber\\
&=&\frac{1}{\sqrt{N}}\sum_{\boR}\mex^{-i\bok\left(\boR+\lambda\bos\right)}\Phi_{\boR\bos}D_s^{\Phi\theta}K_{0}\nonumber\\
&=&\Phi_{\bos}^{-\bok}D_s^{\Phi\theta}K_{0}  
\end{eqnarray}
and together with Eqs.~(\ref{eq:TRLOplus},\ref{eq:TRMatEl})
\begin{eqnarray}
\lefteqn{\left\langle \Phi_{\bos^{\prime}}^{\bok}\mid\hat{B}\mid\Phi_{\bos}^{\bok}\right\rangle 
=\left\langle \Phi_{\bos^{\prime}}^{\bok}\mid\theta^{-1}\theta\hat{B}\theta^{-1}\theta\mid\Phi_{\bos}^{\bok}\right\rangle }\nonumber\\
&&=\left(D_{s^{\prime}}^{\Phi\theta+}\left\langle \Phi_{\bos^{\prime}}^{-\bok}\mid\hat{B}\mid\Phi_{\bos}^{-\bok}\right\rangle D_s^{\Phi\theta}D^{B\theta}\right)^{*}  \label{eq:TRMatelbloch}
\end{eqnarray}
If the operation is a product $\theta{}g$ this gets modified by
carefully inserting Eq.~(\ref{eq:gMatElBloch}) for $-\bok$ into
Eq.~(\ref{eq:TRMatelbloch}) which is not spelled out due to the
unwieldy phase factors. Essentially we let
$D^{\Phi}\to{}D^{\Phi}D^{\Phi\theta}$, $D^{B}\to{}D^{B}D^{\Phi\theta}$
and $\alpha\bok\to-\alpha\bok$ in Eq.~(\ref{eq:gMatElBloch})
and apply complex conjugation to the whole rhs.

In order to symmetrize the matrix elements when needed we make use of
the fact that the sum of the rhs. of Eq.~(\ref{eq:gMatElRealSpace}) over
all group operations divided by the number of group elements is a
projector. So, for all pairs of lattice sites we perform this sum over
all transformed pairs. In case if time reversal is contained in the
operation the corresponding formulas Eq.~(\ref{eq:TRMatElRealSpace})
or a mix of this with Eq.~(\ref{eq:gMatElRealSpace}) must be used.

On a final note, in relativistic case the ambiguity of choosing the
rotation or rotation$+2\pi$ as pointgroup part of $g$ cancels out
in all formulas which transform matrix elements, since they are
bilinear in the orbitals.

\subsection{Bloch theorem}\label{sec:blochtheorem}

Eq.~(\ref{eq:gBloch}) can be specialized to pure translations by
using $\alpha=1$, $\boldsymbol{\tau}=\boR$
i.e.~$g=\left\{ E\mid\boR\right\}$ for which $g\bos=\bos+\boR$ and
$D_{s}^{\Phi}\left(\alpha\right)=1$ which gives the Bloch theorem
\begin{equation}
T_{\boR}\Phi_{\bos}^{\bok}=\Phi_{\bos}^{\bok}\mex^{-i\bok\boR}\label{eq:BlochTheorem}
\end{equation}
which can be applied to matrix elements of translational invariant
operators Eq.~(\ref{eq:transB}) between Bloch sums
Eq.~(\ref{eq:LOblochsum}) of different $\bok$-vectors (the most
general case) to get (dropping all non-essential indices)
\begin{eqnarray}
\left\langle \Phi^{\boq}\mid\hat{B}\mid\Phi^{\bok}\right\rangle 
&=&\left\langle T_{\boR}\Phi^{\boq}\mid T_{\boR}\hat{B}T_{\boR}^{-1}\mid T_{\boR}\Phi^{\bok}\right\rangle \nonumber\\
&=&\left\langle \Phi^{\boq}\mid\hat{B}\mid\Phi^{\bok}\right\rangle \mex^{-i\left(\bok-\boq\right)\boR}\label{eq:blochBmat}
\end{eqnarray}
For $\boq\ne\bok$ in the first BZ the phase cannot become
one and hence
\begin{equation}
\left\langle
  \Phi^{\boq}\mid\hat{B}\mid\Phi^{\bok}\right\rangle
=
\delta_{\boq,\bok}\left\langle \Phi^{\bok}\mid\hat{B}\mid\Phi^{\bok}\right\rangle  \label{eq:matelblochtheoram}
\end{equation}
For the position operator which fulfills
\begin{equation}
T_{\boR}\bor T_{\boR}^{-1}=\bor-\boR  
\end{equation}
Eq.~(\ref{eq:blochBmat}) becomes
\begin{eqnarray}
\lefteqn{
\left\langle \Phi^{\boq}\mid\bor\mid\Phi^{\bok}\right\rangle 
}\nonumber\\
&&=\left(\left\langle
    \Phi^{\boq}\mid\bor\mid\Phi^{\bok}\right\rangle
  -\delta_{\boq\bok}S^{\bok}\boR\right)\mex^{-i\left(\bok-\boq\right)\boR}  
\label{eq:rqk}
\end{eqnarray}
which specializes to 
\begin{equation}
\left\langle
    \Phi^{\bok}\mid\bor\mid\Phi^{\bok}\right\rangle=
\left\langle
    \Phi^{\bok}\mid\bor\mid\Phi^{\bok}\right\rangle
  -S^{\bok}\boR \label{eq:rkk}
\end{equation}
for the $\bok$-diagonal terms, which shows that the limit
$\boq\to\bok$ is badly defined and that the operator is
essentially $\bok$-non-diagonal.

The Berry operator Eq.~(\ref{eq:berryoperator}) transforms like the
position operator itself:  
\begin{eqnarray}
  T_{\boR}\bobek T_{\boR}^{-1}
&=&\mex^{i\bok\bor}\mex^{-i\bok\boR}i\nabla_{\bok}\mex^{-i\bok\bor}\mex^{i\bok\boR}\nonumber\\
&=&\mex^{i\bok\bor}i\nabla_{\bok}\mex^{-i\bok\bor}+\mex^{-i\bok\boR}\left(i\nabla_{\bok}\mex^{i\bok\boR}\right)\nonumber\\
&=&\bobek-\boR  \nonumber
\end{eqnarray}
which then gives
\begin{eqnarray}
T_{\boR}\left|\bobek\Phi^{\bok}\right\rangle 
&=&\left|T_{\boR}\bobek{}T_{\boR}^{-1}T_{\boR}\Phi^{\bok}\right\rangle \nonumber\\
&=&\left|\left(\bobek-\boR\right)\Phi^{\bok}\mex^{-i\bok\boR}\right\rangle \nonumber\\
&=&\left|\bobek\Phi^{\bok}\mex^{-i\bok\boR}\right\rangle -\left|\Phi^{\bok}\right\rangle \mex^{-i\bok\boR}\boR  \nonumber
\end{eqnarray}
Applying Eq.~(\ref{eq:betachainrule}) to the first term of the
rhs. yields
\begin{equation}
T_{\boR}\left|\bobek\Phi^{\bok}\right\rangle =\left|\bobek\Phi^{\bok}\right\rangle \mex^{-i\bok\boR}  \label{eq:berryOpPsiBlochTheo}
\end{equation}
and hence Eqs.~(\ref{eq:blochBmat},\ref{eq:matelblochtheoram}) apply
and $\bobek$ is $\bok$-diagonal.

\subsection{Berry connection/curvature}\label{sec:symmetryBerry}

The transformation of the vector valued functions
Eq.~(\ref{eq:vectorbaluedPhi}) is obtained by using
Eq.~(\ref{eq:ginv}) and $gf\left(\bor\right)=f\left(g^{-1}\bor\right)$
as in Eq.~(\ref{eq:transPhi}) which results in
$\left(g^{-1}\bor-\boR-\bos\right)=\left(\bor-\alpha\boR-g\bos\right)\alpha$
for the $\bor$-factor and hence in
\begin{equation}
  g\left(\bor\Phi\right)_{\boR\bos}=\left(\bor\Phi\right)_{g\left(\boR\bos\right)}D_{s}^{\Phi}
  \left(\alpha\right)
  \alpha  \label{eq:gonrPhi}
\end{equation}
where the last $\alpha$ acts on the vector structure. Hence, the
reduced position operator matrix elements
Eq.~(\ref{eq:LOrelative_r_matels}) transform as
Eq.~(\ref{eq:gMatElRealSpace}) with $D^B\to\alpha$, i.e.~as LO matrix
elements in the orbital indices and as a polar vector in the position
operator indices.  Bloch sums of Eq.~(\ref{eq:gonrPhi}) lead to the
transformation of Eq.~(\ref{eq:reducedrblochsums}) according to
\begin{equation}
\left\langle \Phi^{\bok}\mid\left(\bor\Phi\right)^{\bok}\right\rangle =D^{\Phi\bok+}\left\langle \Phi^{\alpha\bok}\mid\left(\bor\Phi\right)^{\alpha\bok}\right\rangle D^{\Phi\bok}\alpha  \label{eq:gtransreducedr}
\end{equation}
with $D^{\Phi\bok}$ from Eq.~(\ref{eq:DBloch}).

Eqs.~(\ref{eq:basiconnectionfromrelativer},\ref{eq:gtransreducedr})
then lead to
\begin{equation}
  \boA_{\Phi}^{\bok}-\blambda S^{\bok}\bos=D^{\Phi\bok+}\left(\boA_{\Phi}^{\alpha\bok}-\blambda S^{\alpha\bok}\bos\right)D^{\Phi\bok}\alpha\label{eq:gtransAminusS}
\end{equation}
where $\bos$ is the diagonal matrix containing the site vectors of the
corresponding orbitals.  By using Eq.~(\ref{eq:DBloch}) and
Eq.~(\ref{eq:gMatElBloch}) with $\hat{B}=1$ this can be transformed
into
\begin{equation}
\boA_{\Phi}^{\bok}=D^{\Phi\bok+}\boA_{\Phi}^{\alpha\bok}D^{\Phi\bok}\alpha+\blambda S^{\bok}\left(\bos-\bos_{g}\alpha\right)  \label{eq:gtransAPhi}
\end{equation}
From this it is clear that for WFs
($S^{\bok}=1$) approximation Eq.~(\ref{eq:approxA}) fulfills the
transformation law Eq.~(\ref{eq:gtransAminusS}) trivially while
setting the basis connection to zero $\boA_{\Phi}^{\bok}=0$ will in
general not.

Acting with $g$ on $\bor$ in Eq.~(\ref{eq:berryoperator}) and using
$\nabla_{\bok}=\nabla_{\alpha\bok}\alpha$ we get
\begin{equation}
  g\bobek g^{-1}=\left(\boldsymbol{\beta}_{\alpha\bok}-\boldsymbol{\tau}\right)\alpha\nonumber
\end{equation}
which allows to derive the general transformation properties of
$\boA_{\Phi}^{\bok}$ directly when assuming a general law
$g\Phi^{\bok}=\Phi^{\alpha\bok}D^{\Phi\bok}$ with some representation
matrices $D^{\Phi\bok}$ 
\begin{eqnarray}
\boA_{\Phi}^{\bok}
&=&\left\langle \Phi^{\bok}\mid\bobek\Phi^{\bok}\right\rangle
    \nonumber\\
&=&\left\langle
    \Phi^{\alpha\bok}D^{\Phi\bok}\mid\left(\boldsymbol{\beta}_{\alpha\bok}
-\boldsymbol{\tau}\right)\alpha\Phi^{\alpha\bok}D^{\Phi\bok}\right\rangle
    \nonumber\\
&=&\left\langle
    \Phi^{\alpha\bok}D^{\Phi\bok}\mid\left(\boldsymbol{\beta}_{\alpha\bok}\alpha\Phi^{\alpha\bok}\right)D^{\Phi\bok}+\Phi^{\alpha\bok}i\nabla_{\alpha\bok}\alpha
    D^{\Phi\bok}\right\rangle \nonumber\\
&&-S_{\Phi}^{\bok}\boldsymbol{\tau}\alpha\nonumber\\
&=&\left\langle
    \Phi^{\alpha\bok}D^{\Phi\bok}\mid\boldsymbol{\beta}_{\alpha\bok}\alpha\Phi^{\alpha\bok}
\right\rangle
    D^{\Phi\bok}+D^{\Phi\bok+}S_{\Phi}^{\alpha\bok}i\nabla_{\bok}D^{\Phi\bok}\nonumber\\
&&-S_{\Phi}^{\bok}\boldsymbol{\tau}\alpha\nonumber\\
&=&D^{\Phi\bok+}\left(\boA_{\Phi}^{\alpha\bok}\alpha+S_{\Phi}^{\alpha\bok}i\nabla_{\bok}\right)D^{\Phi\bok}-S_{\Phi}^{\bok}\boldsymbol{\tau}\alpha   \label{eq:gtransAPsi}
\end{eqnarray}
which specializes to Eq.~(\ref{eq:gtransAPhi}) by
inserting the gradient of Eq.~(\ref{eq:DBloch}), in other words the local
basis connection transforms as a general Berry connection matrix
would.

Now, we introduce the basis change $\Psi^{\bok}=w^{\bok}C^{\bok}$,
which leads to the Berry connection matrix Eq.~(\ref{eq:Aaux}). Then
we can show that if $\boA_{w}^{\bok}$ transforms as
Eq.~(\ref{eq:gtransAPsi}) with $\Phi\to{}w$, also $\boA_{\Psi}^{\bok}$
(Eq.~(\ref{eq:Aaux})) transforms like Eq.~(\ref{eq:gtransAPsi}) with
$\Phi\to{}\Psi$.  Using the abbreviations $b^{\bok}\to{}b$,
$b^{\alpha\bok}\to{}b^{\alpha}$ and
$C^{+}b C\to{}\left\langle{}b\right\rangle_{C}$ one gets the
transformation properties of $\Psi$, $w$ and $C$
\begin{eqnarray}
gw
&=&w^{\alpha}D^{w}\nonumber\\
\Psi&=&wC\nonumber\\
g\Psi&=&\Psi^{\alpha}D^{\Psi}\nonumber\\
D^{w}C&=&C^{\alpha}D^{\Psi}  \nonumber
\end{eqnarray}
and
\begin{eqnarray}
\left\langle S_{w}^{\alpha}i\nabla\right\rangle _{D^{w}C}
&=&\left\langle \left\langle S_{w}^{\alpha}i\nabla\right\rangle
    _{D^{w}}\right\rangle _{C}+\left\langle \left\langle
    S_{w}^{\alpha}\right\rangle _{D^{w}}i\nabla\right\rangle _{C}\nonumber\\
&=&\left\langle \left\langle S_{w}^{\alpha}i\nabla\right\rangle
    _{D^{w}}\right\rangle _{C}+\left\langle S_{w}i\nabla\right\rangle
    _{C}  
\end{eqnarray}
which leads to 
\begin{eqnarray}
\boA_{w} 
&=& \left\langle \boA_{w}^{\alpha}\alpha\right\rangle
    _{D^{w}}+\left\langle S_{w}^{\alpha}i\nabla\right\rangle
    _{D^{w}}-S_{w}\boldsymbol{\tau}\alpha
\text{ Eq.~(\ref{eq:gtransAPsi})}
  \nonumber \\
\boA_{\Psi}
&=&\left\langle \boA_{w}\right\rangle _{C}+\left\langle
    S_{w}i\nabla\right\rangle _{C}
\text{ Eq.~(\ref{eq:Aaux})}
\nonumber\\
&=&\left\langle \left\langle \boA_{w}^{\alpha}\alpha\right\rangle
    _{D^{w}}\right\rangle _{C}+\left\langle \left\langle
    S_{w}^{\alpha}i\nabla\right\rangle _{D^{w}}\right\rangle
    _{C}-S_{\Psi}\boldsymbol{\tau}\alpha+\left\langle
    S_{w}i\nabla\right\rangle _{C}\nonumber\\
&=&\left\langle \boA_{w}^{\alpha}\alpha\right\rangle
   _{D^{w}C}+\left\langle S_{w}^{\alpha}i\nabla\right\rangle
   _{D^{w}C}-S_{\Psi}\boldsymbol{\tau}\alpha\nonumber\\
&=&\left\langle \boA_{w}^{\alpha}\alpha\right\rangle
   _{C^{\alpha}D^{\Psi}}+\left\langle
   S_{w}^{\alpha}i\nabla\right\rangle
   _{C^{\alpha}D^{\Psi}}-S_{\Psi}\boldsymbol{\tau}\alpha\nonumber\\
&=&\left\langle \boA_{w}^{\alpha}\alpha\right\rangle
   _{C^{\alpha}D^{\Psi}}+\left\langle \left\langle
   S_{w}^{\alpha}i\nabla\right\rangle _{C^{\alpha}}\right\rangle
   _{D^{\Psi}}+\left\langle \left\langle S_{w}^{\alpha}\right\rangle
   _{C^{\alpha}}i\nabla\right\rangle
   _{D^{\Psi}}\nonumber\\
&&-S_{\Psi}\boldsymbol{\tau}\alpha\nonumber\\
&=&\left\langle \boA_{\Psi}^{\alpha}\alpha\right\rangle
   _{D^{\Psi}}+\left\langle S_{\Psi}^{\alpha}i\nabla\right\rangle
   _{D^{\Psi}}-S_{\Psi}\boldsymbol{\tau}\alpha\label{eq:symtransApsi}
\end{eqnarray}
Of course, usually Wannier and eigenfunctions are orthonormal, which
means $S_{w,\Psi}=1$.
Eq.~(\ref{eq:symtransApsi}) explicitly shows that the basis change
again leads to the generic transformation law Eq.~(\ref{eq:gtransAPsi}).
All we need to show now is that this law leads to the proper
transformation behavior of the curvature.

We start with the cross-product term, which we write in it's most
general form (including the correction for non-orthonormal $\Psi$)
using Eq.~(\ref{eq:symtransApsi})
\begin{eqnarray}
\lefteqn{i\boA_{\Psi}^{+}\frac{1}{S_{\Psi}}\times\boA_{\Psi}}\nonumber\\
&&=
\left(\boldsymbol{a}-S_{\Psi}\boldsymbol{\tau}\alpha\right)^{+}\frac{1}{S_{\Psi}}\times\left(\boldsymbol{a}-S_{\Psi}\boldsymbol{\tau}\alpha\right)
\end{eqnarray}
with
$\boldsymbol{a}=\boA_{\Psi}+S_{\Psi}\boldsymbol{\tau}\alpha
=\left\langle \boA_{\Psi}^{\alpha}\alpha\right\rangle
   _{D^{\Psi}}+\left\langle S_{\Psi}^{\alpha}i\nabla\right\rangle$.
Using the fact that $\boldsymbol{\tau}\alpha$ is constant diagonal in
matrix space and hence commutes with any matrix one gets
$\left(\boldsymbol{\tau}\alpha\right)
\times S_{\Psi}\left(\boldsymbol{\tau}\alpha\right)=0$ and
$\boldsymbol{a}\times\left(\boldsymbol{\tau}\alpha\right)
+\left(\boldsymbol{\tau}\alpha\right)\times\boldsymbol{a}=0$.
Furthermore, Eq.~(\ref{eq:Akhermitianconj}) gives
$\boldsymbol{a}^{+}=\boldsymbol{a}-i\nabla S_{\Psi}$ and hence
$\boldsymbol{a}^{+}\times\boldsymbol{\tau}\alpha+\boldsymbol{\tau}\alpha\times\boldsymbol{a}=-\left(i\nabla
  S_{\Psi}\right)\times\boldsymbol{\tau}\alpha$.
which ultimately yields
\begin{equation}
i\boA_{\Psi}^{+}\frac{1}{S_{\Psi}}\times\boA_{\Psi}=i\boldsymbol{a}^{+}\frac{1}{S_{\Psi}}\times\boldsymbol{a}-\left(\nabla
  S_{\Psi}\right)\times\left(\boldsymbol{\tau}\alpha\right)\label{eq:transiAtimesA}
\end{equation}

Next the transformation of $\Psi$ yields
$S_{\Psi}=D^{\Psi+}S_{\Psi}^{\alpha}D^{\Psi}$ and since all matrices are
invertable also
\begin{equation}
1=D^{\Psi}\frac{1}{S_{\Psi}}D^{\Psi+}S_{\Psi}^{\alpha} \nonumber
\end{equation}
which by insertion gives
\begin{eqnarray}
i\nabla D^{\Psi}
&=&D^{\Psi}\frac{1}{S_{\Psi}}D^{\Psi+}S_{\Psi}^{\alpha}i\nabla
    D^{\Psi}\nonumber\\
&=&D^{\Psi}\frac{1}{S_{\Psi}}\left\langle S_{\Psi}^{\alpha}i\nabla\right\rangle _{D^{\Psi}}  \label{eq:igradDPsi}
\end{eqnarray}
Furthermore, for any two vector valued quantities
$\left(\boldsymbol{b}\alpha\right)\times\left(\boldsymbol{c}\alpha\right)=\left(\boldsymbol{b}\times\boldsymbol{c}\right)\overline{\alpha}$
holds, where $\overline{\alpha} =I \alpha$ if the point group
operation is improper and $\alpha$ otherwise. The gradient transforms
as $\nabla_{\alpha\bok}\alpha=\nabla_{\bok}$, which leads to
$\nabla\times\boA_{\Psi}^{\alpha}\alpha=\left(\nabla_{\alpha}\times\boA_{\Psi}^{\alpha}\right)\overline{\alpha}$
and with Eqs.~(\ref{eq:igradDPsi})
\begin{eqnarray}
\lefteqn{\nabla\times\left\langle \boA_{\Psi}^{\alpha}\alpha\right\rangle
  _{D^{\Psi}}}\nonumber\\
&&=\left\langle
   \left(\nabla_{\alpha}\times\boA_{\Psi}^{\alpha}\right)\overline{\alpha}\right\rangle
   _{D^{\Psi}}+i\left\langle S_{\Psi}^{\alpha}i\nabla\right\rangle
   _{D^{\Psi}}^{+}\frac{1}{S_{\Psi}}\times\left\langle
   \boA_{\Psi}^{\alpha}\alpha\right\rangle _{D^{\Psi}}\nonumber\\
&&+i\left\langle \boA_{\Psi}^{\alpha}\alpha\right\rangle
   _{D^{\Psi}}\frac{1}{S_{\Psi}}\times\left\langle
   S_{\Psi}^{\alpha}i\nabla\right\rangle _{D^{\Psi}}  
\label{eq:curlAalpha}
\end{eqnarray}
Also, by differentiation and Eqs.~(\ref{eq:igradDPsi})
\begin{eqnarray}
\nabla\times\left\langle S_{\Psi}^{\alpha}i\nabla\right\rangle
  _{D^{\Psi}}\nonumber\\
&=&i\left(\left\langle
                 S_{\Psi}^{\alpha}i\nabla\right\rangle
                 _{D^{\Psi}}^{+}-\left\langle \left(i\nabla
                 S_{\Psi}^{\alpha}\right)\right\rangle
                 _{D^{\Psi}}\right)\frac{1}{S_{\Psi}}\nonumber\\
&&\times\left\langle
                 S_{\Psi}^{\alpha}i\nabla\right\rangle _{D^{\Psi}}\label{eq:curlSifrad}
\end{eqnarray}
Eqs.~(\ref{eq:Akhermitianconj},\ref{eq:curlAalpha},\ref{eq:curlSifrad})
then yield the curl of Eq.~(\ref{eq:symtransApsi}) after some term-sorting
\begin{eqnarray}
\nabla\times\boA_{\Psi}
&=&\left\langle
    \left(\nabla_{\alpha}\times\boA_{\Psi}^{\alpha}\right)\overline{\alpha}\right\rangle
    _{D^{\Psi}}\nonumber\\
&&-i\left\langle \boA_{\Psi}^{\alpha}\alpha\right\rangle^{+}
    _{D^{\Psi}}\frac{1}{S_{\Psi}}\times\left\langle
    \boA_{\Psi}^{\alpha}\alpha\right\rangle _{D^{\Psi}}\nonumber\\
&&+i\boldsymbol{a}^{+}\frac{1}{S_{\Psi}}\times\boldsymbol{a}-\left(\nabla S_{\Psi}\right)\times\left(\boldsymbol{\tau}\alpha\right)  
\end{eqnarray}
which together with Eq.~(\ref{eq:transiAtimesA}) results in 
\begin{eqnarray}
\nabla\times\boA_{\Psi}
&=&\left\langle
    \left(\nabla_{\alpha}\times\boA_{\Psi}^{\alpha}\right)\overline{\alpha}\right\rangle
    _{D^{\Psi}}\nonumber\\
&&-i\left\langle \boA_{\Psi}^{\alpha}\alpha\right\rangle^{+}
   _{D^{\Psi}}\frac{1}{S_{\Psi}}\times\left\langle
   \boA_{\Psi}^{\alpha}\alpha\right\rangle _{D^{\Psi}}\nonumber\\
&&+i\boA_{\Psi}^{+}\frac{1}{S_{\Psi}}\times\boA_{\Psi}  \label{eq:transcurlA}
\end{eqnarray}
This is the most general case. 

If $\Psi$ is orthonormal and an eigenbasis the representation matrices
$D^{\Psi}$ are now unitary $D^{\Psi+}D^{\Psi}=1$ and block diagonal
with blocks for each degenerate subspace and hence the subspace
projector $P$ commutes $\left[D^{\Psi},P\right]_{-}=0$ and hence
\begin{eqnarray}
Q\left\langle i\nabla\right\rangle _{D^{\Psi}}P
&=&QD^{\Psi+}i\nabla D^{\Psi}P\nonumber\\
&=&D^{\Psi+}iQP\nabla D^{\Psi}\nonumber\\
&=&0  
\end{eqnarray}
and
$Q\boldsymbol{a}P=Q\left\langle
  \boA_{\Psi}^{\alpha}\alpha\right\rangle _{D^{\Psi}}P$
and together Eq.~(\ref{eq:transiAtimesA}) specializes to
\begin{eqnarray}
\lefteqn{iP\boA_{\Psi}Q\times Q\boA_{\Psi}P}\nonumber\\
&&=iP\left\langle \boA_{\Psi}^{\alpha}\alpha\right\rangle
   _{D^{\Psi}}Q\times Q\left\langle
   \boA_{\Psi}^{\alpha}\alpha\right\rangle _{D^{\Psi}}P\nonumber\\
&&=i\left\langle P\boA_{\Psi}^{\alpha}\alpha Q\times
   Q\boA_{\Psi}^{\alpha}\alpha P\right\rangle _{D^{\Psi}}\nonumber\\
&&=i\left\langle P\boA_{\Psi}^{\alpha}Q\times Q\boA_{\Psi}^{\alpha}P\right\rangle _{D^{\Psi}}\overline{\alpha}
\end{eqnarray}
which shows that the $A\times{}A$-term of the non-Abelian Berry
curvature alone  transforms as a matrix in orbital indices and as pseudo
vector in vector indices.
Subtracting $iPAP\times{}PAP$ from Eq.~(\ref{eq:transcurlA}) one
finally gets for the full non-Abelian Berry curvature
$P\boF_{nA}P=P\left(\nabla\times\boA_{\Psi}-i\boA_{\Psi}P\times
  P\boA_{\Psi}\right)P$
the expression
\begin{equation}
\boF_{\Psi nA}^{\bok}=D^{\Psi\bok+}\boF_{\Psi nA}^{\alpha\bok}D^{\Psi\bok}\overline{\alpha}
\end{equation}
Now, since both the full curvature as well as the $A\times{}A$-term
transform the same proper way, so also the difference
$\left\langle \boldsymbol{f}\right\rangle _{C}-i\left\langle
  \boA\right\rangle _{C}Q_i\times{}Q_i\left\langle \boA\right\rangle
_{C}$ in Eq.~(\ref{eq:BerryCurvature}) must transform this way.

 The discussion of this section shows
that the approximation Eq.~(\ref{eq:approxA}) leads to proper behavior
of the connection and hence curvature and consequently, that the
leading term
$\left\langle Si\nabla\right\rangle\times\left\langle
  Si\nabla\right\rangle$
alone cannot transform properly in the periodic gauge.

\subsection{Bloch sum gauge invariance}\label{sec:blochsumgaugeinvariance}
We show the influence of the Bloch sum gauge choice ($\lambda$) on
various expressions used in the main text. From
Eq.~(\ref{eq:LOblochsum}) with $\Phi\to{}w$ one gets
\begin{eqnarray}
w^{\bok\lambda}
&=&w^{\bok\overline{\lambda}}\Lambda^{\bok}\nonumber\\
\Lambda^{\bok}
&=&\mex^{i\bok\bos\left(\lambda-\overline{\lambda}\right)}\label{eq:defLambda}\nonumber\\
0
&=&\left[\Lambda^{\bok},\bos\right]_{-}  \label{eq:comutLs}\\
S^{\bok\lambda}
&=&\Lambda^{-\bok}S^{\bok\overline{\lambda}}\Lambda^{\bok}\label{eq:gaugetransS}
\end{eqnarray}
with $\overline{\lambda}=1-\lambda$ and
$\left(\lambda-\overline{\lambda}\right)^{2}=1$. Together with
$\Psi^{\bok}=w^{\bok}C^{\bok}=w^{\bok\lambda}C^{\bok\lambda}$ this
leads to 
\begin{subequations}
\label{eq:gaugetransCS}
\begin{eqnarray}
C^{\bok\overline{\lambda}}
&=&\Lambda^{\bok}C^{\bok\lambda}\label{eq:gaugetransC}\\
1&=&C^{\bok\lambda+}S^{\bok\lambda}C^{\bok\lambda}\\
&=&C^{\bok\overline{\lambda}+}S^{\bok\overline{\lambda}}C^{\bok\overline{\lambda}}
\end{eqnarray}
\end{subequations}
and altogether to
\begin{eqnarray}
\boA_{w}^{\bok\lambda}&=&\Lambda^{-\bok}\left\langle
                   w^{\bok\overline{\lambda}}\mid\bobek
                   w^{\bok\overline{\lambda}}\Lambda^{\bok}\right\rangle
  \nonumber\\
&=&\Lambda^{-\bok}\left\langle
        w^{\bok\overline{\lambda}}\mid\bobek
        w^{\bok\overline{\lambda}}\right\rangle
        \Lambda^{\bok}\nonumber\\
&&+\Lambda^{-\bok}\left\langle
        w^{\bok\overline{\lambda}}\mid
        w^{\bok\overline{\lambda}}\right\rangle
        \left(i\nabla_{\bok}\Lambda^{\bok}\right)\nonumber\\
&=&\Lambda^{-\bok}\boA_{w}^{\bok\overline{\lambda}}\Lambda^{\bok}-\left(\lambda-\overline{\lambda}\right)\Lambda^{-\bok}S_{w}^{\bok\overline{\lambda}}\Lambda^{\bok}\bos\nonumber\\
&=&\Lambda^{-\bok}\boA_{w}^{\bok\overline{\lambda}}\Lambda^{\bok}-\left(\lambda-\overline{\lambda}\right)S_{w}^{\bok\lambda}\bos \label{eq:gaugetransAw}
\end{eqnarray}
so, the basis connection does not transform like a simple operator
(as the overlap ~Eq.~(\ref{eq:gaugetransS})).
Eqs.~(\ref{eq:comutLs},\ref{eq:gaugetransCS},\ref{eq:gaugetransAw}) then give
\begin{eqnarray}
\left\langle \boA\right\rangle _{C}^{\lambda}
&=&C^{\bok\lambda+}\boA_{w}^{\bok\lambda}C^{\bok\lambda}\nonumber\\
&=&C^{\bok\lambda+}\left[\Lambda^{-\bok}\boA_{w}^{\bok\overline{\lambda}}\Lambda^{\bok}-\left(\lambda-\overline{\lambda}\right)S_{w}^{\bok\lambda}\bos\right]C^{\bok\lambda}\nonumber\\
&=&C^{\bok\overline{\lambda}+}\boA_{w}^{\bok\overline{\lambda}}C^{\bok\overline{\lambda}}-\left(\lambda-\overline{\lambda}\right)C^{\bok\lambda+}S_{w}^{\bok\lambda}\bos
    C^{\bok\lambda}\nonumber\\
&=&\left\langle \boA_{w}\right\rangle _{C}^{\overline{\lambda}}-\left(\lambda-\overline{\lambda}\right)\left\langle S\bos\right\rangle _{C}^{\lambda}\label{eq:gaugetransAC}
\end{eqnarray}
and in a similar way
\begin{eqnarray}
\left\langle Si\nabla\right\rangle _{C}^{\lambda}
&=&C^{\bok\lambda+}S^{\bok\lambda}i\nabla_{\bok}C^{\bok\lambda}\nonumber\\
&=&C^{\bok\lambda+}S^{\bok\lambda}i\nabla_{\bok}\Lambda^{\bok+}C^{\bok\overline{\lambda}}\nonumber\\
&=&C^{\bok\lambda+}\left[S^{\bok\lambda}\Lambda^{\bok+}i\nabla_{\bok}+S^{\bok\lambda}i\left(\nabla_{\bok}\Lambda^{\bok+}\right)\right]C^{\bok\overline{\lambda}}\nonumber\\
&=&\left\langle Si\nabla\right\rangle _{C}^{\overline{\lambda}}+\left(\lambda-\overline{\lambda}\right)\left\langle S\bos\right\rangle _{C}^{\lambda}  
\end{eqnarray}
Consequently, Eqs.~(\ref{eq:defAC},\ref{eq:defSgradC}) are not
invariant under gauge change of the Bloch sums, even though they are
expressions in the eigenbasis. However, their sum and hence the Berry
connection matrix in eigenbasis Eq.~(\ref{eq:Aaux}) and the
$\boA\times\boA$-term in Eq.~(\ref{eq:BerryCurvature}) are invariant.
In the following we drop the $\bok$-index for simplification.
The gradient of Eq.~(\ref{eq:gaugetransS}) then reads
\begin{equation}
\nabla S^{\lambda}=\Lambda^{+}\left(\nabla S^{\overline{\lambda}}+i\left(\lambda-\overline{\lambda}\right)\left[S^{\overline{\lambda}},\bos\right]_{-}\right)\Lambda  
\end{equation}
The curl of the basis connection together with
Eqs.~(\ref{eq:gaugetransAw},\ref{eq:Akhermitianconj}) and $S\bos\times\bos=0$ 
then reads 
\begin{eqnarray}
\nabla\times\boA_{w}^{\lambda}
&=&\nabla\times\left(\Lambda^{+}\boA_{w}^{\overline{\lambda}}\Lambda\right)-\left(\lambda-\overline{\lambda}\right)\left(\nabla
    S_{w}^{\lambda}\right)\times\bos\nonumber\\
&=&\Lambda^{+}\left(\nabla\times\boA_{w}^{\overline{\lambda}}\right)\Lambda  \nonumber\\
&&-i\Lambda^{+}\left(\lambda-\overline{\lambda}\right)\left(\bos\times\boA_{w}^{\overline{\lambda}}+\boA_{w}^{\overline{\lambda}+}\times\bos\right)\Lambda  \nonumber\\
&&+\Lambda^{+}\left(i\bos S^{\overline{\lambda}}\times\bos\right)\Lambda  
\end{eqnarray}
which bracketed between $C^{\bok}$, by inserting $CC^{+}S=1$ where
needed and using Eq.~(\ref{eq:gaugetransC}), gives
\begin{eqnarray}
\left\langle \nabla\times\boA_{w}\right\rangle
  _{C}^{\lambda}
  &=&\left\langle
                    \nabla\times\boA_{w}\right\rangle
                    _{C}^{\overline{\lambda}}\nonumber\\
  &&-i\left(\lambda-\overline{\lambda}\right)\left\langle\left(\bos\times\boA_{w}+\boA_{w}^{+}\times\bos\right)\right\rangle
                    _{C}^{\overline{\lambda}}\nonumber\\
&&+i\left\langle \bos
   S\times\bos\right\rangle
   _{C}^{\overline{\lambda}}
\end{eqnarray}
Finally,
$\left\langle S\bos\right\rangle _{C}^{\lambda}=\left\langle
  S\bos\right\rangle _{C}^{\overline{\lambda}}$ and application of $CC^{+}S=1$ 
and Eq.~(\ref{eq:gaugetransAC}) yields
\begin{eqnarray}
\left\langle \boA\right\rangle _{C}^{\lambda+}\times\left\langle
  \boA\right\rangle _{C}^{\lambda}
&=&\left\langle \boA_{w}\right\rangle
    _{C}^{\overline{\lambda}+}\times\left\langle \boA_{w}\right\rangle
    _{C}^{\overline{\lambda}}\nonumber\\
&&-\left(\lambda-\overline{\lambda}\right)\left\langle
    \bos\times\boA_{w}+\boA_{w}^{+}\times\bos\right\rangle
    _{C}^{\overline{\lambda}}\nonumber\\
&&+\left\langle \bos
    S\times\bos\right\rangle _{C}^{\overline{\lambda}}
\end{eqnarray}
from which follows
\begin{eqnarray}
\lefteqn{\left\langle \boldsymbol{f}\right\rangle _{C}^{\lambda}-i\left\langle
  \boA_{w}\right\rangle _{C}^{\lambda+}\times\left\langle
  \boA_{w}\right\rangle _{C}^{\lambda}}\nonumber\\
&&=\left\langle
  \boldsymbol{f}\right\rangle _{C}^{\overline{\lambda}}-i\left\langle
  \boA_{w}\right\rangle _{C}^{\overline{\lambda}+}\times\left\langle
  \boA_{w}\right\rangle _{C}^{\overline{\lambda}}
\end{eqnarray}
and hence the full Berry connection Eq.~(\ref{eq:BerryCurvature})
is invariant too.

\section{Crystal structures}\label{sec:structures}

\subsection{CaCuO2}\label{sec:CaCuO2}
CaCuO2 forms a tetragonal lattice with spacegroup 123 (P4/mmm). We
used lattice parameters $a_0=7.29434$$a_{\text{B}}$ and
$c_0=6.04712$$a_{\text{B}}$ with Wyckoff: positions Ca at
$1d=\left(\frac{1}{2}\frac{1}{2}\frac{1}{2}\right)$, Cu at
$1a=\left(000\right)$ and O at $2f=\left(\frac{1}{2}00\right)$.  We
performed a non spin polarized calculation in scalar relativistic mode
within the local (spin) density approximation
L(S)DA\cite{PerdewWang92} (PW92) in FPLO version 19.00-60.  The
self-consistent $\bok$-mesh contains $12^3$ points in the primitive
reciprocal unit cell.

\subsection{bcc Fe}\label{sec:bccFe}
We performed a full relativistic spin polarized calculation for bcc
iron with space group 229 (Im$\bar3$m), Wyckoff position $2a=(000)$
and lattice parameter $a_0=5.4$$a_{\text{B}}$ in FPLO version 19.00-60.  The
exchange and correlation functional is LSDA (PW92) and the
magnetization axis is (001) with resulting spin moment of
2.19$\mu_\mathrm{B}$. The self consistent $\bok$-mesh contains $16^3$
points in the primitive reciprocal unit cell.

\subsection{B2 FeAl}\label{sec:B2FeAl}
We performed a full relativistic spin polarized calculation for B2
FeAl with space group 221 (Pm$\bar3$m) with Fe at Wyckoff position
$1a=(000)$ and Al at $1b=(\frac{1}{2}\frac{1}{2}\frac{1}{2})$
and lattice parameter $a_0=5.364$$a_{\text{B}}$ in FPLO version 19.00-60.  The
exchange and correlation functional is LSDA (PW92) and the
magnetization axis is (001) with resulting spin moment of
0.662$\mu_\mathrm{B}$. The self consistent $\bok$-mesh contains $12^3$
points in the primitive reciprocal unit cell.

\subsection{HgS}\label{sec:HgS}
We performed a full relativistic non spin polarized calculation for
HgS with space group 216 (F$\bar4$3m)
with Hg at Wyckoff position $4a=(000)$
and S at $4c=(\frac{1}{4}\frac{1}{4}\frac{1}{4})$
and lattice parameter $a_0=5.85$$a_{\text{B}}$
in FPLO version 19.00-60.  The exchange and correlation functional is
LSDA (PW92). The self consistent $\bok$-mesh contains $12^3$ points in
the primitive reciprocal unit cell.

\subsection{MgB$_2$}
We performed a scalar relativistic non spin polarized calculation for
MgB$_2$ with space group 191 (P6/mmm)
with Mg at Wyckoff position $1a=(000)$
and B at $2d=(\frac{1}{3}\frac{2}{3}\frac{1}{2})$
and lattice parameters $a_0=3.078$$a_{\text{B}}$, $c_0=3.552$$a_{\text{B}}$
in FPLO version 19.00-60.  The exchange and correlation functional is
LSDA (PW92). The self consistent $\bok$-mesh contains $12^3$ points in
the primitive reciprocal unit cell.

\subsection{H$_2$O}\label{sec:H2O}
We performed a scalar relativistic non spin polarized calculation for
H$_2$O
in the group Pmm2 
with H at Wyckoff position $2e=(1.457,0,1.127)a_{\text{B}}$ and O at 
$1a=(000)$ in FPLO version 19.00-60 within LDA (PW92).

\subsection{Related citations}\label{sec:relcite}

For the interested reader we give an (incomplete) list of recent publications,
which used the FPLO Wannier function module:
\cite{Thi21}
\cite{Rie19}
\cite{Mazin19}
\cite{Iako19}
\cite{Mash19}
\cite{Shim20}
\cite{Shimi18}
\cite{Nawa17}
\cite{Shi21}
\cite{guin20212d}
\cite{le2020ab}
\cite{zhang2020spin}
\cite{yao2020observation}
\cite{xu2020descriptor}

\protect\vspace*{\fill}



\bibliography{main_text}

\end{document}